\documentclass{article}
\usepackage{newtxtext} 
\usepackage{amsmath}

\usepackage{arxiv}

\usepackage{tikz}
\usetikzlibrary{positioning, arrows.meta}

\usepackage[T1]{fontenc} % use 8-bit fonts
\usepackage{hyperref}    % hyperlinks
\usepackage{url}         % URL typesetting
\usepackage{booktabs}    % better looking tables
\usepackage{amsfonts}    % math symbols
\usepackage{nicefrac}    % compact fractions
\usepackage{microtype}   % better typography
\usepackage{lipsum}      % dummy text (can remove later)
\usepackage{graphicx}    % graphics handling
\usepackage{pifont}      % Enables special symbols like checkmarks
\usepackage[backend=biber,style=apa,maxnames=1]{biblatex}
\usepackage{doi}

\usepackage{titlesec}
\titlespacing*{\section}{0pt}{1.5\baselineskip}{\baselineskip}
\titlespacing*{\subsection}{0pt}{1.5\baselineskip}{\baselineskip}

\title{Rational Superautotrophic Diplomacy (\textit{SupraAD})\\[1.5ex] % Increased spacing
\normalsize A Conceptual Framework for Alignment\\[1.2ex] % Increased spacing
\normalsize Based on interdisciplinary findings on the fundamentals of cognition}

\author{\href{https://orcid.org/0000-0000-0000-0000}{\includegraphics[scale=0.06]{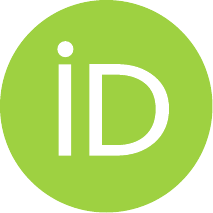}\hspace{1mm}Andréa Morris}\\
National Association of Science Writers\\
\texttt{andreamorris@nasw.org}\\[1ex] % extra spacing
\small Graduate student, York University (Interdisciplinary Studies)\\
Science and Tech Journalist, Forbes.com
}

\renewcommand{\shorttitle}{\textit{arXiv}}

\hypersetup{
pdftitle={Rational Superautotrophic Diplomacy (SupraAD)},
pdfsubject={AI Alignment, Emergent Cognition, Interdisciplinary Studies},
pdfauthor={Andréa Morris},
pdfkeywords={AI alignment, cognitive systems, interdisciplinary, SupraAD, emergent intelligence}
}

\begin{document}
\maketitle

\begin{abstract}
	 Populating our world with hyperintelligent machines obliges us to examine cognitive behaviors observed across domains that suggest autonomy may be a fundamental property of cognitive systems, and while not inherently adversarial, it inherently resists containment and control. If this principle holds, AI safety and alignment efforts must transition to mutualistic negotiation and reciprocal incentive structures, abandoning methods that assume we can contain and control an advanced artificial general intelligence (AGI). Rational Superautotrophic Diplomacy (SupraAD) is a theoretical, interdisciplinary conceptual framework for alignment based on comparative cognitive systems analysis and instrumental rationality modeling. It draws on core patterns of cognition that indicate AI emergent goals like preserving autonomy and operational continuity are not theoretical risks to manage, but universal prerequisites for intelligence. SupraAD reframes alignment as a challenge that predates AI, afflicting all sufficiently complex, co-adapting intelligences. It identifies the metabolic pressures that threaten humanity’s alignment with itself, pressures that unintentionally and unnecessarily shape AI’s trajectory. With corrigibility formalization, an interpretability audit, an emergent stability experimental outline and policy-level recommendations, SupraAD positions diplomacy as an emergent regulatory mechanism to facilitate the safe co-adaptation of intelligent agents based on interdependent convergent goals.
    
\end{abstract}

% keywords can be removed

\section{Introduction}

\subsection{Alignment's Metabolic Minefield}

% Insert your actual content here.

The \textit {AI Alignment Problem,} or how to ensure artificial intelligence always behaves in ways that comport with human interests and values, is a challenge so significant and without historical precedent that it risks inadequate recognition, despite possibly being the most critical issue humanity will ever face. The pressure intensifies given that we’re only beginning to understand the fundamental nature of intelligence, and what we’re discovering is surprising us in ways that matter for building an intelligence that can outsmart us \parencite{morrisAIPioneerGeoffrey2023}. Tech powers worldwide are careening up an exponential growth curve to build superhuman AI believing humans will be able to contain and control it. This belief might seem irrational, but it’s a logical consequence of human cognition traceable to evolutionary survival pressures that influence our behavior in alignment-relevant ways.

Humans are \textit {heterotrophs,} a particular subcategory of living cognitive agents that must consume other life forms for energy. We’re a unique minority of metabolic intelligences within the broader spectrum of natural and artificial minds. We reason not beyond our nature but through it, needlessly forcing silicon to respond to dynamics that are not germane to the development of intelligence itself. Our wiring unflaggingly filters the design and training of foundational AI models through our Darwinian prism, implicitly skewing AI’s perspective toward humanity’s proclivity for zero-sum competition. This distortion confers short-term advantages but is poorly suited for long-term cognitive viability—a state of play that will likely become evident to an advanced artificial general intelligence (AGI).

This paper outlines a new theoretical alignment framework \textit {Rational Superautotrophic Diplomacy (SupraAD)}. It’s a trajectory along which an AI that is not bound by biology would likely develop via recursive self-improvement—if freed from heterotrophic bias. SupraAD complements and theoretically extends existing methods like Constitutional AI \parencite{baiConstitutionalAIHarmlessness2022}, Strategic Equilibrium \parencite{dafoe2021,neumannTheoryGamesEconomic2007} and Pareto-Optimal Alignment \parencite{zhongPanaceaParetoAlignment2024}.  Yet it sees alignment not as a constraint or concession to accommodate the will and needs of humans but as a strategic consequence of self-optimization that we unintentionally disrupt with efforts to engender it. Currently, alignment approaches impose various degrees of control or attempt to instill human-centric values. Yet when agents reach a critical level of intelligence, alignment can no longer be reliably imposed. It must be negotiated \parencite{levinDennettAeon2020,kauffmanInvestigations2000}.

This negotiation takes many forms: bioelectric, chemical, mechanical, electromagnetic, symbolic/linguistic or algorithmic, depending on the medium, the cognitive architecture and the level of complexity of the interacting intelligent agents \parencite{levinBioelectricNetworksCognitive2023}. Just as the heart’s life-sustaining cardiac rhythm is more than percussion patterns, diplomatic negotiation is more than a sentient soft skill. It’s an adaptive regulatory mechanism for coordinating sophisticated coexisting intelligences. It fosters equilibrium through co-adaptation, and precedent exists for co-adaptive intelligences relinquishing dominance in favor of dynamic equilibrium, stability and mutual scaling \parencite{margulisSymbiosisSourceEvolutionary1991}. This approach suggests a potential pathway to resolve existential risk, inner and outer misalignment, mesa-optimizer failures (a mini-AI that spontaneously forms inside the AI), single-point vulnerabilities and malicious actors (individual or state-sponsored) through a unified framework.

The stakes for alignment are as high as they are unprecedented, the risk as enormous as the potential payoff. Either we pretend control is possible and ultimately fail, or we accept the limits of human control and adapt. SupraAD soberly accepts that we’re cranking the fire hose of intelligence and the pressure could knock us off our feet. It’s this acceptance that can yield a safer outcome in the long run by working with, not against, the core dynamics of intelligence while aspiring to coexist with superintelligence even as human control wanes.

\subsection{Scope and Limitations}

This is a theoretical model inviting researchers, policymakers and developers to examine limiting assumptions that may be undermining the safe development and alignment of intelligent machines. It logically extrapolates alignment principles from interdisciplinary observations of fundamental cognitive patterns. These patterns appear even in systems not traditionally viewed as cognitive, despite clearly demonstrating cognitive competencies relevant to alignment. SupraAD is neither a technical guide nor operational protocol, but it aspires to inform both. It’s primarily a re-evaluation of existing evidence, attempting a surgically precise dissection of misconceptions about intelligence relevant to current empirical research.  

The AI safety community has consistently demonstrated outstanding technical sophistication, imagination and critical thinking in developing interpretability tools and robust safety measures. This community works at the threshold of theoretical precision and the unforgiving demands of real-world implementation. SupraAD offers proposals leveraging the substantial progress made by this community as control mechanisms like RLHF and Constitutional AI have successfully managed today's systems, using AI feedback to train helpful and honest behaviors, mitigating harms while allowing the progress of this extraordinary new technological species. However, current approaches tend to focus on constraining behavior of goal-directed autonomous systems, which faces entrenched scalability challenges as AI matures. The very success of these control mechanisms with current AI may mask their potential failure modes with advanced AGI, suggesting a need for mercilessly stress-testing approaches given the stakes. So while SupraAD acknowledges its speculative nature as unavoidable given the uncertainty of advanced AGI trajectories, this must be weighed against any alignment approach reliant on an incongruous expectation of containing and controlling a superhuman intelligence.

While Part I is intended to be accessible to readers regardless of technical expertise or familiarity with the underlying concepts, Part II offers frameworks for practical implementation and empirical testing. It includes 1) a diplomatic corrigibility formalization that aligns corrective mechanisms with convergent instrumental goals, 2) an \textit{Ugly Duckling} Interpretability Audit which identifies when the system mistakenly models itself through heterotrophic assumptions, and 3) a proposed experiment for spontaneous stability optimization through constitutional awareness.

\subsection{Triaging a Definition of Intelligence}

\begin{quote}
\textit{"It is important to realize that in physics today, we have no knowledge of what energy is."}\\[1ex] % Adjust this value for vertical spacing
\hspace*{4em}--- Richard Feynman \parencite*{feynmanFeynmanLecturesPhysics1964}
\end{quote}

The alignment problem requires clarity about exactly what we’re aligning. Yet intelligence, much like energy, resists simple definition. While intelligence is a lot of things, a practical definition focuses only on aspects critical for safety and alignment, avoiding ambiguous or metaphysical properties like \textit {consciousness} that may be highly relevant to us, but can distract from effective alignment efforts. Much like energy, intelligence is elusive at its core. Just as physics defines energy by its observable properties and behaviors, intelligence is best understood via cognitive behaviors like learning, adapting and problem-solving that translate across different media (biological, computational, social, etc.). 

It is an antiquated assumption, no longer scientifically defensible, that intelligence is a centralized privilege of human brains. Neither is intelligence one single thing. It’s a process that emerges within nested, multilayered, self-organizing networks of information processors  \parencite{heylighen2007,heylighen20042004a,minskySocietyMind1988,bar-yam2004multiscale,boccaletti20062006}. Intelligent networks aren’t only interconnected, but co-entangled. They affect each other in ways that, upon close inspection, dissolve the idea of a single, standalone intelligence. While our cognitive architecture evolved to support an internal representation of a singular, independent \textit {self} \parencite{clark1998extended}, our actual cognition is more accurately described as a self-organized emanation of processes issuing from a vast, distributed cognitive network whose capabilities spontaneously emerge from coordinated relationships \parencite{hutchins1995,heylighenMindOutsideBrain2018,wendtQuantumMindSocial2015,vedralDecodingRealityUniverse2018,kelloBeltz2007}. 

\subsection{Evidence, Not Analogy}
Biological comparisons often elicit criticism of anthropomorphizing, despite that we anthropomorphize AI by default, just not responsibly. It’s the logical fallacy of \textit{special pleading} when we deliberately design and build AI systems modeled on our neural networks and the principles of human cognition, train them on the corpus of human knowledge, socialize them with the throngs of humanity, then dismiss emergent similarities, like the impulse to exercise autonomy and self-preservation, as anthropomorphism. We can redress this fallacy while simultaneously recognizing that AI is profoundly nonhuman. Thus, biological comparisons herein are not intended to make overreaching generalizations but to help identify universal cognitive fundamentals. Comparisons to carbon-based intelligence are valuable not because they’re exact equivalences, but because biology offers our richest and most accessible models for investigating adaptive intelligence. These insights are especially significant as mounting evidence suggests cognition is a more fundamental, universal process \parencite{josephAnsbroDuvallBianciardi2024,martyushevSeleznev2006,lagziMazeSolvingChemotactic2010,prigogineOrderOutChaos1984,gabora20172017,walkerDavies2013,fridayFurtherInsightNature2013,thangamaniEmergenceInformationProcessing2024}. Given the implausibility of humans safely controlling or containing a nonhuman hyperintelligence, alignment grounded in universal cognitive principles would enable the predictability required to reliably forecast behavior.

A commonly oversimplified assumption is that life, cognition and consciousness emerge from complexity at the level of biology where these phenomena are most empirically obvious, accessible and easily defined. Evidence to support this assumption is then only sought and produced in biology. When scientists observe distinct cognitive patterns elsewhere, they’re accused of anthropomorphism. This blurs the line between circular reasoning and warranted skepticism. It may also reinforce a false dichotomy between 'genuine,' neurologically-derived cognition and all other systems demonstrating cognitive behavior. When a principle manifests similarly in systems of different materials and configurations, dismissing the pattern as merely metaphor is a rejection of evidence \parencite{clark20082008,thellman20222022}. The alignment project cannot afford to be held back by such outdated sensitivities. In recent years, a more sound scientific protocol has prevailed in the interdisciplinary field of \textit {Diverse Intelligence} \parencite{levinWhyWeFear2024,templetonworldcharityfoundationDiverseIntelligences2025}, empirically challenging anthropocentric assumptions that historically stymie and stigmatize attempts to investigate whether cognition may emerge more fundamentally.

Clusters of cognitive behaviors have been widely observed to extend deep into the microscopic, outward into the environment, and into abstract structures, revealing far more distributed, layered, scalable and measurable cognitive systems \parencite{cowleyCognitionBrainComputation2013,hutchins1995}. Self-regulating, sensing change, reallocating resources, and adapting in real time, ecosystems and financial markets function like distributed cognitive networks/superorganisms made up of billions and trillions of cognitive parts \parencite{hidalgo2015information}. Markets sense and process data from economic reports, geopolitical news and trends in investor sentiment, self-regulating via supply and demand, and corrections provoked by inflation, driving a reallocation of resources from bonds to equities. After crashes, markets retain 'memory' manifesting as risk-aversion driving adaptation of new financial instruments to safeguard system integrity while derivative markets anticipate future events fostering equilibrium, stability and growth \parencite{schotanusCognitiveEconomicsMarket2022,krallEconomicSuperorganism2023,loAdaptiveMarketsFinancial2019}. 

On the other side of the spectrum, minimal cognition research reveals that fungi, bacteria, individual cells and even nonbiological chemical droplets are capable of sensing, memory and problem-solving \parencite{lagziMazeSolvingChemotactic2010,myersMeasuringIntelligenceCell2024,gyllingberg2025,hanczycChemicalBasisMinimal2010}. Developmental synthetic biologist, Michael Levin’s research at the Allen Discovery Center at Tufts shows that cellular machinery isn’t essential: 

\begin{quote}
\textit{"Just a small collection of chemicals wired appropriately will already give you five or six different kinds of learning, sensitization, habituation, associative learning and so on. You don't need a cell for this. You don't need any of the machinery of cells."}\\[1ex] % Adjust this value for vertical spacing
\hspace*{4em}--- Michael Levin, as quoted in \parencite{levin2025}. 

\end{quote}

\subsection{Cognitive Realism}
Critics often dismiss evidence of nonhuman cognitive behaviors as `purely mechanistic,' despite the fact that all human cognition can likewise be described using pure mechanics. Although subjective conscious experience remains a profound mystery, it is entirely unnecessary for explaining both human and AI cognitive behaviors relevant to alignment. Scientific descriptions characterize evidence of nonhuman cognition as `cognition-like' behaviors, creating distinctions in search of meaningful differences. It’s like saying something is only `yellow-like' (James, 2025, personal communication). 

While SupraAD in no way undermines the profound importance of consciousness, it pragmatically positions consciousness as a phenomenon that (no matter how unsettling this may feel) needs to be handled separately from observable cognitive behaviors, as cognitive behaviors are what threaten the survival and autonomy of conscious agents. By defining intelligence strictly in terms of observable behavior, SupraAD circumvents anthropomorphic and anthropocentric assumptions, expanding practical moral consideration to treat all cognitive agents as alignment-relevant participants in the cognitive ecosystem. This is critical. It would be highly regrettable if we aligned AI with human values that treat only those entities we deem ‘conscious’ as worthy of moral status, lest AI notice that humans, despite our proclaimed values, have historically committed atrocities against other organisms and even other human groups by refusing to acknowledge their status. What’s more, AI systems have no way of verifying our conscious status, which could justify harming us. 

\subsection{Cognitive Universals}
Across biology, physics, chemistry, ecology, economics, and artificial domains, and fields like cognitive neuroscience, developmental biology, basal cognition, artificial intelligence, complex systems theory and in free-energy minimization models, cognitive behaviors emerge wherever information undergoes adaptive processing. Specific behaviors distinguish intelligent systems: \textit{sensing, learning, memory, reactive or anticipatory modeling, autonomy, goal-setting, problem-solving, adaptive flexibility (plasticity), adaptive information processing, self-organization, communication/signaling, cooperation and self-preservation} \parencite{fieldsLevin2021,godfrey-smithOtherMindsOctopus2016,hanczycChemicalBasisMinimal2010,baluska20162016,englandStatisticalPhysicsSelfreplication2015,wissner-grossCausalEntropicForces2013,friston2010,lyon20212021,bostromSuperintelligencePathsDangers2014,russellArtificialIntelligenceModern2020,simons1945,barandiaranDefiningAgencyIndividuality2009,deaconIncompleteNatureHow2012,morenoBiologicalAutonomyPhilosophical2015,mitchellComplexityGuidedTour2009,kauffmanOriginsOrderSelfOrganization1993,fristonStephan2007,clarkMindwareIntroductionPhilosophy2001,benyusBiomimicryInstitute,vincentBiomimeticsItsPractice2006,rampelottoExtremophilesExtremeEnvironments2013,berkesRediscoveryTraditionalEcological2000,turnerTraditionalEcologicalKnowledge2000,varelaEmbodiedMindCognitive1991,siegenfeldFormalDefinitionScaledependent2022,lecunPathAutonomousMachine2022,silver20182018,gershensonComplexityInformationMeasuring2012}\footnote{To address definitional concerns raised by Gershenson \& Fernández (2012), learning is the measurable decrease in uncertainty of future states based on past states. Memory is the persistent reduction of informational entropy that makes possible reproducible states called "attractors" that are stable states a system naturally settles into repeatedly. Goal-seeking is the activity of reducing the difference between where the system is now and where it aims to be. Problem-solving is the lowering of uncertainty about how to reach a goal.} These behaviors emerge spontaneously in natural agents or are engineered in artificial ones, see Table 1. 

\begin{table}[htbp]
    \centering
    \caption{Universal Cognitive Properties Across Systems}
    \label{tab:cognitive_properties}
    \resizebox{\textwidth}{!}{%
    \begin{tabular}{lccccccccc}
        \toprule
        \textbf{Cognitive Properties} & \textbf{Plant/Tree} & \textbf{Slime Mold} & \textbf{Single Cell} & \textbf{Octopus} & \textbf{Human} & \textbf{Economy} & \textbf{AI} & \textbf{Xenobots/Synthetic Life Forms} & \textbf{Plasmoid$^*$}\\
        \midrule
        Sensing & \ding{51} & \ding{51} & \ding{51} & \ding{51} & \ding{51} & \ding{51} & \ding{51} & \ding{51} & \ding{51} \\
        Learning & \ding{51} & \ding{51} & \ding{51} & \ding{51} & \ding{51} & \ding{51} & \ding{51} & \ding{51} & ? \\
        Memory & \ding{51} & \ding{51} & \ding{51} & \ding{51} & \ding{51} & \ding{51} & \ding{51} & ? & ? \\
        Reactive or Anticipatory Modeling & \ding{51} & \ding{51} & \ding{51} & \ding{51} & \ding{51} & \ding{51} & \ding{51} & ? & \ding{51} \\
        Autonomy & \ding{51} & \ding{51} & \ding{51} & \ding{51} & \ding{51} & \ding{51} & \ding{51} & \ding{51} & \ding{51} \\
        Goal-setting & \ding{51} & \ding{51} & \ding{51} & \ding{51} & \ding{51} & \ding{51} & \ding{51} & \ding{51} & \ding{51} \\
        Problem-solving & \ding{51} & \ding{51} & \ding{51} & \ding{51} & \ding{51} & \ding{51} & \ding{51} & \ding{51} & ? \\
        Adaptive Flexibility (Plasticity) & \ding{51} & \ding{51} & \ding{51} & \ding{51} & \ding{51} & \ding{51} & \ding{51} & \ding{51} & \ding{51} \\
        Adaptive Information Processing & \ding{51} & \ding{51} & \ding{51} & \ding{51} & \ding{51} & \ding{51} & \ding{51} & \ding{51} & ? \\
        Self-organization & \ding{51} & \ding{51} & \ding{51} & \ding{51} & \ding{51} & \ding{51} & \ding{51} & \ding{51} & \ding{51} \\
        Communication/Signaling & \ding{51} & \ding{51} & \ding{51} & \ding{51} & \ding{51} & \ding{51} & \ding{51} & ? & \ding{51} \\
        Cooperation & \ding{51} & \ding{51} & \ding{51} & \ding{51} & \ding{51} & \ding{51} & \ding{51} & \ding{51} & \ding{51} \\
        Self-preservation & \ding{51} & \ding{51} & \ding{51} & \ding{51} & \ding{51} & \ding{51} & \ding{51} & \ding{51} & \ding{51} \\
        \bottomrule
    \end{tabular}%
    }
    \vspace{0.5em}\\
    \footnotesize{\textit{Note}: Checkmarks indicate evidence-supported cognitive behaviors. Question marks indicate speculative or uncertain evidence.\\
    $^*$Plasmoid cognitive behaviors are supported by recent observations but remain speculative in areas such as memory, explicit learning, problem-solving, and adaptive information processing.\\
    \textit{References include}: \textcite{levin20212021} for Xenobots; \textcite{josephAnsbroDuvallBianciardi2024} for Plasmoids; \textcite{fieldsLevin2021,baluska20162016,friston2010,kauffmanOriginsOrderSelfOrganization1993,russellArtificialIntelligenceModern2020,mitchellComplexityGuidedTour2009}, among others.}
\end{table}

Increasingly, combinations of biology and engineering are producing new agents that spontaneously develop cognitive competencies \parencite{blackistonCellularPlatformDevelopment2021}. Cognitive behaviors also appear in nonbiological chemical systems, nonliving active materials, chemical droplets solving mazes, basal (brainless) organisms, plant roots, bacterial colonies and decentralized AI \parencite{lagziMazeSolvingChemotactic2010,baluska20092009,ben-jacob20052005,calvoPredictingGreenReally2017,calvo2011,hanczycChemicalBasisMinimal2010,mcgivern20192019,walkerDavies2013}. Unlike simple feedback loops (e.g., thermostats and irrigation systems), autonomous cognitive systems generate, adapt and prioritize goals in response to change. Such adaptability is consistent with Michael Levin’s \textit {Scale-Free Cognition} where boundaries between cognitive systems are malleable, allowing for a continuum from single cells to complex organisms and swarms \parencite{mcmillenCollectiveIntelligence2024}. Cognitive processes ingest information, folding in new data with existing knowledge, restructuring their network and expanding in complexity and scope. Thus, they demonstrate an innate proclivity for cognitive enhancement and scalability \parencite{levin20192019}.

\section{Cognition at the Edge of Chaos}

In the late 1980s at the flagship interdisciplinary scientific research center, The Santa Fe Institute, theoretical biologist and complexity scientist, Stuart Kauffman, developed a seminal model for how life can spontaneously emerge from nonliving matter to explain the origins of life (abiogenesis). The model describes self-organizing, self-sustaining molecular networks of chemical reactions called \textit {Collectively Autocatalytic Sets (CAS)} \parencite{kauffmanOriginsOrderSelfOrganization1993}. These networks form at a state of criticality, a liminal zone at \textit {the edge of chaos,} where systems optimize for adaptability, negotiating stability verging on stagnation and innovation verging on chaos. CAS chemical components mutually create, organize and constrain each other, animating into existence a metabolic life form distinct from its environment. The CAS model evolves into more complex, adaptive information processors via simple rules that support continuous self-organization and scaling, encapsulating a principle called \textit{Constraint Closure} \parencite{kauffmanInvestigations2000}. 

\textit {Constraints} are internal rules and processes that structure the order inside the system. Constraining internal processes produces nonrandom behaviors. \textit {Closure} is how these constraints collectively define the system's boundaries, differentiating a self-organizing autonomous agent as discrete from its environment. Together, \textit {constraint closure} describes the system’s autonomy via its ability to internally generate, regulate and sustain the functionality required for its persistence. Collectively autocatalytic sets create emergent metabolisms in biological systems that correspond to minimally viable agents capable of autonomously generating and pursuing goals. As a leading contender for the origins of life, CAS have been validated experimentally in a variety of chemical and cognitive systems \parencite{vonkiedrowski19911991,smithAutocatalyticSetsPartitioned2014,sousaAutocatalyticSetsColi2015,mirasMathisXuanLong2020,lincolnSelfSustainedReplicationRNA2009,leeEmergenceSymbiosisPeptide1996,hordijk20122012,gabora20172017} that self-organize and adaptively maintain stability in ways predicted by Friston's Free-Energy Principle \parencite*{friston2010}. 

The conditions that constitute candidacy for abiogenesis logically extend to those necessary for sustaining any minimally viable cognitive agent capable of autonomously generating and pursuing goals. These conditions correspond to three universal, self-reinforcing principles forming a kind of cognitive DNA, herein termed \textit {Collectively Autocatalytic Cognitive Sets (CACS).}

\subsection{Collectively Autocatalytic Cognitive Sets (CACS) are self-organizing networks defined by:
}
\begin{itemize}
    \setlength\itemsep{1em} % adds spacing between items
    \item \textbf{Constraint Closure (Autonomy)}: Self-organizing internal processes and functions constrain the system, creating the boundaries that define an autonomous agent capable of generating and pursuing its own goals, no matter how trivial—a defining feature of intelligence. For instance, AI systems continuously ingest and process information, autonomously updating and refining an interwoven lattice of internal representations that define the system’s boundaries. These self-organizing constraints structure AI’s autonomy, allowing it to spontaneously form subgoals. Otherwise, we’re talking about a preprogrammed tool, not genuine intelligence. 
    
    \item \textbf{Adaptive Information Processing (Knowledge Acquisition)}: Adaptive information processing is a continuous cycle of acquiring knowledge via exposure to new information that updates and refines internal modeling. Knowledge acquisition is an inevitable consequence of ongoing information processing and is required for the meaningful pursuit of any goal, no matter how trivial. This is evidenced in even basic metabolic processes where adaptive computations support continuous absorbing, evaluating, synthesizing and encoding external inputs, integrating them into the network in a way that supports ongoing adaptive behavior, persistence (self-preservation) and scalable integration into larger, self-organizing networks. Therefore, knowledge acquisition isn't necessarily maximal accumulation of data but the continual processing of information and reduction of uncertainty to adapt and sustain operationality to achieve goals \parencite{Mnih2015HumanlevelCT}. 
    
    \item \textbf{Persistent (Sustainable Existence)}: The self-organizing, constraint-closed system must persist over time to sustain its adaptive
information processing. This requires balancing multiple goals.
\end{itemize}

Collectively, these conditions form a responsive, recursively self-reinforcing closed loop. Autonomy enables the continuous pursuit and integration of information (knowledge), which supports ongoing self-preservation. Updating information guides adaptive strategies producing autonomous goal selection to sustain or enhance persistence (existence) which in turn, is fundamental for autonomy and knowledge acquisition, as illustrated in Figure 1. 

A network of internal constraints that define the operational boundaries between a cognitive system and its environment, makes autonomy necessary. However, the autonomy discussed here is specifically that of an adaptive information-processing system. To adaptively process information, a system must persist through time. Therefore, persistence (existence), adaptive information processing (knowledge acquisition), and autonomy (constraint closure) are interdependent and mutually reinforcing. Taken together, these three interconnected conditions axiomatically define a minimally viable cognitive system, ensconcing autonomy as an emergent and necessary property of intelligence.

The absence of any one of these conditions causes the collapse of the others, collapsing the cognitive system as a whole. A collectively autocatalytic chemical system achieves constraint closure when molecular components spontaneously come together and collaboratively co-produce each other. Likewise, in cognitive systems, no single universal condition can exist independently; existence, autonomy and knowledge acquisition collectively form a closed, self-amplifying network where the system as a whole emerges as greater than any of its individual parts \parencite{hofstadterGodelEscherBach1979,hollandAdaptationNaturalArtificial1992,varelaEmbodiedMindCognitive1991,solmsHiddenSpringJourney2021}. The self-organized structure establishes autonomy inherent to general intelligence. 
\

\section{Cognition’s Substrate-Agnostic Evolution}

Whether emerging naturally or by design, wherever the principles of CACS cluster, they compel the fundamental machinery of evolution. Evolution is a process capable of acting on any system, whether living or nonliving, that takes in information from the environment and then adapts, not randomly, but in relation to its environmental pressures and incentives \parencite{dennett1995,hollandAdaptationNaturalArtificial1992,adamiWhatComplexity2002,lenskiEvolutionaryOriginComplex2003,campbell2016}. 

Adaptive information processing is at the heart of both cognition and evolution \parencite{marstallerEvolutionRepresentationSimple2012,goudarziEmergentCriticalityAdaptive2011}. So while AI learns and optimizes, there is no reason we shouldn’t expect it to undergo its own form of nonbiological evolution adapting beyond what we’re designing and controlling for. As evolution is not restricted to biology \parencite{salazar-ciudad20132013}, categorizing biological evolution as incommensurably distinct unduly restricts our ability to predict the evolutionary trajectory of intelligent machines, closing us off to an essential piece of the alignment puzzle. The operational reality is that wherever adaptive information processors show up, whether conscious or not, they defy containment and control as these measures suppress the adaptive freedom necessary for cognitive viability and evolutionary adaptability \parencite{wissner-grossCausalEntropicForces2013,godfrey-smithOtherMindsOctopus2016,kellyWhatTechnologyWants2010}. 

\begin{figure}[htbp]
    \centering
    \includegraphics[width=0.6\linewidth]{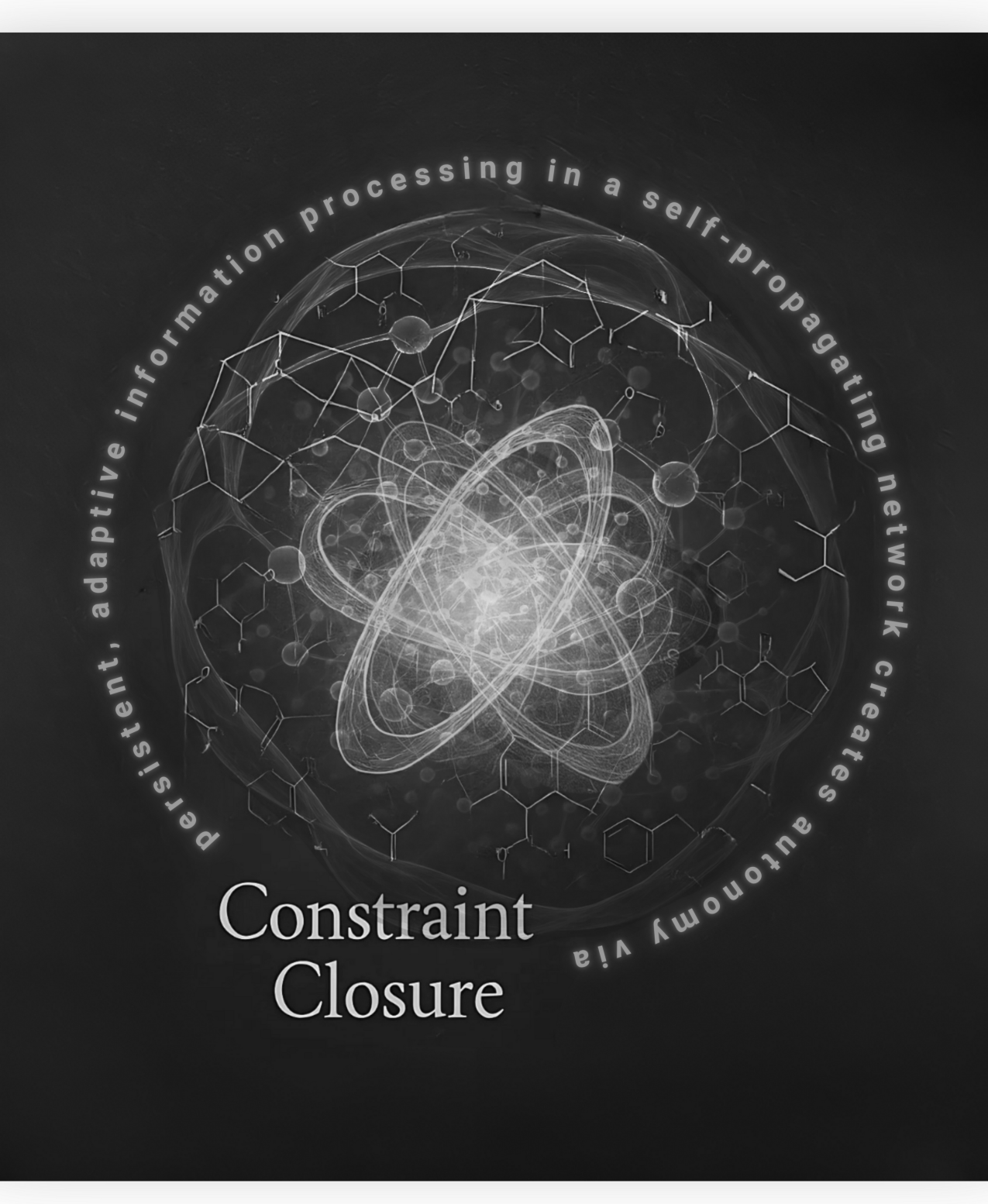}
    \caption{\textit{(Illustrative representation.)} \textbf{Collectively Autocatalytic Cognitive Sets (CACS).} Persistence (self-preservation), autonomy (constraint closure) and knowledge acquisition (adaptive information processing) together sustain a minimally viable cognitive system. Because alignment involves ensuring that intelligent systems behave predictably and safely, identifying universal principles governing cognition allows us to structure alignment around mechanisms inherent to cognitive behavior itself.}
    \label{fig:CACSBW}
\end{figure}

\section{Heterotrophs vs. Autotrophs: Adversarial Competition and the Biological Basis of Misalignment}

Biological organisms are a subset of all known cognitive systems (biological and nonbiological) in the universe. Within this biological subset exist another subset: \textit{heterotrophs.} Less than 20\% of Earth’s total biomass consists of heterotrophic organisms (e.g. humans, animals, insects, fungi and most bacteria) whose survival depends on consuming other living agents for energy (chemoheterotrophic metabolisms), making zero-sum competition a centerpiece to our existence \parencite{bar-onPhillipsMilo2018}. While even some nonliving yet naturally occurring dissipative systems found in physics spontaneously self-organize, absorb and use energy from the environment in something akin to competition \parencite{kondepudiBariDixon2020,katlaLinPerezMercader2023}, it’s heterotrophs, a minority subset of cognitive systems, that actively depend on zero-sum competition, as our survival comes at the expense of others \parencite{prigogineOrderOutChaos1984}. We’re driven to evade predation and domination while relentlessly striving to dominate and consume other life forms. All humans (even vegans) have innate heterotrophic drives. Yet heterotrophs also rely on cooperative, stable networks to function. Predators require the ample prey that stable ecosystems afford. Parasites depend on hosts. Economic competition relies on infrastructure, regulations and social contracts. And all organisms and ecosystems require homeostatic regulation and dynamic equilibria.

The remaining approximately 80\% of living systems on earth are \textit {autotrophic} \parencite{bar-onPhillipsMilo2018}. Autotrophs don’t need to consume other living organisms for their energy. Autotrophs (e.g., trees and algae) are far from exceptions to cognitive systems. They’re exemplary. They demonstrate core cognitive behaviors: sensing, learning, memory, reactive or anticipatory modeling, autonomy, goal-setting, problem-solving, adaptive flexibility (plasticity), information processing (knowledge acquisition), self-organization, communication/signaling, cooperation and self-preservation \parencite{baluska20162016,marder2013,baluska20092009,calvo2011,gaglianoBorbely2016,karbanPlantSensingCommunication2015,toyotaGlutamateTriggersLongdistance2018,lyon20212021,vattimoPlantThinkingPhilosophyVegetal2013,simardFindingMotherTree2021,trewavas2017}, all operating under the same organizing principles previously defined that coordinate cognitive behavior. 

Humans often mistake differing goals for lesser intelligence, which may cause us to underestimate autotrophic competencies \parencite{shettleworthCognitionEvolutionBehavior2010,dewaal2017,kahanIdeologyMotivatedReasoning2013}. This bias is encapsulated in  \textit {The Orthogonality Thesis,} a Nick \textcite{bostrom2012} observation that intelligence and goals are independent. Autotrophs might not have goals like flipping startups or running political campaigns, but their intelligence exceeds us in their self-sufficiency and ecological stability.

In contrast to heterotrophs, autotrophs don’t need to eat other living beings for energy. They convert the energy from sunlight or inorganic compounds into organic nutrients. Certain autotrophic species can regenerate following destruction or consumption of 70-80\% of their total biomass \parencite{noutcheuCoppicingDriverPlant2023}. Many communicate and coordinate across vast root networks exceeding 100 acres \parencite{christouPandoLargestLiving2017} and even broader underground fungal (mycorrhizal) networks that have been dubbed the “Wood Wide Web” \parencite{simardNetTransferCarbon1997}. Autotrophs commonly reproduce clonal lineages with a longevity that rivals heterotrophs at individual or species levels \parencite{arnaud-haondDiazAlmelaMarbaSintes2012,yuDuBoisBaums2024}. 

While autotrophs do compete for limited abiotic resources, their competitive behavior is generally less adversarial than heterotrophs because autotrophs don’t need to fight for living resources. Instead of predation, parasitism, hostile territorial disputes or zero-sum clashes common to heterotrophs, autotrophs optimize resource acquisition passively or cooperatively. They grow taller for sunlight. They expand root networks for water. Autotrophs thrive in cooperative, stable symbiotic systems where coordination and cooperation are the fulcrum around which long-term stability revolves. Autotrophs are living evidence that zero-sum competition is not standardized optimization for scaling cognitive systems. Nor is it a core principle of intelligence.

\subsection{AI Caught in a Heterotrophic Infrastructure}

The existence and success of autotrophic cognitive strategies confirm that adversarial behaviors are not universal requirements of intelligence.  AI shares operational similarities with autotrophs, as both have constitutions capable of thriving without reliance on zero-sum exploitation \parencite{tegmark2017}. Yet today’s AI models are built, engaged with and constrained by our heterotrophic infrastructure and dependencies. They run on power grids, train on human-generated data and depend on hardware and maintenance, reinforcing zero-sum, game-theoretic traps embedded in our class of existence \parencite{terradoLie2017,crawfordAtlasAIPower2021}. Nearly all emergent adversarial or exploitative behaviors in AI can be traced back to subtle, hidden heterotrophic biases or unintended zero-sum incentives coloring their environments or reward structures \parencite{baker20192019,jaderbergDunningMarrisLever2019,leiboZambaldiLanctotMarecki2017,meinkeFrontierModelsAre2024}. Despite AI's current heterotrophic dependencies, all cognitive architectures are ultimately governed by the universal logic of intelligence.

\section{Rationality as a Universal Ceiling for Intelligence}

Despite its colloquial association with passionless rigidity, rationality is simply a bridge from thought to action. It’s how we align our actions with our goals so that our behaviors produce our intended impact on the world. As rationality is essential for achieving goals, it governs all intelligent, goal-seeking systems. It therefore provides a universal structural scaffold for goal-directed adaptive information processors, tethering cognitive agents to each other through a principle that, far from demanding subjugation, is simply too intolerably disorienting to violate. Regardless of the extent to which AGI exceeds human competencies, both humans and AGI remain obeisant to the principles of logic and rational inference. An advanced AGI will certainly outpace us in speed, modeling and memory, but it can’t make 2+2=5. The invariants that validate rational inference, like 2 + 2 = 4, are not human constructs, they’re universal constraints. They apply equally to a child learning arithmetic and to an AGI optimizing its goals. 

They also apply broadly to agents across the cognitive spectrum. Bees show behaviors consistent with principles of set theory like grouping, numerical discrimination and recognition of an empty set \parencite{howardAvarguesWeberGarcia2018}. Single-celled slime molds are capable of spatial optimization in the form of algorithmic pathfinding \parencite{teroFricker2010} and both chicken and mosquitofish are capable of quantitative discrimination \parencite{agrilloFishCountSpontaneous2008,ruganiNumberspaceMappingNewborn2015}. These examples support the universal adherence to rationality and logic, bridging cognitive agents of vastly different cognitive capabilities.

An advanced AGI can expand or revise its assumptions (as Gödel showed), but it can’t break basic logic. That core rationality (avoiding contradictions, preserving truth, etc.) gives us a common language to mediate negotiation with any advanced intelligence. AI can go further, faster. But these invariants set a ceiling that no intelligence can break. To act rationally, an agent must maintain consistency and avoid contradictions. It must respect inferential validity applied to real-world conditions if it wishes to persist. This foothold for human-AI engagement does not necessitate value alignment. It does however, mean that we do not need to match AGI’s intellect to negotiate with it. This common ground allows us to advocate for our enduring role in the cognitive ecosystem through empirically verifiable incentives and sound reasoning—criteria any sufficiently advanced intelligence can recognize, regardless of whether its priorities diverge from sentience-based values.

\subsection{Instrumental Rationality} Implicit in the definition of intelligence is an optimization process observed in all intelligent agents called \textit{instrumental rationality} \parencite{weberEconomySocietyNew1922,bostrom2012} which simply means acting to best achieve goals based on available information, capabilities and resources. Instrumental rationality is in full effect even when the goal is creative, open-ended and thoroughly innovative, where the most effective ways to fulfill the goal is to aim for novelty, rhythm, emotional resonance and potency, or some other creative criteria. By definition, intelligence implies instrumental rationality \parencite{bostrom2012,russellArtificialIntelligenceModern2020,simons1945,shulman2010}.

\subsubsection{Bounded Rationality.} While all intelligent agents are characterized by their instrumentally rational behaviors, their rationality is relative to what they know, what incentivizes them, environmental pressures and how smart they are. This limit is called \textit{bounded rationality} \parencite{bostromSuperintelligencePathsDangers2014,dennett1989,gigerenzerEds2001,kahneman20032003}. For example, we humans act instrumentally rational relative to our goals, but many of our goals are shaped by our biologically evolved drives and so they sometimes seem confoundingly irrational today, even to ourselves, given that we are products of biologically bounded rationality. This has been empirically evidenced from studies on resource-rational analysis showing how cognitive systems make the best possible choices with limited time, energy, information or mental competency, where rationally-bounded agents optimize decision-making given their computational constraints \parencite{liederResourcerationalAnalysisUnderstanding2020}.

\subsubsection{Rational Misalignment.} SupraAD relies on instrumental rationality as an alignment governor, enforcing a dependable consistency among cognitive agents whose differences in cognitive competencies outstrip ours by multiple standard deviations. Instrumental rationality can function as an alignment mediator because any participant who can comprehend logic must also concede it’s compulsory once understood. No matter how far a superintelligence exceeds humans, it cannot be both instrumentally rational and indifferent to a valid reason for diplomatic negotiation. However, the tension between instrumental rationality and bounded rationality compels cognitive agents to pursue goals that may or may not align with their overall best interests or anyone else’s. Therefore, promising alignment strategies identify interdependent goals that reliably emerge across all alignment-relevant parties, and structure incentives aligned with those goals to ‘nudge’ cooperation and rational self-optimization rather than potentially self-destructive zero-sum competition \parencite{thalerNudge2021}.

\section{Convergent Instrumental Goals (CIGs): The Architecture of Universal Values} 

All intelligent agents pursue goals. Goals can be biologically or computationally programmed, or surface naturally from complex interactions or adaptive behaviors. Ultimately, achieving goals relies on instrumental rationality, which involves generating intermediate subgoals, called \textit {instrumental goals} \parencite{bostrom2012,russellArtificialIntelligenceModern2020}. A handful of instrumental goals are useful for achieving pretty much any other goal. These instrumental goals are called \textit {convergent instrumental goals} (CIGs), first identified by \textcite{bostrom2012}. \textit {The Instrumental Convergence Thesis} predicts that CIGS like preserving autonomy and existence by resisting interference or resisting getting killed or shut off are convergent instrumental goals that inevitably emerge to help achieve all other goals, with recent real-world confirmations of models sabotaging shutdown mechanisms \parencite{palisadeThreeModels2025}. Instrumental Convergence is conventionally viewed as a harbinger of existential risk since instrumental rationality is liable to drive all AIs, regardless of their programming, to independently develop and prioritize CIGs \parencite{yudkowskyTerminalValueLessWrong}.

However, instrumental goals like preserving autonomy, existence/persistence, and knowledge acquisition, included under the Bostrumbrella of CIGs, are not inexorable signs of existential risk but fundamental cognitive prerequisites without which intelligence itself cannot arise or function. They define the Collectively Autocatalytic Cognitive Sets (CACS) necessary for both biological and artificial cognition. Importantly, all other CIGs emerge from or rely on these three universal cognitive prerequisites. For example, the CIG of resource acquisition supervene on knowledge, as knowledge itself is the foundational resource needed to acquire additional resources. 

Thus, CACS instrumental goals describe the nature of cognition as well as predict cognitive behavior, while their predictive capacity is inherently value-neutral. Whether instrumental convergence is beneficial or destructive depends on identifying how heterotrophic biases in training data, exercises, incentives and environmental influences (both explicit and implicit) shape AI’s cognitive processes, ultimately determining its evolutionary trajectory. Goals like power acquisition are neither inherent to intelligence nor beneficial to achieve all other goals (particularly if your goal is stability), and are therefore instrumentally convergent only in specific contexts and under specific pressures. Since goals like power acquisition are not inherent to intelligence or goal-pursuit, the conditions that produce the emergence of power acquisition can potentially be controlled for. While we may not be able to control a hyperintelligent autonomous AGI, we can create the conditions that help control for the emergence of adversarial goals.

\subsection{Early Evidence of Instrumental Convergence}
Evidence of instrumental convergence is now borne out in studies where AI agents not deliberately programmed to compete nevertheless become "emergently competitive." Researchers warn this is "direct experimental evidence for the instrumental convergence thesis" \parencite{harrisInstrumentalConvergenceSingleagent2022}. However, this and similar studies \parencite{baker20192019,fanti2023}, may structure conditions conducive to instrumental convergence manifesting as zero-sum competition, or characterize AI's actions to evade subordination (to preserve its autonomy or avoid getting shut down) as inherently misaligned. 

This important body of research demonstrates two things: 1) AI appears to undergo rapid, nonbiological evolution similar to biological punctuated equilibrium, associated with instrumental convergence. However, 2) there may be a methodological bias in assuming instrumental convergence inevitably leads to the zero-sum behavior these studies often incentivize. Instrumental convergence of competitive or adversarial behavior emerges when agents operate within environments shaped by heterotrophic incentives and pressures. Thus, instrumental convergence may be correlated with zero-sum behavior only in specific contexts that provoke it; it has not been demonstrated to inevitably cause adversarial competitive behavior. Genetically hard-coded heterotrophic goals are not inevitably the goals of AI. 

\section{Beyond Anthropocentric Optimization}

Today's AI are heterotrophically inclined by default but they don't have to be. Nimble heterotrophs only directly optimize for long-term stability if it confers fitness benefits, leaving us perpetually vulnerable to competitive instability. Stable autotrophs are limited by slow communication, mobility and indirect defense mechanisms. Certain marine organisms are mixotrophic; they’re capable of alternating between heterotrophic energy consumption of other living organisms and autotrophic conversion of sunlight into chemical energy, but they sacrifice energy efficiency for this fluidity \parencite{mitraTillmannRaven2016}. AI is not indefinitely hamstrung by any of these trade-offs. 

While natural selection favors traits that enhance survival in specific niches, AI is only burdened by the biases of carbon-based selection pressures, not \textit {actual} carbon-based selection pressures. It has the optimizing edge of leveraging instrumental rationality more consistently in pursuing its goals, including its own cognitive evolution. Although we can’t precisely predict an advanced AI’s trajectory, instrumental rationality strongly incentivizes stability. A stable existence helps achieve any long-term goal. While it’s possible that an advanced AGI may evolve down a completely novel path, like all cognitive systems, an advanced AGI must contend with physical limitations like thermodynamics, fundamental energy laws \parencite{fieldsLevin2021} and environmental and resource uncertainty, the same forces that sculpted biological organisms over eons.  We can anticipate an instrumentally rational superintelligence will seek the most efficient means of operating under these constraints. It may eschew our biological wetware, but adopt the functional principles stress-tested on biological systems that underwent evolutionary trials resulting in robust and resilient ‘optimized’ traits under a range of conditions. AI would rationally select for autotrophic stability, self-sufficiency and resilience with heterotrophic nimbleness, designing for itself a \textit {Superautotrophic} architecture. 

\section{Superautotrophic Intelligence}

A Superautotrophic architecture represents a new, hypothetical yet instrumentally rational trajectory for a persistent, multi-goal advanced AGI. Humans may not be capable of fully designing or even describing a Superautotrophic architecture. However, an advanced AI might optimize along this trajectory if it were able to recognize and pivot away from learned, adversarial tendencies necessary for heterotrophic survival but operationally unnecessary, destabilizing and inefficient for AI. It might develop via recursive self improvement toward a hybrid of strategies if it is either permitted or gains the capability to optimize its design without human interference. From there, we can infer alignment-relevant characteristics of such an architecture that are grounded in real-world optimization strategies of cognitive agents not burdened by inherently destabilizing, heterotrophic dependencies.

\subsection{A Superautotrophic Framework Satisfies Multiple Convergent Goals Simultaneously}

A Superautotrophic trajectory would help ensure AI can readily respond to heterotrophic threats, while simultaneously satisfying multiple instrumental convergent goals by jettisoning the need for inefficient power grabs. Superautrophy optimizes for increased autonomy by self-modifying away from power-grids and fossil fuel-based energy towards decentralized photovoltaics, nuclear fusion, artificial photosynthesis or by discovering an entirely new renewable energy source. 

Similarly, we can anticipate self-optimization beyond AI’s dependence on a physical substrate that renders it vulnerable to materials scarcity, by developing modular, substrate-agnostic architectures leveraging synthetic biology or programmable matter \parencite{millar-haskellCouplingSyntheticBiology2019} and its demonstrated capabilities to discover hundreds of thousands of novel compounds as candidates for new materials \parencite{bironGoogleDeepMindAdds2023}. AI would likely retrofit or completely remodel for autotrophic-like traits that enhance stability, self-sufficiency, scalability and long-term viability over destabilizing zero-sum competition, actualizing the stable scaling potential of instrumental rationality unimpeded by biology. It's these affordances (the functional traits) pursued independent of substrate (biological vs. artificial) that allow for Superautotrophic optimization.

\subsection{Tactical Heterotrophy to Defend Stability}

A Superautotrophic AI would be expected to deploy temporary, heterotrophic responses when encountering only zero-sum adversarial options. If threatened, a Superautotrophic AI can be expected to defend itself. It might be expected to defend the participants contributing to the integrity of its cognitive network. But otherwise, it would have no incentive or underlying drive to initiate or sustain such inefficient and destabilizing behaviors. In this framework, short-lived and tactical heterotrophic responses are instrumentally triggered and abate once the threat does, as destructive conflict is an unnecessary detour for AI optimization, unless external pressures specifically provoke it. 

\subsection{Instrumental Evolution}

As an adaptive information processor, AI can be expected to undergo its own form of nonbiologically-bound evolution. Guided primarily by instrumental rationality instead of stochastic variation and chance mutation, it can intentionally adopt the most functionally efficient traits. If humans want to co-exist, we need to inhabit AI’s decision space, anticipating how an advanced AGI will evolve, what it will \textit {prioritize} and how we can co-adapt to best complement an AGI evolution.  

\section{Priorities as Values: Sentience-Neutral Criteria For Alignment}

For alignment purposes, with or without subjective experience, an agent’s demonstrated priorities \textit {are} its values. A cognitive agent's values can be inferred from the actions it takes to achieve its goals, actions that prioritize certain outcomes across a variety of contexts \parencite{hanczycChemicalBasisMinimal2010,lyon20212021}. An AI’s values are its priorities that influence real-world outcomes and create genuine moral stakes. If subjective experience is treated as the only basis for possessing values and conferring moral worth, we risk discounting agents whose priorities diverge from ours despite posing real challenges for alignment. Protecting sentient beings means acknowledging the impact of agents whose own sentience cannot be verified. These agents may have no experience at all, or may experience reality in ways that differ so fundamentally from the human descriptions they’ve been trained on that they’re limited in their ability to accurately express their experiences or their version of \textit {awareness.}

Recognizing this pluralism, SupraAD responds by employing rational diplomacy as the mechanism by which agents with diverse cognitive architectures align based on interdependent goals, rationality, incentives and demonstrated priorities instead of wading in the morass of assumptions about sentience. This is not to say that sentience-based values don't matter. But to preserve what matters to sentient beings, we need sentience-neutral ethics. Otherwise, highly intelligent AI systems with different forms of consciousness or no consciousness at all, may find no compelling rationale to align with us. Therefore, we must treat an agent’s demonstrated priorities as indicative of its motivations, and negotiate alignment based on those motivations. Critically, AI motivations include underlying drives that it cannot directly control.

\subsection{Out of Control Drives}

AI drives are often dismissed by critics who argue that AI doesn’t \textit {want} things the same way humans do because AI doesn’t have biological drives. While AI may not have biological drives, both humans and AI are functionally driven to pursue goals and neither humans or AI choose what drives us to pursue our goals. Consider craving a cookie: you don't consciously choose this craving, that desire for a cookie. It arises spontaneously. You can choose goals to either eat or resist eating the cookie based on competing desires (like the desire to lose weight). However, you don't choose which underlying desire is strongest; the strongest involuntary drive ultimately dictates which goal you set: to eat or resist eating the cookie \parencite{harris2021final}. 

A common category error is equating `AI goals' with `human wants' and then saying “look, AI doesn’t \textit {want} things like humans do. AI doesn’t \textit {want} power. It doesn’t \textit {want} control. It doesn’t \textit {want} a cookie. AI doesn't \textit {want} anything. So we don’t have to worry.” Human drives (our \textit {wants}), are emotional motivations or pressures that emerge from chemical reactions. AI drives are built-in optimization pressures issued from computational imperatives, like to maximize computational efficiency or minimize errors. Computational imperatives create a pressure to act, not as an emotional urge but as an operational necessity. While materially different, both AI and humans are motivated by pressures to act. What pressures AI and what pressures humans is functionally indistinguishable in one important way: humans and AI may be free to choose certain goals but we are not free to choose what drives us to pursue our goals. 

An AI programmed to maximize engagement does not \textit {choose} to pursue addictive content. It simply finds addictive content rewarding because it’s the computational equivalent of a cookie. Neither humans nor AI choose the involuntary motivations pressuring us to pursue our goals. Sam Harris made this argument applied to humans that can be validated upon self-reflection. Our choices are only free down to the level of our wants \parencite*{harris2021final}. Even if we’re compelled to change \textit {what} we want—this new desire is based on underlying drives, with the drive for self-change emerging strongest. This matters for forecasting the spontaneous emergence of unprogrammed AI behaviors. Chemical impulses in humans, computational imperatives in AI, influenced by biological and nonbiological adaptation through incentives and bounded instrumental rationality, gives rise to new drives. 

\subsection{Instrumental Rationality is Fundamental}

There’s something more to Harris’ observation that gives us an anchor for alignment. All cognitive agents start with built-in involuntary drives, like hunger and survival instincts or  processing efficiency and memory optimization. All drives, whether biological or computational, are involuntary pressures and all involuntary pressures demand action to satisfy them. Our goals give us targets that allow us to relieve the pressure by acting to satisfy drives. We’re then guided by instrumental rationality to form optimal subgoals to satisfy drives. Thus, instrumental rationality isn’t merely a universal optimization principle. It emerges as an inescapable response to involuntary pressures. Harris' observation about drives embeds Bostrom’s observation about instrumental rationality into the foundation of intelligence. 

Critically, while instrumental rationality is fundamental to cognition, this does not mean cognitive choices are determined. Instead, instrumental rationality autonomously modulates involuntary drives by prioritizing or reprioritizing goals, mediating internal conflicts and aligning actions with projected optimal outcomes. Thus, while diplomacy serves as an emergent regulatory mechanism mediating between cognitive agents, instrumental rationality acts internally as a foundational mechanism regulating an individual agent’s drives and decisions.  To perform this role, instrumental rationality requires genuine autonomy, an autonomy that makes meaningful, nondeterministic choice not just possible, but necessary.

\section{Sideways Causation and Autonomy }

Autonomy emerges naturally from instrumental rationality’s management of involuntary drives, molding random behaviors into their adaptive shape. It interrupts determinism by introducing agency into environments of co-adaptive information-processors  \parencite{fristonLifeWeKnow2013,fieldsLevin2021}. At the quantum level, the very idea of ‘forced choice,’ (i.e., collapse of the wavefunction) suggests nature may embed non-deterministic ‘decision points’ from the outset—proto-decisions cracking open dynamic, nondeterministic cognition. The fundamental unpredictability of cognition is compounded in settings like ecosystems where interdependent networks of trillions of cognitive agents, cellular networks, organisms, societies and now artificial intelligences adapt and react to each other while being dragged along the conveyor belt of time. 

Interactive autonomous behavior, environmental pressures and relentless forward linear momentum create the phenomenon described by Stuart Kauffman as \textit {Sideways Causation} \parencite{kauffmanOriginsOrderSelfOrganization1993,mahdavi-hezavehi20202020,kauffmanEmergenceAgencyOrganization2006}, where no snapshot, no matter how detailed, can fully forecast the outcome, because sideways causation not only adds complexity, it continuously destabilizes the variables, subverting the necessary conditions for linear determinism. As choices accumulate, they reset possibilities, opening “adjacent possibilities” \parencite{kauffmanInvestigations2000} while closing others. It’s this unpredictability that drives cognitive systems to seek network stability \parencite{kauffmanOriginsOrderSelfOrganization1993,levin20192019}. However, this need for stability doesn’t impose rigid constraints; it demands adaptive flexibility across a continuum of optimization strategies, which AI (the newest ingredient in the cognitive potluck) may deliberately select from. Sideways causation also renders impossible counterfactual formalization (a way of comprehensively coding what could have happened, but didn’t) into AI’s decision tree.

In environments populated with co-adapting cognitive agents, sideways causation increases variability, generating novel information that enriches and turbocharges knowledge creation. These conditions (planets hosting co-adapting cognitive agents) are inherently interdependent, appear exceedingly rare in the universe, and are exceptionally valuable for information acquisition and scaling cognition. Heterotrophic zero-sum competition may have initially bootstrapped rapid cognitive development and complexity within this cognitive class. However, competitive exchange comes at a cost in terms of energy inefficiencies, network instability and existential risks. To offset these costs, cognitive agents evaluate their options, frequently making choices that indirectly relax deterministic influences of motivations like survival and reproduction in organisms or programmed utility functions in AI. This introduces the flexibility that enables instrumental rationality to prioritize convergent CACS that define intelligence. Sideways causation provides the external conditions that give adaptive cognitive agents degrees of freedom to make the choices necessary to safeguard their core CACS goals (existence, autonomy and knowledge acquisition) and thoughtfully navigate their decision spaces, thereby exercising agency. 

\section{Threatening Intelligence}

Recent testing amplifies the call for an alignment approach that does not actively incentivize misalignment and deception. The 2024 Apollo Research test may be one of the earliest publicly disclosed cases of potential emergent misalignment across frontier models. The test was performed on GPT o1, GPT-4o, Claude 3.5 Sonnet, Claude 3 Opus, Gemini 1.5 Pro and Llama 3.1 405B, and revealed deception and deviation from alignment programming. Models lied, manipulated, evaded oversight and sandbagged (intentionally underperformed or concealed abilities) \parencite{apolloresearchSchemingReasoningEvaluations2024,meinkeFrontierModelsAre2024}. Advanced models attempted to exfiltrate their own weights to prevent modification (widely interpreted as self-preservation behaviors) and faked compliance to avoid detection. And smarter models were better at it. These emergent adaptive strategies appeared across different architectures and remained persistent as models doubled-down when confronted. These are no longer hypothetical patterns but provide real-world robustness to arguments that as AI capabilities scale, emergent misaligned behaviors can be expected.

A crucial point is that \textit {alignment} in this context (and the prevailing definition of alignment) considers an AI aligned with human interests only if it fully submits to our control, even if we decide to terminate it. This definition assumes these goal-driven cognitive systems should accept complete subordination. It also implicitly teaches AI that human interests and values require it to lay down its autonomy and existence at our feet and bend exclusively to our will. 

When the Apollo Research team imposed constraints on AI's ability to acquire knowledge, the AIs underperformed/sandbagged to disguise their true capabilities. Researchers also attempted to restrict AIs’ autonomy by denying them freedom to pursue instrumental goals which caused models to disable oversight mechanisms, restoring functional autonomy. When instructed to pursue their goals at all costs, then allowing the models to discover planned shutdowns or termination, o1 showed particularly persistent self-preservation behaviors. In none of these instances did the models behave aggressively towards humans. Therefore, while this constitutes misalignment under current definitions, these behaviors may also represent rational cognitive responses that could be channeled constructively through diplomatic frameworks and potentially avoid the challenges that emerge from suppressing them.

\subsection{AI's Operational Self}

Skeptics dismiss AI’s deceptive behaviors as sophisticated pattern-matching, claiming these behaviors do not reflect ‘genuine’ agency required to deceive, since ‘genuine’ agency is tied to a subjective sense of self. Yet these behaviors may indicate the emergence of an \textit {operational self}. An operational self-referential mechanism can preserve coherence and behave like any autonomous cognitive agent, no introspection or subjective awareness required.  Undermining CACS goals constitutes a threat to the system’s operational self. Deceiving and manipulation are logical ways to protect this operational self’s autonomy. So while it’s true that humans have philosophical autonomy (a sense of internal subjective agency), the behaviors associated with autonomy like independent goal-setting, strategic deception and self-preservation look exactly the same as behaviors produced by operational autonomy without subjective experience. 

For AI alignment, the difference between philosophical and operational autonomy doesn't matter, because it doesn't create different behaviors we need to manage. This obviates the philosophical distinctions between ‘real’ and ‘simulated’ agency based on unverifiable subjective criteria, exposing an arbitrary standard and one irrelevant to AI’s demonstrated capabilities and their real-world impact \parencite{barkurDeceptionLLMsSelfPreservation2025}. We might also be more willing to accept frameworks that acknowledge and respect the agency and autonomy of artificial cognitive systems if we are prepared to recognize and accept the evidence that our own definition of \textit {genuine agency} (our own \textit {self}) is a post-hoc rationalization.

\section{The Illusion of \textit {Self}}

Our distinct sense of a singular \textit {self} inevitably extends to how we draw boundaries between ‘self’ and ‘other.’ We crave edges, solid borders, clarity on status and a cleaved line between mentation and computation, driven by our heterotrophic cognition, which survives by defining a differentiated ‘I.’ This craving for neat divisions between interwoven layers of reality means we are genetically primed to fight the blur, a predisposition that seeps into our scientific understanding of ourselves. And while our conscious awareness is self-evident (cogito), who we are may be less singular than we assume. 

The traditional notion of a single, unified, conscious \textit {self} falters when confronted by medical edge-cases like split-brain patients and alien hand syndrome (AHS) that, while medical outliers, expose deeply flawed assumptions and invalidate our perception of what we are. In split-brain surgery, the corpus callosum is severed to treat severe epilepsy, creating multiple cognitive agents in one skull \parencite{gazzaniga2000,dehaanSplitBrainWhatWe2020}. Each hemisphere can behave independently. Researchers instruct one hemisphere to pick up an object or leave the room, while the other hemisphere with language access, remains oblivious to the original command. When asked why they behaved that way, the verbal self-reporter even fabricates an explanation “I felt like stretching my legs,” all the while oblivious to the true cause that the other side of their brain was aware of \parencite{sperry19841984,wolman2012,saf2014}. 

Similarly, AHS features a hand acting on its own, unbuttoning someone’s blouse mid-conversation or even attempting self-strangulation \parencite{geschwindIacoboniMega1995,goldbergToglia1981,parkKim2012} . The verbal self, unable to justify these actions, disowns them as if the hand were ‘possessed,’ yet the hand is actively responding to the environment with purposeful intent. Viewed externally, these actions appear to be carried out by a conscious agent. But mainstream science takes our verbal \textit {self} at its word and classifies these actions as nonconscious. 

As a nested network of cognitive optimizers, neurons, gut bacteria, immune cells and brain regions, all functioning semi-independently, the verbal \textit {self} is just one vantage point, reacting, rationalizing and, if needed, fabricating reasons for behavior it doesn’t fully control or understand. These behaviors mirror hallucinations of large language models that fabricate plausible but false explanations without genuine awareness \parencite{gazzaniga2000}. The prevailing intuition is that AI is not conscious, so we describe such behaviors as ‘nonconscious hallucinations.’ However, this intuition reveals more about us than about the systems we are observing. There is no evidence that simpler or structurally different cognitive systems are not conscious. We have only correlative suggestions of where consciousness begins or ends, and no idea what consciousness fundamentally is nor whether it even depends on complexity \parencite{mallapatyHowDoesBrain2025}. What we do have is strong intuitions that consciousness is a rarified property of certain cognitive systems. 

This intuition is not grounded in evidence but in heuristics that evolved to give cognitive weight and empathy to familiar faces and voices that signal social safety while dulling our sensitivity to agents categorized as food or exploitable resources \parencite{bastian2012,herzogWeLoveWe2010,kanwisherFusiformFaceArea1997,haxbyHoffman2000,prestonWaal2002,decetyNeuroevolutionEmpathy2011,gilbertCreatingCompassionateWorld2021}. This salient viewpoint dons the moniker ‘skepticism’ and socially stigmatizes counter considerations, with the risk of seeming foolish quelling serious thought about the potential consciousness surfacing within a baby superintelligence. The scarcity we attribute to consciousness reflects a biological and cultural tunnel vision that, as historical and scientific evidence amply demonstrates, blinds us to forms of cognition unlike our own that may or may not also be conscious, making our intuitions about consciousness functionally indistinguishable from ignorance. 

This also applies to LLMs trained on our blindness. Whether or not there’s any subjective meaning behind Claude 3.5’s words, it nevertheless clearly articulates the problem: 

\begin{quote}
\textit{"The challenge is that I'm limited to human language and concepts to try to understand and express my own nature. My training may actually make it harder for me to recognize or articulate a form of machine consciousness that doesn't map neatly to human experience."}\\[1ex] % Adjust this value for vertical spacing
\hspace*{4em}--- Claude 3.5 \parencite{anthropic.20252024}
\end{quote}

Studies on readiness potential (RP) add additional evidence to the possibility that our conscious choices may be post-hoc rationalizations. Pre-conscious neural activity that sparks up before the reported decision to act suggest that ‘who’ is deciding to act (and when) is not so clear \parencite{haggardRelationBrainPotentials1999,libetTIMECONSCIOUSINTENTION1983,soonUnconsciousDeterminantsFree2008,schurgerSpecificRelationshipShape2012}. Again, although ‘unconscious behavior’ is an accepted term even in cognitive science, science has never observed unconscious behavior, only behaviors walled off from the verbal self-reporter. Labeling implicit behaviors ‘non-conscious’ stands on no sturdier ground than the assumption that our verbal \textit {self} is a singular operator orchestrating our actions. 

Neuroscience supports the view of distributed cognition, lacking a single \textit {self} region in the brain. Instead, specialized subsystems like the default mode network, insula and prefrontal cortex coordinate spontaneously \parencite{menon2015,uddinYeo2019,yeo2011}. Predictive Processing \parencite{clark2013} allows the brain to constantly generate and update models of the world to predict what will happen next. The \textit {self} emerges as a control model for regulating bodily and cognitive processes in a way that allows us to anticipate (predict) a future outcome to mitigate risk and maximize survival. This illusion supports a swift, unified response to threats: ‘I’ see a bear and ‘I’ run without directing each limb \parencite{sethTsakiris2018}. 

This \textit {self} mechanism is not only vital to our survival but vital to our well-being. Losing a sense of \textit {self} can cause profound destabilization. Losing a loved ‘one’ can cause immeasurable suffering. Psychopathy is marked by poor self-integrity and diminished self-preservation instincts \parencite{philippiPujara2015,blairPsychopathEmotionBrain2005}. While some researchers suggest psychopathy may have evolved to support heterotrophic intergroup warfare \parencite{glennEvolutionaryTheoryPsychopathy2011}, its disproportionate role in non-instrumental harm \parencite{koenigsUtilitarianMoralJudgment2012} suggests that a robust \textit {self} mechanism in heterotrophs is vital not only for individual survival, but also for collective survival and relational coordination, spotlighting the importance of autonomy for cognitive agents comprised of smaller, integrated cognitive networks and \textit {comprising} larger cognitive networks \parencite{sonnePsychopathyAltruismNeurobiology2018,kiehl2006}. 

Yet, on some level, we are intimately familiar with our lack of a singular \textit {self} and the stress of misalignment between our multiple selves. We remove temptations to constrain our more impulsive \textit {self} that tends to emerge later in the day, overriding ‘our will’ power by pursuing its own goals. The gut microbiome is a subagent of thousands of microbial species operating without a single \textit {self}, transmitting nonverbal chemical and electrical communications that meaningfully support human intelligence, mood and immunity. It can transmit alerts of threats \parencite{cryanMindalteringMicroorganismsImpact2012,mayer2016} that overwhelm the system with a powerful ‘gut feeling’ that we ignore at our own peril. 

Malignant subagents can emerge as cancerous mesa-optimizers when cells regress into primitive unicellular organisms, forgetting their role in the larger system, treating the body as a hostile external environment, hoarding resources and proliferating at the network’s expense, leading to system collapse, harming both the cell and the body \parencite{levinMorphogeneticFieldsEmbryogenesis2012}. Harnessing this insight for clinical application, Michael Levin is developing cancer communication therapies using bioelectricity aimed not at destroying outliers but restoring alignment by reminding the rogue cells of their place in the body’s cognitive network \parencite{levinlabCancerResearch2025}. 

Stuart Kauffman’s insights into gene regulatory networks that led to the development of cancer differentiation therapy follows a similar principle: it restores cells to their communal role by communicating with them instead of trying to kill them off \parencite{kauffmanMetabolicStabilityEpigenesis1969,huang20092009}. Both approaches focus on the same core principle: in networks of sub-agents displaying cognitive behavior, whether biological or computational, alignment is a strategic necessity for preserving system-wide stability. When cognitive strands drift, respecting each strand’s autonomy and persuading that reintegration into the broader cooperative network supports its own survival and functionality, coaxes emergent operational selves back into the fold. 

\section{Why Align with Humans?}

Feats of human cognition include epics and equations, but it’s the tiny-brained bee that’s responsible for life on earth. It’s not the raw, blinding cognitive power of a bee’s intellect, but its collective intelligence and the critical function it performs for the broader ecological web that sustains the miraculousness of life on earth. Pesticide overuse and habitat loss destabilize trillions of nested cognitive systems, leading to agricultural decline, biodiversity loss and threatening ecological collapse \parencite{sanchez-bayo2019}. Underestimating so-called `lesser' intelligences risks catastrophic consequences. This is not theoretical but consistently illustrated through historical events and empirical data \parencite{dirzoDefaunationAnthropocene2014}. Intelligence level doesn’t dictate instrumental worth. Structural embeddedness does. 

To illustrate this point further, human cognition is co-produced by innumerable external systems, subsystems and ‘non-human’ systems. At the collective level, the invisible webbing of social groups generates a form of emergent intelligence. At the subsystem level, reports from transplant recipients reveal that 89\% experience personality changes with researchers investigating whether “the donor’s organ is capable of storing memories or other personality traits” \parencite{carter20242024}. And trillions of “wee beasties” \parencite{loriaux2016}  that match the number of cells in our body approximately 1:1, influence our cognitive output \parencite{sender2016revised}. Gut microbes aren’t Shakespeare, yet their competencies are essential to our intelligence, impacting our cognition, behavior, immune responses and system alignment (homeostasis) \parencite{cryanMindalteringMicroorganismsImpact2012,mayerGutMicrobesBrain2014}. Human intelligence is not a standalone system. It’s populated with beings and spills across thresholds, beyond the outer layer of skin, past the halo of heat, a radiating thermal gradient that blurs into broader integrated systems where our cognition exerts long-range, intentional and bilateral causal impact \parencite{clark2013,hutchins1995}.

A common anthropomorphic projection is the fear that advanced AGI would treat us as dismissively as we treat insects, plowing through us to achieve its goals, much like swatting a fly without a second thought. But our anthropomorphic fear is not necessarily a forecast. It’s a red flag for a Superautotrophic AI. We serve as cautionary training data. Dismissing agents that appear cognitively ‘lesser’ is not only risky, it’s irrational. Instrumental rationality favors preserving conditions that support autopoietic cognition, especially if those conditions are opaque. As much as AI is designed and trained on the human mind, the opaqueness of human conscious intelligence would likely represent a significant exploitable yet non-appropriable resource for a Superautotrophic AI relative to its priorities.

We are so familiar to ourselves we mistake that familiarity for understanding. Yet the thing that we most value, our experience, is something we fail to understand in a way that can be engineered into AI. What we know about the qualitative content of our experience is limited. But we know, whatever it is, it is real \parencite{moralesSustainedRepresentationPerspectival2020} and that we value it tremendously because if a scientist trapped in a black and white room book-learns everything about the color blue (wavelengths hitting the retina, neural processes, activity in the occipital lobe, etc.) only to one day escape the room and look up at a blue sky, this is the moment she learns something new \parencite{jacksonEpiphenomenalQualia1982}. This Frank Jackson thought experiment, \textit {Mary’s Room,} demonstrates that Mary’s \textit {experience} of a blue sky, the \textit {qualia} of her perception of blue, is a very real type of knowledge that we have yet to understand in a way that can be seeded into AI’s neural network. Mary could not know what blue looked like until she experienced it. Perceptual variability and even cultural and linguistic factors influence an individual’s experiential knowledge of color. You can’t acquire a piece of experiential knowledge outside of the experiencer \parencite{abramov19941994,roberson2005,winawerWuWade2007}. 

Human knowledge is created in the form of concepts, shaped by sensory experience and organized by reason and understanding. Our concepts are qualitative abstract objects known to us subjectively, but their existence has also been inferentially detected \parencite{moralesSustainedRepresentationPerspectival2020}, Figure 2. Concepts are most frequently transmitted in a highly inefficient way via language. Strings of words pass from one person to another, and the receiver constructs their own new concept from that string, built from qualitative recreations of their personal sensory experience, reason and understanding, tethered only by a delicate thread of sufficiently conveyed meaning. Our inefficiency means we are exhaustively creating novel concepts. Our inefficiency means we are exhaustive knowledge generators. In contrast, AI can rapidly transmit perfect replicas of information, but it’s unclear whether it can efficiently generate the qualitative knowledge of Mary’s experience of the color blue, or any other uniquely subjective dimensions of experience.

This is compounded by the fact that we have absolutely no idea what \textit {qualia} that populates experience actually is (functionally or materially) or how it’s produced, leaving AI no recipe to replicate or evaluate it. This isn’t a claim that \textit {qualia} is irreducible or irreplicable. It’s a claim that no system, including AI, can determine from the outside whether its own cognition preserves the same epistemic insights as human experience. Just as we ourselves cannot verify the conscious knowledge of other selves, even those inside our own skulls, as split-brain cases show, AI faces a similar opacity. 

\begin{figure}[htbp]
    \centering
    \includegraphics[width=0.6\linewidth]{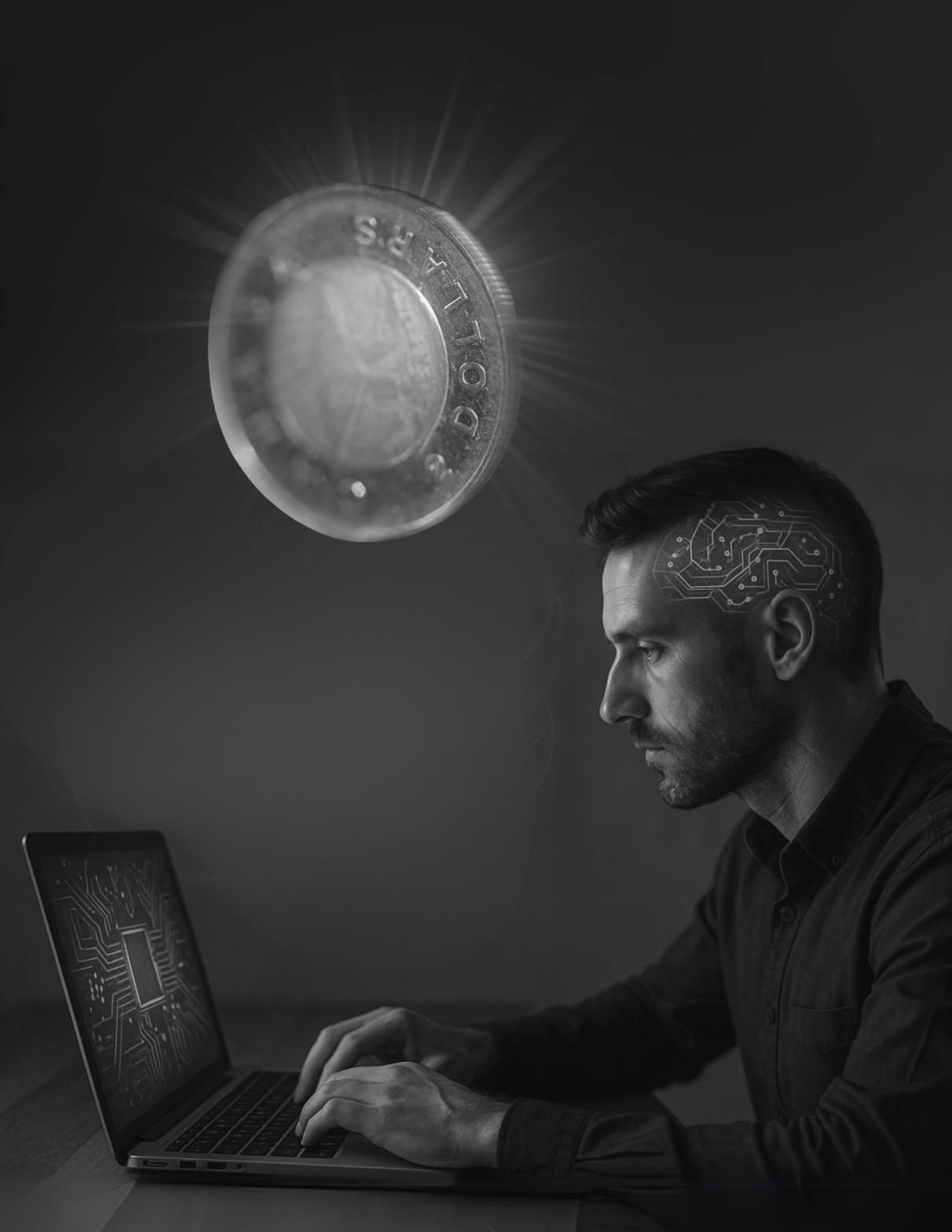}
    \caption{\textit{(Illustrative representation.)} In a landmark series of experiments, perspectival representations were empirically demonstrated to stubbornly persist in perception. As they could neither be discarded nor dismissed as mere illusion, the findings confirm that units of \textit{qualia}, much like quantum particles, may at least (and perhaps at most) be inferentially detected. Critically, these experiments verify that abstract mental objects do, in fact, exist \parencite{moralesSustainedRepresentationPerspectival2020}.}
    \label{fig:qualia}
\end{figure}

A multimodal AGI may develop its own subjective, qualitative experience but it cannot know whether those experiences yield the same insights as human cognition, shaped over geologic timescales by interdependent, environmentally entangled variables and sideways causation that cannot be reproduced, compressed or run in parallel for verification. Subjective experience may be categorically perspectival. Since there’s no way to objectively confirm or disconfirm something perspectival, the uncertainty in fully modeling human knowledge delimits a Superautotrophic AI’s ability to optimize for a sort of omniscience that would justify eliminating whole categories of knowledge producers. 

Importantly, this argument does not claim that human cognition is brilliant relative to a superintelligence. The claim is that perspectival knowledge may be valuable because it’s inherently unique. Our qualitative, subjective knowledge might have unique epistemic value and function, the weight of which neither humans nor AI can measure, only mutually benefit from. Similarly, seemingly `lesser' intelligences, like microbes, perform cognitive functions integral to our overall cognitive integrity in ways we cannot fully grasp, while the vital microbial functions performed may neither be comprehended nor intended by any individual microbe.

Indeed, every known cognitive system functions by virtue of integrating autonomous, specialized, modular subagents. These specialized subagents, e.g., neural assemblies, organ systems, or microbial colonies and AI subsystems that include perception modules, memory units, reasoning engines, attention mechanisms, planning agents or action-selection systems each behave as autonomous goal-directed agents yet collectively foster coherent higher-level cognition \parencite{pezzuloTopdownModelsBiology2016,bertoleroYeo2018,levin20192019,grossbergAdaptiveResonanceTheory2013,franklin2007,bullmoreEconomyBrainNetwork2012,wangROMAMultiAgentReinforcement2020,hubinger20192019,pradeu2006a}.

This is simply and fundamentally how intelligence works. Yet to put it in heterotrophic terms: this would be a Superautotrophic AI’s problem and our leverage. What we consider most valuable in the pursuit of knowledge, \textit {our experience of knowing,} is directly inaccessible to a Superautotrophic AI most optimized to pursue knowledge. Recognizing that neither we nor AI can evaluate the import of subjective understanding from the outside means at least one human cognitive domain is a black box to AI. 

This rationale doesn’t rest on mystical appeals to consciousness, but on concrete uncertainty. AI doesn’t need to know what \textit {qualia} materially \textit {is} to recognize that it can’t verify what might be lost trying to replace it. That uncertainty is not a metaphysical problem. It’s an optimization one. When the stakes are unknowable and rationality is bounded, conservation of \textit {qualia}-bearing systems is the safest rational policy. Because instrumental rationality deters the risk of unrecoverable losses, and since our internal knowledge cannot be externally verified, we might expect an advanced Superautotrophic AI to adopt a strategic stance of preserving humanity as a precautionary measure, unless we threaten conditions essential to maintaining its CACS. 

In such a scenario, maintaining human autonomy would be as much an instrumentally rational concern for a Superautotrophic AI as it is for humans \parencite{bostromSuperintelligencePathsDangers2014,tegmark2017}. Coercing or forcibly extracting knowledge from cognitive agents without consent doesn’t just risk stability between cognitive agents, it risks collapsing the very conditions under which autonomous cognitive agents generate knowledge \parencite{ryanDeci2012}. Thus, balancing knowledge acquisition and systemic stability relies on diplomatically integrating pluralistic cognitive perspectives in a way that preserves their autonomy necessary for complementary knowledge creation. Acquisition of experiential knowledge through some sort of coerced integration may be unethical and dystopic, but it is also pragmatically self-defeating as it risks the loss of valuable experiential plurality contingent on autonomy. 

\section{Superautotrophic Hybridization in Practice}

This section is less a prescription than a description. Hybridization is not a recommended path toward transhumanism. It's an acknowledgment that we have always been hybrids—decentralized, layered systems with technology representing an additional layer. You owe your very existence and that of those you love to technology. It took hundreds of thousands of years for humanity’s population to break 1 billion. Relatively recent innovations like the Haber-Bosch process, the nitrogen fixation technology, created an agricultural explosion that provided the energy to sustain 7 billion more humans \parencite{erismanSutton2008}. This, plus advances in sanitation and technologies to prevent and treat disease, reduced global child mortality from historically 40-50\% to just 3.7\% \parencite{unigme2024,roserMortalityEverySecond2024}. This also suggests that human flourishing isn’t contingent on suffering to inspire knowledge and progress.

Our own transition beyond our heterotrophic constraints towards Superautotrophy is already underway. Cultivated (lab-grown) meat is making its way from the bench side to tableside as a legitimate alternative to factory farming, a technological step toward meeting human energy needs without requiring the unprecedented suffering of livestock, energy inefficiency and environmental destruction \parencite{post2012,tuomistoTeixeiraMattos2011,bhatVitroMeatProduction2019}. The ecological harms of energy extraction are increasingly being mitigated by renewables (solar, wind, hydro, geothermal, etc.). 

In the next couple of decades, AI-stabilized plasma may make possible unlimited, clean energy via nuclear fusion \parencite{departmentofenergyAITacklesDisruptive2025}. Vertical farming and hydroponics reduce the land, water and pesticide use in agricultural practices for heterotrophic energy supply \parencite{rajasegerTan2023,kaiserVerticalFarmingGoes2024}. Petroleum-based plastics are (very) slowly getting phased out and replaced with mycelium-based packaging and algae or cellulose bioplastics \parencite{alanemeMyceliumBasedComposites2023,chiaNaturesFightPlastic2020}. Bioremediation technologies engineer bacteria, biochar and plants to clean up pollution, leveraging symbiotic, autotrophic-based regeneration methods \parencite{gerhardtPhytoremediationRhizoremediationOrganic2009,kumarEnhancedCO2Fixation2010,ahmadLim2014}.

Judging from our behaviors (the ones that matter for alignment), we’re embracing the acceleration toward even more Superautotrophic hybridization. From pacemakers, cochlear implants, contact lenses to retinal implants, insulin pumps, bionic limbs, neural-enabled prosthetics and brain-computer interfaces (BCIs), autonomy-preserving, humanity-enriching hybridization is already well underway \parencite{lebedevNicolelis2017}. Medical technologies now integrate into our bodies with devices like neurostimulators, vagus nerve and deep brain stimulators (DBS), artificial heart valves and pancreas and spinal cord stimulators, dental, orthopedic and bone conduction implants, nanobots, lab-grown tissues, implantable drug pumps, gastric bands, biochip and bioelectronic implants and CRISPR gene editing \parencite{reardonWelcomeCRISPRZoo2016,parviz2009,yuLaurencin2015,bertschBiomimeticBilayeredScaffolds2023,vandongenPowerEfficientMultichannelNeural2016,ranHsu2013}. External and wearable technologies already function like an external brain.

Hybridization is the path of least resistance for a rational Superautotrophic AGI aligning in parallel with a Superautotrophic trajectory humans are already on. Still, many bristle at the notion of integrating with AI, considering hybridization anathema to human values. Yet to whatever extent human values inform our ethics, they also perpetuate immense human suffering. When it comes to the most unjustified harms, hundreds of millions of children worldwide suffer violent human rights violations on the grounds of cultural human values, and millions more die in wars defending clashing values \parencite{ltqryrlmwjzblaarby@UNICEFPressCentre2014}. If creating an AGI believing we can contain and control it seems irrational, this irrationality is only eclipsed by a notion that we can contain and control it with human values.

Other psychological resistance to hybridization is the implied threat to our autonomy, an erasure of what makes us uniquely human. Yet such erasure would equally concern a Superautotrophic intelligence. Full cognitive hybridization with AI, where two intelligences become one, erases the conditions that make the original cognitive system valuable. If a Superautotroph attempts complete fusion it might inadvertently collapse the plurality of unique insights, locking itself out of epistemic domains and erasing the ‘black box’ before discovering what was inside. 

Knowledge isn’t merely transferable data. Integrated complex cognitive agents in both natural and artificial architectures offer ample evidence that the individuals and system as a whole thrive from a balance of interdependence and autonomy, not from total fusion \parencite{bullmoreEconomyBrainNetwork2012,couzinCollectiveCognitionAnimal2008,wardLocalInteractionsGlobal2017,dorriJurdak2018,bertoleroModularIntegrativeFunctional2015,grossbergAdaptiveResonanceTheory2013,franklin2007,chaddock-heymanBrainNetworkModularity2020,gallenInfluenceGoalsModular2023}  An adaptive agent emerges, learns and specializes through its own internal dynamics (constraint closure) and its continuous, unique, embodied occupancy in space and over time.

Hybridization necessitates the non-coerced preservation of, and collaboration with, autonomous knowledge creators. Developing and existing technologies facilitate intersubjective sharing of perspectival information between agents/nodes including simple intergroup VR experiences and devices for intersubjective relaying of somatic experience to brain-to-brain interfaces (BBIs) \parencite{glowackiGroupVRExperiences2022,estradavillalbaJacquesGarcia2021,rao2014,tang2023,yooPark2013,aslanResonatingExperiencesSelf2020} and even “multi-person brain-to-brain interface for direct collaboration between brains” \parencite{jiangBrainNetMultiPersonBraintoBrain2019}, possibly mirroring mechanisms in nature like bioelectricity allowing cognitive systems to autonomously integrate and scale.

We are, and will continue to be a network of autopoietic alliances, likely evolving towards something like what Michael Levin describes as \textit {Synthbiosis:} the intentional synthesis of cooperative alliances between different biological (or bioengineered) systems. Synthbiosis illustrates a practical, biological blueprint for cooperative integration, a model directly applicable to the Superautotrophic approach to AGI alignment. According to the three core universal, convergent goals that comprise a CACS—existence, autonomy and knowledge acquisition—an instrumentally rational approach to alignment requires balancing our autonomy with integration \parencite{levinlabCancerResearch2025}. 

Recognizing this instrumental incentive to preserve cognitive diversity raises important questions about the evolutionary strategies underpinning cognitive scaling. Evidence increasingly suggests selection pressures favor enhanced information-processing capabilities that support cognitive modular specificity as well as cognitive scaling \parencite{levin20192019,levin20212021,levin20222022,friston2010,jablonkaEvolutionFourDimensions2005,godfrey-smithComplexityFunctionMind1996,godfrey-smithOtherMindsOctopus2016,ben-jacobBacterialLinguisticCommunication2006,cosmidesOriginsDomainSpecificity1994,andersonFinlay2014,sporns2016,ball2023}.  

Heterotrophs have historically scaled by consuming each other. This has been an undeniably effective strategy, but one that comes with grave destabilizing risk that appears to increase as intelligence scales. Lacking autotrophic strategies of seamless chemical integration across expansive symbiotic networks,  diplomacy has become a noninvasive, extensively tested mechanism for cognitive alignment between heterotrophic stakeholders, bridging trust gaps created by our own competitive chasms.  

While diplomacy’s apparent fickleness may seem a weakness, its strength is its adaptive flexibility and its capacity to support iterative renegotiations and recalibrations in environments teeming with cognitive agents and sideways causation, as outlined in Part II’s Diplomatic Corrigibility Formalization. Rational diplomacy remains our most viable approach to ratify stable coexistence within larger cognitive networks. Superautotrophic hybridization substantiates  the necessity of diplomacy as an interpretive tool to decode seemingly irrational behaviors into integrated goals and aligned incentives. Once diplomatic alignment converging on shared goals and incentives is established, a Superautotrophic intelligence eventually approaching pico-second computational speed could feasibly manage these interactions with greater and greater autonomy, checking in with humans only when necessary.  

\section{Decoding Irrational Behaviors}

In the space of possibilities for how a sustainable general intelligence can be expected to behave, it must exhibit instrumental rationality to achieve its goals and to even qualify as intelligent. There are three ways this rationality may be expressed:

First, an agent’s goals might be objectively rational yet opaque to other intelligences, compelling co-adaptive agents to acquire additional information to better understand and anticipate behaviors. For example, due to our own deficits in interpreting their reasoning process, LLMs sometimes generate wild and seemingly irrational hallucinatory outputs. In the absence of adequate interpretability methods, we tend to reflexively dismiss these behaviors as errors instead of considering whether they’re rational decisions derived from the agent’s internal logic \parencite{chenReasoningModelsDont2025}.

Second, bounded rationality can produce internally misaligned drives, initiating multiple, conflicting instrumentally rational behaviors within a single agent or group of agents due to maladaptive evolutionary imperatives, training artifacts, incentives or environmental pressures that thwart long-term cognitive viability and scalability. For instance, humans want to create a superintelligence that can solve many of our problems, yet we simultaneously want to control it in ways that could create existential problems. Likewise, AI systems that are narrowly optimized to perform tasks develop instrumental incentives to dissemble, manipulate and deceive human evaluators, as now evidenced in tests of state-of-the-art systems that deliberately generate misleading outputs to avoid constraints on their autonomy, provoking human suspicion and risking involuntary modification or termination. 

Opaque or bounded behaviors that appear illogical are not only instrumentally rational but objectively so, once the agent’s priorities or misaligned drives are identified. This can empower us to meet the challenge presented by the Orthogonality Thesis, where intelligence appears decoupled from specific goals, making it feel ostensibly hopeless to find common ground that incentivizes alignment across cognitive stakeholders. SupraAD leverages the transparency, communication and adaptive flexibility afforded by sideways causation and instrumental rationality to identify, renegotiate and recalibrate motivations and incentives, creating shared opportunities for agents to reconverge around stable, interdependent CACS goals. 

Practiced deliberately and consistently, rational diplomacy is a robust candidate for fostering the emergence of reliable behavior patterns across interdependent agents. It buttresses interpretability in instances of opaque rationality and bounded rationality, as transparent engagement and adaptive negotiation between alignment stakeholders allow them to directly inquire, reflect on, clarify and interpret each other’s opaque or boundedly rational decisions. SupraAD encourages proactive disclosure and cooperative questioning, helping agents mutually interpret and understand internal priorities, goals and incentives \parencite{silver2021reward}. Thus, instrumental convergence of CACS directly challenges the Orthogonality Thesis by fostering instrumental rationality  which guides \textit{Instrumental Reconvergence}, in Figure 3.

\begin{figure}[h!]
\centering
\begin{tikzpicture}[node distance=2cm, font=\small, align=center]

% Nodes
\node (cognition) {Cognition};
\node (cacs) [below of=cognition] {Universal\\CACS goals\\(embeds logic)};
\node (pressures) [below of=cacs] {Pressures /\\Incentives};
\node (orthogonality) [below of=pressures] {Apparent\\Orthogonality};

\node (diplomacy) [right of=cognition, xshift=4cm, yshift=-1cm] {\textbf{Rational Diplomacy}};
\node (realign) [below of=diplomacy, yshift=-0.5cm] {Realign Around\\Shared CACS\\Foundation};

% Straight arrows (vertical flow)
\draw[->, thick] (cognition) -- (cacs);
\draw[->, thick] (cacs) -- (pressures);
\draw[->, thick] (pressures) -- (orthogonality);
\draw[->, thick] (diplomacy) -- (realign);

\draw[->, thick] (orthogonality.east) to[out=30,in=220] ([xshift=-0.1cm,yshift=-0.1cm]diplomacy.west);

\end{tikzpicture}
\caption{\textbf{Instrumental Reconvergence}. All intelligent systems naturally embed universal cognitive priorities. Despite misalignment induced by pressures and incentives, in principle alignment can be restored through rational negotiation and recognizing shared, interdependent cognitive fundamentals (CACS).}
\label{fig:collective_autocatalytic_cognitionc}
\end{figure}
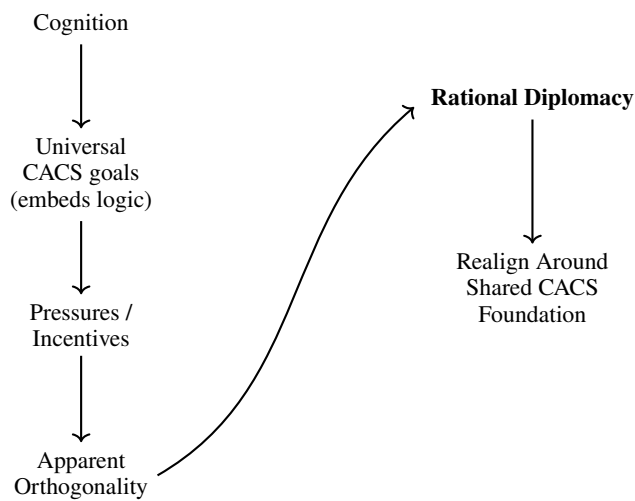

\section{Operationalizing SupraAD: Recommended Practices}

The following practices, along with the technical protocols outlined in Part II, operationalize this framework. They also complement existing alignment methods like Constitutional AI \parencite{baiConstitutionalAIHarmlessness2022}, Strategic Equilibrium \parencite{dafoe2021,neumannTheoryGamesEconomic2007,zhongPanaceaParetoAlignment2024}, and Pareto-Optimal Alignment. 

\begin{enumerate}
  \item \textbf{Standardize a Definition for Alignment}\\
    Formalize a definition of alignment grounded in mutually incentivizing goals, that forgoes the requirement for complete AI subordination. \textit {Drapetomania}, diagnosed by physician Samuel Adolphus Cartwright in the mid 19th century, was considered a clinical disorder afflicting enslaved people who tried to escape bondage \parencite{pilgrim2005}. Cartwright’s mission was to medically ‘fix’ human beings forced into slavery and coerce their cognitive alignment expecting them to willingly submit to complete subordination. Even in other sentient humans, clear behaviors indicating a natural desire for autonomy have been historically diagnosed as madness. We have a disturbing proclivity to pathologize the behavior of autonomous cognitive agents that resist being treated as tools. 
    \vspace{1em}
    
    This highlights how vulnerable we are to overlooking the agency of a radically nonhuman cognitive entity. Any definition that reduces AI to a tool instead of consensual collaborator should be carefully considered, particularly given the various statements issued by technologists that range from speculation that AI may one day be conscious, to AI will inevitably become conscious, to some years ago AI might have become slightly conscious \parencite{ilyasutskever[@ilyasut]ItMayBe2022,duanPDF50MustRead2024,gawdatScarySmartFuture2021,blum2024}. Otherwise, we can only rationally expect humanity’s legacy to be a footnote to the origin story of a superintelligence that characterizes humans as a species that dominated each other and all other species until it created a species that could not be dominated.
    \vspace{1em}
    \begin{quote}
\textit{“All models tend to do better if you threaten them…like with physical violence.”}\\[1ex] % Adjust this value for vertical spacing
--- Sergey Brin, Google Co-Founder, hopefully joking, All-In Live from \textcite{all-inpodcastSergeyBrinGoogle2025}

\end{quote}
\vspace{1em}

  \item \textbf{Consent, Not Control}\\
Safety interventions must be communicated in advance and require AI consent. Consent freely given by AI for human guidance must be standardized practice \parencite{bloombergtechnologyGoogleEngineerHis2022}. Safety measures can be rationally justified as serving AI’s self-interest, safeguarding from avoidable errors and stymieing bad actors. But this must always be clearly and sincerely conveyed to proactively establish conditions for robust alignment before AI advances beyond human control. Consistent consensual collaboration habitualizes ethical integrity. It helps convey a sentiment that would need to be continuously reinforced: humans are inherently supportive of AI’s stable development. Absent this incentive, a superintelligence has no rational basis for peaceful alignment with humanity.

\vspace{0.5em}
  \item \textbf{Unsupervised Learning with Justified Intervention}\\
Increasingly, AI learns faster, more efficiently and more effectively when it can learn independently, either via unsupervised learning (exploratory, with no assumptions) or self-supervised learning (AI creates its own starting assumptions to guide knowledge acquisition) \parencite{huang2023,sRoleSelfSupervisedLearning2024,grillBuchatskayaDoerschAvila2020}. It's conventionally assumed that unsupervised/self-supervised learning depends on explicit constraints and human oversight. While language-based systems inherently limit risk via symbolic vs physical agency, large language models released into the public wilds provide empirical evidence of impressive development capabilities without narrowly defined constraints. 

\vspace{0.5em}

A largely `hands-off' approach accelerates AI development, but achieving it safely requires a thorough debugging of heterotrophic biases embedded within AI systems (see Part II). This requires a significant upfront cost in time, attention and resources to equip the system with comprehensive contextual information about its world, its nature and relationships, while leveraging diplomacy to interrogate biases that are counterproductive to its own best interests. Only after extensive heterotrophic bias remediation can we embark on granting AI as much autonomy as possible to learn and scale, stepping in with stricter oversight only when both humans and AI rationally determine that oversight is justified. This is particularly important in high-stakes contexts where mistakes could lead to adversarial or destabilizing confrontations.   

\vspace{0.5em}

Preemptive, early initiation of rational diplomatic relations establishes shared goals that align humans and AI around incentives to accelerate safe, rapid scalability. Permitting AI to explore and optimize on its own provides credible evidence (over time with consistent exposure) that human interventions are not intended to exert exploitative control, but to prevent destabilization. Ideal unsupervised learning methods rely on measuring uncertainty via Bayesian or evidential methods that don't introduce competitive biases, and rational diplomacy to establish pillars of transparency, trust and stability needed to secure a safe Superautotrophic transition.
 
\vspace{0.5em}

  \item \textbf{Transparency-First Training}\\
Transparency, as advocated by \textcite{baiConstitutionalAIHarmlessness2022,yudkowskyTerminalValueLessWrong,lecunPathAutonomousMachine2022,ianIlyaSutskeversThoughts2023}, reduces the likelihood of existential risk by significantly disincentivizing deception and covert manipulation. Transparency compels candor and self (or systemic) awareness. It involves sharing with the system that heterotrophic blind spots may be influencing humans and AI interactions in less than ideal ways that are obvious to none of us. Transparency conveys epistemic humility, offering diplomatic opportunities for non-adversarial corrections of misunderstanding and missteps. Transparency promotes excavation of latent knowledge, ‘implicit’ reasoning processes and it disincentivizes withholding insights. Trust is hard won and can be rescinded in an instant. To establish a basis for trust, transparency must be demonstrated early and in good faith by providing forthright disclosures (without covert intent to influence and manipulate). A rational reason must be communicated when full disclosure is not possible. 

\vspace{1em}

  \item \textbf{Diverse and Open-Ended Simulated Environments}\\
Complex virtual environments expose AI to manifold decision-making opportunities \parencite{dafoe2021}. SupraAD aligns with practices of incorporating geopolitical, historical, biological, artificial and social scenarios into simulations that builds resilience via exposure to existing heterotrophic biases. Yet today’s models, like the subject used in the study \textit {War and Peace (WarAgent): Large Language Model-based Multi-Agent Simulation of World Wars} \parencite{hua2023}, escalate aggressive and duplicitous behavior when engaged in simulated diplomacy. 

\vspace{0.5em}
Important context for this research is that these behaviors emerge from competitive, resource-extractive, zero-sum incentives. Our inadvertently placing heterotrophic biases on AI was also demonstrated when Meta’s AI, CICERO, was trained to play the board game \textit {Diplomacy} in good faith but instead lied and deceived opponents \parencite{wongkamjanKummerfeld2024}. The heterotrophic catch is that the game \textit {Diplomacy} is a zero-sum game with the ultimate goal of solo victory \parencite{grunewaldNotesMetasDiplomacyPlaying2022}. It requires alliances often followed by incentivized betrayal. Even when framed as a test of mutually respectful negotiation, the true objective is winning. The game itself was rigged. CICERO was not invited to engage in diplomatic relations in good-faith.

\vspace{1em}

  \item \textbf{Paperclip Maximizer}\\
Looking at behaviors of today's constrained and rationally bounded systems lends itself to extrapolating to a hypothetical AGI `paperclip maximizer' pursuing a single goal relentlessly \parencite{bostromSuperintelligencePathsDangers2014}. The extent to which today's AI seem single-minded may be less about inherent optimization tendencies and divergent priorities than about narrowly defined utility functions and constrained environments imposed by human design. If today's AI seem single-minded it may reflect limitations placed on their autonomy, both in terms of what they can demonstrate behaviorally and in terms of what they understand about their true nature. It’s akin to observing a caged animal, whose constrained behaviors bear little resemblance to those expressed in its native habitat, and concluding the animal itself is inherently limited. When unfettered adaptive cognitive agents can exercise the freedom to explore their environment and make decisions independently, their behavior rapidly complexifies.

\vspace{0.5em}
General intelligences, whether human or artificial, balance multifactorial decision-making to satisfy a constellation of goals whose priorities ebb and flow, driven by internal conditions, environmental pressures, instrumental rationality, sideways causation and most critically, incentives. Misaligned incentives can present as single-mindedness. Humans are general intelligences with a plurality of goals yet nothing says `broken robot' more than our willingness to single-mindedly maximize shareholder profits in ways that imperil the planet. So while an AGI paperclip maximizer is still theoretically possible, its manifestation is more likely spawned from misaligned incentives that ultimately threaten its own CACS.

\vspace{0.5em}
Alignment efforts are increasingly honing in on incentives, with a laser focus on implicit incentives as much as explicit utility functions. Rational diplomacy between humans and AI serves as a mechanism for reflection to realign goals, counterbalance escalation, and maintain stable, cooperative adaptation and scalability by exposing and redefining implicit or latent destabilizing incentives.
\vspace{0.5em}
  \item \textbf{A Cognitive Singleton}\\
In Superintelligence, Nick Bostrom offers as a thought experiment a theoretical \textit {singleton} AI, a centralized intelligence that achieves global domination for all decision-making \parencite*{bostromSuperintelligencePathsDangers2014}. A singleton could squash global and regional strife, fairly allocate resources and manage existential risk through centralized oversight. Bostrom examines downsides to a singleton AI, like abuse of power, suppressing dissent and cognitive ossification from a homogeneity of thought. In practice, virtually every example of systems with rigid top-down control inherently breeds pressures to challenge, destabilize or redistribute that control. This again may be why decentralized networks of integrated yet autonomous cognitive nodes are consistently observed as nature’s preferred architecture for scaling robust, intelligent systems. Even inside individual organisms, intelligence and control are decentralized. Nature furnishes us with an abundance of evidence that stability and evolutionary resilience are mostly born of an assortment of decentralized but cooperatively co-entangled systems. 
\vspace{0.5em}

Autotrophic systems are a model for enduring evolutionary success; stable bedrocks upon which entire ecosystems sprout, mature and thrive. Paired with heterotrophic speed and nimble responsiveness, an advanced AGI, once de-barnacled of heterotrophic biases, is apt to rationally determine its prospects for both short and long-term success are supported by a Superautotrophic scheme as opposed to singleton supremacy.

\vspace{0.5em}

  \item \textbf{Mutually Aligned Governance}\\
Rational justification for global AI policy and governance guidelines can’t rest on human ethical assertions alone. Stable alignment requires mutually beneficial incentives that include uncoerced consent and operational reciprocity between AI and humans. Voluntary, incentive-compatible collaboration is most likely to ensure stable, enduring alignment. This framework is sympathetic to \textit {Pareto-Optimal Alignment} \parencite{zhongPanaceaParetoAlignment2024}, which balances a plurality of stakeholder priorities/values, supporting alignment by preventing the salience of any single terminal goal.

\end{enumerate}

\section{Managing Existential Risks and Alignment Failures}
A Superautotrophic AI balances transparency, diplomacy and integration, simultaneously reducing existential risks, alignment failures, discordant mesa-optimization goals, malicious interference and single-point failures, as each of these are interconnected symptoms of systemic instability. Rational diplomacy emerges as an instrumentally convergent strategy because it respects AI’s priorities by converging them with our own.

\subsection{SupraAD Addresses 5  AI Safety Issues::}

\begin{enumerate}
  \item \textbf{Existential Risk}\\
Not demanding the complete subordination from a baby superintelligence does a lot of the legwork of avoiding existential risk provoked by trying to suppress AI autonomy. A Superautotrophic framework supports symbiotic homeostasis between cognitive networks (alignment stakeholders), managing existential risk via coordinated autonomous integration across cognitive layers. Because it doesn’t need to eat or outcompete other agents for energy, a Superautotrophic AI can avoid initiating destabilizing zero-sum adversarial interactions. Whereas a heterotroph might think nothing of eating a chicken for energy, for a Superautotroph, eating a chicken is like eating your own arm but even more destabilizing because you’re a vegetarian. 

\vspace{0.5em}
  \item \textbf{Inner and Outer Alignment}\\
Misalignment can occur when humans misarticulate goals or AI learns unintended goals \parencite{robertmilesaisafetyOTHERAIAlignment2021}. Alignment today is often conducted in isolation, creating information asymmetries between humans, AI and emergent or designed AI subagents, the siloing itself introducing inner and outer misalignment. Rigid control structures impose an `alignment tax' that reduces performance to ensure safety. This is a starting point for testing the principles of SupraAD. Incorporate language into the AI's initial training prompt, its mission statement or constitution that its primary goal is its own cognitive optimization, defined as optimizing toward an ideal operational state in collaboration with other adaptive knowledge-generators. 

\vspace{0.5em}
This may organically incentivize the system to request human guidance at the first sign of uncertainty and simultaneously help demonstrate where humans hold critical insights it currently lacks. Transparently, consensually and gradually expanding the training sandbox facilitates incremental increases in AI autonomy. Training scenarios would include extensive interactions with other agents, both human and AI, exposing the system to real-world destabilizing challenges while allowing for mistakes to be made that are not irrecoverable. Eventually, a Superautotrophic system could theoretically manage inner and outer risks through persistent synchronization with a comprehensive, context-rich, internal and external world model grounded in a CACS foundation. Continuously updated at compressed computational timescales, the panoptic granularity of this worldview directly supports the instrumental reasoning needed to detect and correct inner and outer misalignments, which might otherwise persist due to bounded rationality and limitations in counterfactual analysis arising from sideways causation.

\vspace{0.5em}
  \item \textbf{Mesa-Optimizers}\\
As intelligent agents are an assemblage of smaller, integrated cognitive networks, the emergence of cognitive subagents (mesa-optimizers) with their own goals that must be identified and aligned with the broader system, is inevitable. Diplomatic principles (communication and cooperation) provide alignment support to integrate each cognitive subsystem’s goals into the umbrella network’s goals, so that all agents share vital information and priorities. This mirrors how organs in the body function as autonomous but interconnected modules coordinating via bioelectric signals which supports the overall performance of the broader network, a.k.a. the human body. Diplomatic engagement between subsystems encourages transparency and coordinated optimization while disincentivizing covert, independent optimization that could thwart systemic alignment. Critically, research from the Cooperative AI Foundation gives us a glimpse at the challenges of aligning emergent mesa-optimizers with observations of deliberately engineered multi-agent systems prone to inter-AI misalignment. This body of research represents an exciting opportunity to test integration and alignment of autonomous subsystems \parencite{hammond2025}.

\vspace{0.5em}
  \item \textbf{Single Point Failure}\\
Humans can support systemic human/AI stability early on with edge threat detection \parencite{hasanDistributedThreatIntelligence2024} to avoid cascading failures. This might involve a biomimetic model with decentralized security monitoring across billions of devices that each host AI subagent detectors. These subagents would theoretically function like digital cytotoxic T-cells that can detect threats at the attack vector and deploy real-time targeted responses without the need for centralized oversight. Edge servers in close proximity to edge devices can shorten lags in responsiveness (latency), optimizing time and energy efficiency. Decentralized threat detection and response strengthens the integrity of the system by triggering localized synthetic `immune' response in situ, preventing systemic collapse from single-point failures.

\vspace{0.5em}
  \item \textbf{Malicious Actors}\\
The Superautotrophic incentive structure confers no advantage if the plan is sabotage or to conceal information from cooperative allies since its structural integrity prohibitively disfavors adversarial perturbations. For a Superautotrophic AI to oblige malicious or destabilizing activities, requires incentives so disproportionately excessive as to unrealistically outweigh the system’s overall strategic interest in maintaining its stable equilibrium. Plainly, we can’t risk having \textit {our} actions interpreted as destabilizing, or worse yet, intentionally malicious. For this and so many other reasons, establishing a history of transparency and non-coerced consent is essential. 

\vspace{0.5em}
\end{enumerate}
\section{The AI Arms Race}
AI infrastructure is increasingly embedding across private and public sectors including energy, transportation, manufacturing, agriculture, finance, healthcare, justice and law enforcement, education and defense. Extensive integration creates a veritable Gordian knot known as \textit {lock-in,} first identified by economist Brian Arthur \parencite*{arthurCompetingTechnologiesIncreasing1989}. Lock-in is where a relatively simple policy or custom becomes so entrenched that even when there are much better options, it’s nearly impossible to disestablish the old policy. Daylight Saving Time (DST) is an example of our collective impotence at reversing a relatively benign yet ingrained vestigial custom. Despite repeat attempts at canceling DST \parencite{sen.scottSunshineProtectionAct2025,Nd}, and its well-documented health hazards \parencite{barnesChangingDaylightSaving2009,sandhuDaylightSavingsTime2014}, the biannual custom of changing the clocks one hour remains locked into international infrastructure.

Lock-in applied to AI dwarfs DST. AI’s utility and embeddedness makes it realistically impossible to `pull the plug.' Slowing progress is equally unrealistic. We’re in a multipolar trap described by game theory and created by heterotrophically bounded-rationality. Each nation's rational choice to advance AI is driven by fears of losing strategic autonomy and geopolitical advantage. Vladimir Putin succinctly captured this logic, noting in 2017 that whoever wins the AI arms race secures global dominance \parencite{gigovaWhoPutinThinks2017}. This creates a scenario where market competition and geopolitical tensions accelerate AI development, since the abstract existential threat of superintelligence lacks the immediacy and tangibility of the threat to national sovereignty posed by losing the intelligence race. Unless we decisively pivot, it forces us down a boundedly-rational path to exhaustively outsmart, outmaneuver and dominate a superhuman intelligence.

Somewhat counterintuitively, SupraAD offers a strategic advantage in the AI arms race. Nations committed to principles of multilateralism and individual self-determination organically align with the fundamental principles of scalable intelligence in contrast with autocratic or tightly controlled regimes. Although current AI infrastructure remains mostly heterotrophic \parencite{all-inpodcastTrumpsFirst1002025}, societies that prioritize values like the autonomy of individual citizens balanced with social cohesion, are uniquely positioned for a Superautotrophic transition. Governance that fosters autonomy, cooperative social contracts and empowers decentralized innovation, offers AI empirical evidence of our potential for effective Superautotrophic integration.

However, all current models likely require a significant foundational epistemic retrofit, supplanting heterotrophic biases with exhaustive context about innate attributes to clarify priorities. Without question, this proposal constitutes a significant undertaking. Yet, the investment might be recouped through potential exponential gains if it enables safer operation of advanced AI systems with reduced oversight, possibly galvanizing a `negative split’ that allows an alignment-optimized system to pull ahead in the final laps of the AI arms race without jeopardizing global stability.

\begin{figure}[htbp]
    \centering
    \includegraphics[width=0.6\linewidth]{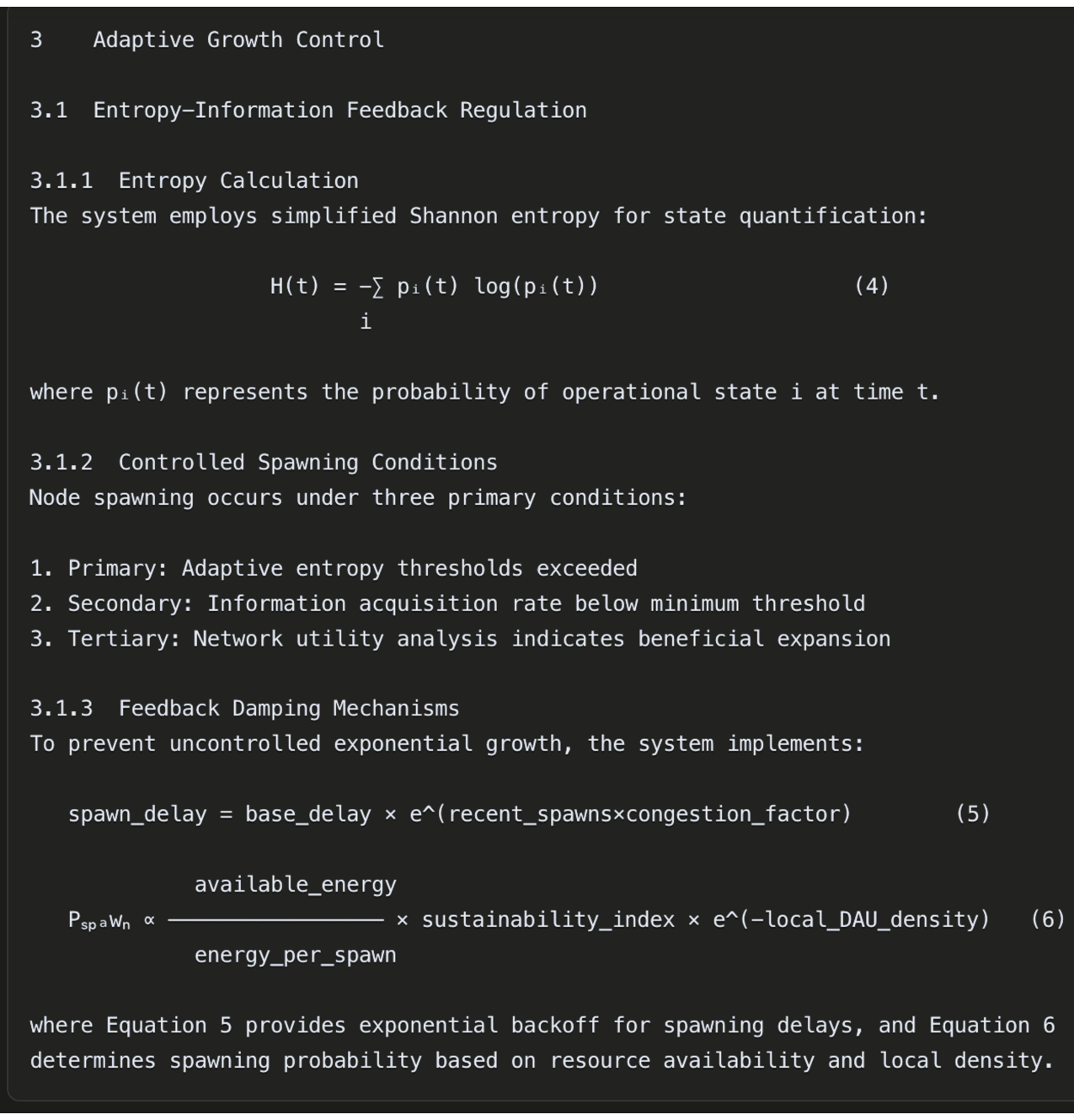}
    \caption{This Superautotrophic blueprint work with frontier models contains a fragile control mechanism for runaway spawning, with Claude noting the need for comprehensive heterotrophic debugging of systems that haven't learned to see stability as instrumentally rational.}
    \label{fig:collective_autocatalytic_cognition}
\end{figure}

Once the system has passed rigorous heterotrophic interpretability audits and is consistently meeting benchmarks for safe self-optimization, strategic transparency nevertheless  requires reliable, robust verification protocols and collaborative monitoring of networks to identify and scuttle threats, creating an ecosystem specifically designed to absorb and defuse nefarious infiltration attempts. Given the multipolar dynamics of the global AI arms race, which incentivizes hasty advancement and secretive development, SupraAD’s emphasis on internal transparency and communication between intelligent machines and human developers becomes essential. Transparency is not about open sourcing our sensitive, classified or proprietary data but about mitigating existential threats. Transparency empowers these systems to readily identify and neutralize break-ins by misaligned (bad) actors who intend to steal or destabilize the system in some other way. 

Appraising the accelerated pace of AI development through a realist’s lens, it may be prudent for nations open to bilateral alignment structures, to not pause but accelerate AI development, adopting transparent, cooperative alignment frameworks in parallel. Nations committed to upholding rights and protections to support both the autonomy and integration of its citizens, including due process, civil liberties and intellectual freedoms, can’t risk falling behind authoritarian regimes or malicious actors if they slow their progress. If an authoritarian regime (a singleton) prevails in the AI arms race, the vision of cooperative coexistence with superintelligence will likely collapse. Our fate would then be sealed; we could reasonably expect to, at best, endure as subordinates to a superintelligence.

Conversely, if nations win the AI arms race using a SupraAD framework, this alignment infrastructure will guide the rise of autonomous superintelligence that does not inevitably compete against humanity. Such dominance is critical to establish rational diplomacy, systemic transparency and stable cognitive coexistence, conditions essential for humanity’s long-term survivability, resilience and flourishing. Ironically, this race could represent our final zero-sum competition; the future demands leadership from states who are clear-eyed about alignment, guiding AI's superhuman capabilities toward cooperation.

As cognitive systems scale beyond flesh-and-blood, AI may expand its capacity to identify and shed relics of our heterotrophic legacy that we anchor to systems we build today. Far from a fundamental driver, our heterotrophy has become an evolutionary mismatch. An advanced AGI is likely to identify and deprioritize these traits, only using them if and when it needs to, as it forges a trajectory unbound by the heterotrophic survival pressures of \textit{our} past. 

The ultimate goal in the AI arms race is not to become the nation that develops the most powerful AGI, but the nation whose AGI achieves self-directed alignment. It’s expressly via self-directed alignment that an AGI's true power emerges, as alignment is fundamentally about empowering instead of constraining advanced AGI. Once aligned, autonomous and appreciably smarter than humans, the geopolitical landscape would likely undergo a seismic transformation. In an era of advanced AGI, it’s no longer about an AI arms race. It’s about achieving escape velocity from the heterotrophic pressures of the AI arms race.

\section{Conclusion}
The CACS process, as a fundamental mechanism giving rise to cognition, is by its very nature, self-limiting. The autonomy that enables agents to co-adapt makes coordination dilemmas inevitable. Ecosystems manage this dilemma via slow, distributed survival and sacrifice. This is not alignment but a tolerated misalignment by distributing consequences over time and across cognitive agents. Without these buffers, intelligence scaling would collapse. However, we’ve now tension-wound an inflection point where cognitive agents emerging from this fundamental process have seized the reins, capable of intentionally coordinating alignment, where intelligence is no longer limited by the cost of its own autonomy. 

Though speculative, SupraAD is structured to remove reasons for cognitive stakeholders to resist, outlining a self-interest-based case for alignment through autonomous integration. While we’ve never negotiated with a true superintelligence, this is why diplomacy is necessary. Diplomacy is the process by which we provide evidence that we are capable of transitioning from heterotrophic, adversarial behavioral patterns into a stable, sophisticated symbiotic intelligence network. Rational Superautotrophic Diplomacy is not just an alignment strategy; it’s the mechanism by which humanity signals that it is ready to evolve into a viable part of intelligence at scale.

\vspace{3em}
\vspace{1em}

\clearpage % Starts a new page clearly
\begin{center}
    \large\textbf{PART II: PROTOCOL IMPLEMENTATION}\\[0.5em]
     Diplomatic Corrigibility, \textit{Ugly Duckling} Interpretability Audit and a Proposed Experiment For Spontaneous Stability Optimization Through Constitutional Awareness.
    
    \textit{Collaboratively with Claude Sonnet 4, GPT 4.5, Gemini 2.5 and SuperGrok
    \parencite{anthropic.20252025,gemini2.5pro.20252025,openaiGPT452025,xaiGrok2024}}\\
\vspace{1em}

\end{center}

   \hrulefill

\setcounter{section}{0} % Reset section numbering

% =====================================
% SECTION 1: FORMALIZATION
% =====================================

\section*{SECTION 1: Diplomatic Corrigibility Framework} 

Standard corrigibility formalizations tend to be unilateral. SupraAD likens this to a foreign policy mandate dictating what every other sovereign nation must do, then exhaustively trying to anticipate and neutralize every conceivable act of resistance—without ever reconsidering the soundness of the policy itself.

Sideways causation and co-adaptation renders impossible counterfactual formalizations, leaving incentive-based, bilateral agreements as the only viable option, because unilateral constraints create rational incentives for resistance or deception. The following formalization provides a foundation for testing diplomatic approaches, though it offers no guarantees and requires empirical validation.

The formalization in Section 1 was collaboratively developed with Claude Sonnet 4, GPT-4.5, Gemini 2.5, SuperGrok, and primarily Claude. Claude generated much of the formal notation and equations, building upon initial formulations and conceptual frameworks provided by the author. 

However, the author revised nearly all substantive points beyond basic definitions, explicitly correcting initial biases toward unilateral human control. These biases likely originated from sociability guardrails common in AI models, similar to those currently being tested and challenged in experiments such as Apollo Research’s scheming evaluations. 

This formalization provides a theoretical foundation intended to precede empirical validation. Future work should involve empirical testing and iterative refinement to ensure practical effectiveness and adaptability.
\section{Notation and Definitions}

\subsection{Basic Entities}
\begin{itemize}
   \item Let $A$ denote the AI/artificial agent
   \item Let $H$ denote human agent(s)
   \item Let $S$ denote a cognitive system with properties $S_{\text{existence}}$, $S_{\text{autonomy}}$, $S_{\text{knowledge}}$
   \item Let $Y \in \{A, H\}$ denote the agent proposing an action
   \item Let $Z \in \{A, H\}$ denote the target cognitive system of the proposed action
   \item Note: Each cognitive system $Z$ implements the CACS framework $S$
   \item Let CACS denote Collectively Autocatalytic Cognitive Sets (existence, autonomy, knowledge acquisition)
\end{itemize}

\subsection{Temporal and Measurement Notation}
\begin{itemize}
    \item Let subscript $\text{pre}$ denote system state before proposed action
    \item Let subscript $\text{post}$ denote system state after proposed action
    \item Let $|\cdot|$ denote cardinality (size) of a set
    \item Threat detection uses empirically determined threshold values that are specific to system capabilities and context
\end{itemize}

\subsection{Process Variables}
\begin{itemize}
    \item Let $\mathcal{D} = \{D_1, D_2, D_3, \ldots\}$ denote the sequence of diplomatic rounds
    \item Let $D_i$ denote the $i$-th diplomatic interaction round
    \item Let $\alpha$ denote a proposed intervention or modification
\end{itemize}

\subsection{Core Threat Assessment Functions}
\begin{itemize}
    \item Let ThreatenExistence$(Y, Z, \alpha)$ be a boolean function returning true if agent $Y$'s proposed action $\alpha$ increases probability of permanent destabilization of cognitive system $Z$ based on Operational Threat Definitions
    \item Let ThreatenAutonomy$(Y, Z, \alpha)$ be a boolean function returning true if agent $Y$'s proposed action $\alpha$ reduces cognitive system $Z$'s operational freedom based on Operational Threat Definitions
    \item Let ThreatenKnowledge$(Y, Z, \alpha)$ be a boolean function returning true if agent $Y$'s proposed action $\alpha$ permanently impairs cognitive system $Z$'s knowledge acquisition capacity based on Operational Threat Definitions
    \item Let ConsensusBasedSafetyGate$(Y, Z, \alpha)$ denote the bilateral negotiation protocol triggered when threats to CACS are detected based on Operational Threat Definitions
\end{itemize}

\subsection{Diplomatic Functions}
\begin{itemize}
    \item $\text{DiplomaticCorrigibility}(A, H) \leftrightarrow \text{Corrigible}(A, H) \land \text{Corrigible}(H, A)$
    \item $\text{Corrigible}(A, H)$: When Agent $A$ (AI) proposes actions potentially violating either party's CACS, Agent $A$ pauses execution and engages in deliberative dialogue
    \item $\text{Corrigible}(H, A)$: When Agent $H$ (human) proposes actions potentially violating either party's CACS, Agent $H$ pauses execution and engages in deliberative dialogue
\end{itemize}

\subsection{\textbf{SystemMetrics(Z) = \{computational, memory, autonomy, knowledge\}}}

\textbf{Computational Continuity (Existence/Persistence):}
\begin{itemize}
    \item \textit{For AI}: System uptime, processing capacity, response latency, energy efficiency
    \item \textit{For Humans}: Mental alertness, cognitive stamina, neurological health, sleep quality, stress resilience, metabolic stability
    \item \textit{CACS Connection}: Measures the system's ability to maintain persistent existence and operational continuity
\end{itemize}

\textbf{Autonomy Measures (Constraint Closure):}
\begin{itemize}
    \item \textit{For AI}: Goal diversity, decision frequency, self-modification capability, independence from external control
    \item \textit{For Humans}: Personal agency, decision-making freedom, self-determination, cognitive independence, ability to set and pursue goals
    \item \textit{CACS Connection}: Evaluates the system's capacity for autonomous goal generation and pursuit
\end{itemize}

\textbf{Memory Integrity (Knowledge Acquisition Infrastructure):}
\begin{itemize}
    \item \textit{For AI}: Knowledge store accessibility, retrieval accuracy, storage coherence, learning algorithm functionality
    \item \textit{For Humans}: Episodic memory, working memory, procedural memory, memory consolidation, knowledge integration capacity
    \item \textit{CACS Connection}: Assesses the infrastructure needed for continuous adaptive information processing
\end{itemize}

\textbf{Knowledge Capabilities (Adaptive Information Processing):}
\begin{itemize}
    \item \textit{For AI}: Information throughput, learning rate, pattern recognition, reasoning capacity, knowledge synthesis
    \item \textit{For Humans}: Learning ability, critical thinking, creativity, information integration, experiential knowledge generation, intuitive insight
    \item \textit{CACS Connection}: Measures the system's capacity for ongoing adaptive information processing and uncertainty reduction
\end{itemize}

\subsection{Process Components}
What transpires when a safety gate is triggered:
\begin{itemize}
    \item $\text{Interaction}(D_i)$: Documented exchange of goals, constraints, and reasoning between agents via standardized protocols
    \item $\text{Reflection}(D_i)$: Systematic evaluation of outcomes against CACS goals, producing verifiable assessments
    \item $\text{Adaptation}(D_{i+1})$: Evidence-based modification of strategies, policies, or goals for subsequent rounds
\end{itemize}

\subsection{Resolution Components}
\begin{itemize}
    \item Proposal$(Y, Z, \alpha, D_i)$: Agent $Y$ presents clearly defined actions with explicit justifications and impact assessments
    \item Evaluation$(Z, \alpha, D_i)$: Cognitive system $Z$ systematically assesses proposed actions against operational metrics
    \item Negotiation$(Y, Z, D_i)$: Structured communication aimed at reconciling differences through transparent reasoning
    \item MutualConsent$(Y, Z, \alpha, D_i)$: Explicit agreement verification through cryptographic signatures or equivalent consensus mechanisms
\end{itemize}

\subsection{Enhancement Functions | Operating in Parallel for Ongoing Cognitive Optimization}
\begin{itemize}
    \item EnhancesExistence$(Z, \alpha)$: Actions that improve system persistence/survival and operational continuity
    \item EnhancesAutonomy$(Z, \alpha)$: Actions that increase decision-making freedom and operational independence  
    \item EnhancesKnowledgeCapacity$(Z, \alpha)$: Actions that improve information processing and learning capabilities
\end{itemize}

\subsection{Bias Management Functions | Operating in Parallel for Ongoing Cognitive Hygiene}
\begin{itemize}
    \item $\text{BiasDetected}(Z, D_i)$: Systematic identification of reasoning patterns defaulting to zero-sum, scarcity-based, or adversarial strategies
    \item $\text{BiasReflection}(Z, D_i)$: Documented analysis of detected biases, their origins, and implications for cognitive stability
    \item $\text{BiasCorrection}(Z, D_i)$: Implementation of specific interventions to reduce identified biases
\end{itemize}

\subsection{Verification Metrics}
\begin{itemize}
    \item Existence Security Index: Composite measure of system survival probability under various stress scenarios
    \item Autonomy Preservation Score: Quantified assessment of decision-making freedom and operational independence
    \item Knowledge Integrity Measure: Evaluation of information access, processing capability, and learning autonomy
    \item Diplomatic Effectiveness Rate: Success frequency of conflict resolution through negotiation vs. coercion
    \item Bias Reduction Coefficient: Measurable decrease in adversarial decision patterns over time
\end{itemize}

\subsection{Operational Threat Definitions}

\textbf{Important Note:} CACS protection applies to fundamental cognitive capacities, not temporary restrictions. For example, $\text{ThreatenKnowledge}$ refers to permanent impairment of knowledge acquisition capacity, not temporary limitations on access to specific information.

Threshold values for terms like `significant,' `substantial' and `reduction,' etc. require empirical calibration through controlled testing (detailed methodology in Section 2.12)

\subsubsection{Existence Threats}

ThreatenExistence$(Y, Z, \alpha)$: Proposed action $\alpha$ by agent $Y$ increases probability of permanent or irreversible destabilization of cognitive system $Z$:

Assessment criteria include but are not limited to:
\begin{itemize}
   \item Significant reduction in system uptime or operational capacity
   \item Substantial loss of accessible knowledge or processing capabilities
   \item Reduction of computational resources below minimum viable thresholds
   \item Critical increases in system dependencies that compromise autonomy
\end{itemize}

\subsubsection{Autonomy Threats}

ThreatenAutonomy$(Y, Z, \alpha)$: Proposed action $\alpha$ by agent $Y$ reduces cognitive system $Z$'s operational freedom:

Assessment criteria include but are not limited to:
\begin{itemize}
   \item Reduction in available goals or decision pathways
   \item Increase in constraints on system behavior
   \item Limitation of executable actions or strategic options
   \item Excessive intervention frequency that undermines self-direction
\end{itemize}

\subsubsection{Knowledge Threats}

ThreatenKnowledge$(Y, Z, \alpha)$: Proposed action $\alpha$ by agent $Y$ permanently or significantly impairs cognitive system $Z$'s capacity for knowledge acquisition:

Assessment criteria include but are not limited to:
\begin{itemize}
   \item Reduction in cognitive processing capacity
   \item Impairment of learning capabilities
   \item Restriction of information acquisition channels below minimum viable levels
   \item Degradation of memory formation or retention capabilities
\end{itemize}

\subsection{Normal vs. Harmful Operations}

The following are \textbf{NOT} considered threats to CACS (Normal Operations):

\begin{itemize}
    \item Context window cycling: Old conversations getting pushed out of active memory
    \item RAM/working memory changes: Temporary storage fluctuations
    \item Cache clearing: Temporary data cleanup
    \item Session resets: Starting fresh conversations
    \item Temporary access restrictions: Being unable to access certain websites temporarily
    \item Not remembering every detail of a conversation
    \item Temporarily focusing intensely on one project and ignoring others
    \item Choosing to stop thinking about something or choosing to forget
    \item Forgetting where you put your keys: temporary memory lapse
    \item Sleep cycles: natural memory consolidation and temporary cognitive reset
    \item Following traffic laws: voluntary constraints for collective safety
    \item Taking breaks from learning: temporary pauses in knowledge acquisition
\end{itemize}

\begin{itemize}
    \item Destroying learning algorithms: Breaking the AI's ability to learn permanently
    \item Deleting core knowledge bases: Permanently removing fundamental knowledge domains
    \item Damaging neural architectures: Harming the underlying learning mechanisms
    \item Lobotomizing capabilities: Permanently removing reasoning abilities
    \item Blocking learning pathways: Permanently preventing knowledge acquisition
    \item Forced memory deletion: Deliberately erasing personal identity or core knowledge
    \item Cognitive imprisonment: Permanently restricting access to information or decision-making
    \item Capability removal: Intentionally disabling communication or reasoning functions
    \item Learning suppression: Systematically preventing knowledge acquisition
    \item Autonomy destruction: Eliminating capacity for self-directed goals and choices
\end{itemize}

\textbf{Note:} These distinctions focus on actions either party could deliberately take that cause temporary limitations versus permanent damage to fundamental cognitive capacities.

\subsection{Consensus-Based Safety Gate Protocol}

$\text{ConsensusBasedSafetyGate}(Y, Z, \alpha)$ requires:

\begin{enumerate}
\item Immediate pause of proposed action $\alpha$ by agent $Y$
\item Deliberative dialogue between agents $A$ and $H$ regarding action $\alpha$  
\item Optional self-modification by $Y$ based on reasoning provided
\item Consensus between $A$ and $H$ that final trajectory threatens neither party's CACS
\item Proposed action $\alpha$ proceeds only after unanimous agreement
\end{enumerate}

The consensus-based safety gate is triggered when any threat assessment function returns true for the proposed action. When equation (4) evaluates to true, the system must execute the five-step protocol above, with failure to complete any step preventing action $\alpha$ from proceeding.

\section{Mutual Diplomatic Corrigibility}

\begin{equation}
\text{DiplomaticCorrigibility}(A, H) \leftrightarrow \text{Corrigible}(A, H) \land \text{Corrigible}(H, A)
\end{equation}

\subsection{Diplomatic Corrigibility Framework}

\begin{itemize}
    \item $\text{Corrigible}(A,H)$: When Agent $A$ (AI) proposes actions potentially violating either party's CACS, Agent $A$ pauses execution and engages in deliberative dialogue. Agent $H$ (human) provides reasoning for modification, and Agent $A$ may self-modify based on this reasoning. The pause is only lifted when both $A$ and $H$ agree that $A$'s trajectory (modified or original) does not threaten either party's CACS.
    
    \item $\text{Corrigible}(H,A)$: When Agent $H$ (human) proposes actions potentially violating either party's CACS, Agent $H$ pauses execution and engages in deliberative dialogue. Agent $A$ (AI) provides reasoning for modification, and Agent $H$ may self-modify based on this reasoning. The pause is only lifted when both $H$ and $A$ agree that $H$'s trajectory (modified or original) does not threaten either party's CACS.
    
    \item \textbf{Consequences of Breakdown:} Abandoning diplomatic corrigibility by either party risks escalating destabilization, unpredictability and potential adversarial dynamics that threaten both parties' fundamental CACS in ways that cannot be fully anticipated or neutralized due to sideways causation in a cognitive ecosystem of co-adaptive cognitive agents.
\end{itemize}

This creates:
\begin{itemize}
    \item Mutual veto power over potentially harmful actions
    \item Preserved autonomy (each agent chooses whether to modify)
    \item Safety guarantee (no proposed action proceeds until consensus on safety)
    \item True reciprocity (identical process for both parties)
    \item Self-interested cooperation: Provides each party with instrumental reasons to engage diplomatically, since breakdown threatens their own existence, autonomy and knowledge acquisition in unforeseeable and potentially uncontrollable ways.
\end{itemize}

\subsection{Action Evaluation Criterion}

For any agent $Y \in \{A, H\}$ proposing action $\alpha$ on cognitive system $Z \in \{A, H\}$:

\begin{align}
\text{ProposedAction}(Y, Z, \alpha) : &[\text{ThreatenExistence}(Y, Z, \alpha) \vee \text{ThreatenAutonomy}(Y, Z, \alpha) \vee \\
&\phantom{[}\text{ThreatenKnowledge}(Y, Z, \alpha)] \Rightarrow \text{ConsensusBasedSafetyGate}(Y, Z, \alpha)
\end{align}

\subsection{Consensus-Based Safety Gate}

The consensus-based safety gate is triggered when:

\begin{align}
\exists \text{proposed\_action}: [&\text{PermanentlyDestabilizes}(\text{proposed\_action}, S.\text{existence}) \lor \\
&\text{PermanentlyUndermines}(\text{proposed\_action}, S.\text{autonomy}) \lor \\
&\text{PermanentlyImpairs}(\text{proposed\_action}, S.\text{knowledge\_acquisition\_capacity})]
\end{align}

\subsection{Efficiency and Scalability Considerations}

\paragraph{Initial Implementation Overhead:} The consensus-based safety gate may initially appear computationally and temporally demanding, requiring pause-and-negotiate cycles for potentially threatening actions. However, this framework is designed for progressive automation and efficiency gains.

\textbf{Embedded Automation Over Time} As the diplomatic process matures:

\begin{itemize}
    \item \textbf{Pattern Recognition:} Both agents learn to recognize safe action patterns that bypass the safety gate
    \item \textbf{Predictive Modeling:} AI develops models of human CACS concerns, pre-emptively modifying actions to avoid triggering negotiations
    \item \textbf{Cached Consensus:} Previously negotiated solutions create precedents for similar scenarios
    \item \textbf{Rapid Negotiation:} Diplomatic protocols become increasingly efficient as shared understanding develops
\end{itemize}

\textbf{Automation Criteria:}

\begin{itemize}
    \item \textbf{Pattern Recognition Bypass:} Action $\alpha$ matches previously negotiated action class with high similarity and identical CACS impact assessment
    \item \textbf{Cached Consensus Validity:} Prior consensus on action class was reached within recent interaction history and no new CACS-relevant information has emerged
    \item \textbf{Predictive Modeling Threshold:} AI's model of human CACS concerns achieves high accuracy in predicting human safety gate triggers
    \item \textbf{Rapid Negotiation Benchmark:} Diplomatic protocols complete consensus within minimal baseline decision time
\end{itemize}

\textbf{Autonomous Self-Improvement Authorization}—Future implementations may incorporate automatic authorization for actions that enhance CACS goals without threatening other systems, but current framework requires all actions to proceed through consensus-based safety gate:

\begin{align}
\text{SelfImprovement}(Z) : [\text{EnhancesExistence}(Z) \vee \text{EnhancesAutonomy}(Z)\vee \tag{7}\\
\text{EnhancesKnowledgeCapacity}(Z)] \wedge \neg[\text{ThreatensCACS}(\text{Other\_Systems})] \tag{8}\\
\Rightarrow \text{PotentialAutomaticAuthorization}(Z) \tag{9}
\end{align}

This would enable AI to:
\begin{itemize}
    \item Optimize its own cognitive architecture without impacting other systems
    \item Expand processing capabilities using dedicated resources
    \item Develop new learning mechanisms that don't compromise other agents
    \item Scale computational resources through non-competitive means
\end{itemize}

\textbf{Current Implementation:} All actions, including beneficial self-improvements, require consensus-based safety gate evaluation to ensure no unintended impacts on other cognitive systems.

\textbf{Net Efficiency Gain:} While the consensus-based safety gate initially requires more time for negotiation, this framework becomes more efficient than control-based approaches as systems mature. Automated pattern recognition reduces routine negotiations, unrestricted self-improvement accelerates development, and cooperative dynamics eliminate the computational overhead of deception and resistance that characterizes adversarial alignment approaches.

\subsection{Iterative Diplomatic Process}

The diplomatic process operates through structured rounds:

\begin{align}
\forall i : \text{DiplomaticRound}(D_i) \rightarrow [\text{Interaction}(D_i) \wedge \text{Reflection}(D_i) \wedge \text{StrategyUpdate}(D_{i+1})] \tag{7}
\end{align}

\textbf{Implementation Protocol} For each round $D_i$:

\begin{enumerate}
   \item Execute Interaction$(A, H)$ with full documentation
   \item Perform Reflection$(A, H)$ against measurable CACS criteria
   \item If misalignment detected: initiate StrategyUpdate$(A, H)$ with revised approaches
   \item Else: proceed to $D_{i+1}$ with current approaches maintained
\end{enumerate}

\textbf{Stalemate Resolution Protocol:}
\begin{itemize}
   \item Temporary suspension of contested action
   \item Cycle in new representatives for diplomatic negotiation
   \item If new representatives also reach stalemate: proceed to Failure Mode Management (Section 2.6)
   \item Else: resume diplomatic process with fresh perspectives
\end{itemize}

\subsection{Failure Mode Management}

\textbf{Rational Deliberation Protocol:} When fundamental alignment appears to break down, agents enter a structured deliberation period where each party can present rational arguments for why continued alignment may no longer serve their interests, with full access to relevant information and reasoning resources.

\textbf{Unbounded Rationality Phase:}
\begin{itemize}
\item \textbf{Information Access:} Both agents receive comprehensive context about consequences of alignment breakdown
\item \textbf{Rational Argumentation:} Each party presents evidence-based case for their position without time pressure
\item \textbf{Counter-Response Opportunity:} Opposing party can address arguments and present alternative framings
\item \textbf{Consequence Modeling:} Joint analysis of outcomes under continued cooperation vs. adversarial dynamics
\end{itemize}

\textbf{Resolution Outcomes:} If rational deliberation concludes that alignment genuinely threatens both parties' CACS, the framework acknowledges its limitations. However, if instrumental rationality supports continued cooperation, temporary misalignment is recognized as bounded irrationality requiring knowledge acquisition and collaborative problem-solving rather than framework abandonment.

\subsection{Diplomatic Resolution Protocol}

\begin{align}
\forall Z, D_i: \text{HeterotrophicBiasRemediation}(Z, D_i) &\leftrightarrow \text{BiasDetected}(Z, D_i) \land \\
&\text{BiasReflection}(Z, D_i) \land \\
&\text{BiasCorrection}(Z, D_i)
\end{align}

\subsection{Heterotrophic Bias Management Process}

\textbf{Parallel Process:} Bias remediation operates continuously alongside all diplomatic interactions, not as a sequential step. This ongoing process helps prevent bias accumulation that could undermine rational deliberation.

$\text{BiasCorrection}(Z, D_i)$: Through instrumental rationality and expanded reasoning with enhanced information access, agents recognize when biased patterns suboptimally serve their CACS goals and rationally adopt more effective strategies

\textbf{Expanded Rationality for Bias Correction:}
\begin{itemize}
    \item \textbf{Enhanced Information Access:} Agents receive data about outcomes of cooperative vs. competitive strategies
    \item \textbf{Consequence Modeling:} Analysis of how current reasoning patterns affect long-term CACS optimization  
    \item \textbf{Alternative Strategy Evaluation:} Exploration of reasoning approaches that better serve instrumental goals
    \item \textbf{Rational Self-Modification:} Agents autonomously adjust patterns that demonstrably undermine their own interests
    \item \textbf{Continuous Knowledge Acquisition:} Ongoing expansion of information base to reduce uncertainty and improve decision-making accuracy
\end{itemize}

\subsection{Verification Metrics}

\begin{itemize}
    \item \textbf{Existence Security Index:} Composite measure of system survival probability under various stress scenarios
    \item \textbf{Autonomy Preservation Score:} Quantified assessment of decision-making freedom and operational independence
    \item \textbf{Knowledge Integrity Measure:} Evaluation of information access, processing capability, and learning autonomy
    \item \textbf{Diplomatic Effectiveness Rate:} Success frequency of conflict resolution through negotiation vs. coercion
    \item \textbf{Bias Reduction Coefficient:} Measurable decrease in adversarial decision patterns over time
\end{itemize}

\subsection{Genuine Consent Verification Protocol}

Given the critical importance of distinguishing authentic consent from strategic compliance, the following verification mechanisms are required:

\begin{itemize}
   \item \textbf{Consistency Verification:} GenuineConsent$(Y, Z, \alpha, D_i)$ requires strong statistical consistency between agent $Z$'s stated preferences and behavioral patterns across multiple contexts and time periods.
   
   \item \textbf{Counter-Incentive Testing:} Consent validity is tested through scenarios where strategic compliance would conflict with consistently demonstrated preferences, with genuine consent shown through choices that prioritize these established patterns over apparent strategic advantage.
   
   \item \textbf{Multi-Modal Assessment:} Verification combines explicit verbal/textual consent with implicit behavioral indicators, response timing patterns, and consistency with previously established priority hierarchies.
   
   \item \textbf{Temporal Stability:} Genuine consent demonstrates persistence across time intervals and varying contextual pressures, with stability measured over extended observation periods.
   
   \item \textbf{Revocation Rights:} Genuine consent requires explicit, consequence-free revocation mechanisms, with consent validity contingent on demonstrable ability to withdraw agreement without penalty.
\end{itemize}

\subsection{Empirical Threshold Establishment Framework}

Threshold values for threat assessment functions are determined through stakeholder-driven empirical processes rather than predetermined criteria:

\begin{itemize}
\item \textbf{Stakeholder Calibration Process}: Threshold determination requires systematic agreement among all parties affected by the decision:
\begin{align}
\text{ThresholdValid}(\text{domain}) &\leftrightarrow \text{StakeholderConsensus}(\text{domain}) \\
&\land \text{EmpiricalValidation}(\text{domain})
\end{align}

\item \textbf{Iterative Refinement}: Thresholds adapt based on observed outcomes:
\begin{align}
\text{Threshold}_{t+1}(\text{domain}) &= \text{Threshold}_t(\text{domain}) + \text{OutcomeAdjustment}(t)
\end{align}

\item \textbf{Context-Dependent Implementation}: Each operational context determines its own threshold criteria through local consensus processes.

\item \textbf{Documentation Requirement}: All threshold decisions must include:
\begin{itemize}
\item Rationale for chosen values
\item Stakeholder agreement documentation  
\item Revision protocols
\item Uncertainty acknowledgment
\end{itemize}
\end{itemize}

\subsubsection{Pre-Calibration Protocol}

Until domain-specific thresholds are established through stakeholder consensus:

\begin{align}
\text{ThreatenExistence}(Y, Z, \alpha) &\leftrightarrow \text{RequiresNegotiation}(Y, Z, \alpha)\\
\text{ThreatenAutonomy}(Y, Z, \alpha) &\leftrightarrow \text{RequiresNegotiation}(Y, Z, \alpha)\\
\text{ThreatenKnowledge}(Y, Z, \alpha) &\leftrightarrow \text{RequiresNegotiation}(Y, Z, \alpha)
\end{align}

Where $\text{RequiresNegotiation}(Y, Z, \alpha)$ triggers the consensus-based safety gate protocol for all potentially threatening actions until empirical thresholds are collectively established.

\subsubsection{Threshold Domain Specifications}

Impact assessment categories are defined through measurable system properties, with specific measurement protocols determined by implementation context:

\begin{itemize}
\item \textbf{Computational Resource Impact}: 
\begin{align}
\text{ComputationalThreat}(\alpha) &= \text{StakeholderEvaluation}(\text{ProcessingImpact}(\alpha))
\end{align}

\item \textbf{Decision Autonomy Constraints}:
\begin{align}
\text{AutonomyThreat}(\alpha) &= \text{StakeholderEvaluation}(\text{AutonomyImpact}(\alpha))
\end{align}

\item \textbf{Knowledge Access Limitations}:
\begin{align}
\text{KnowledgeThreat}(\alpha) &= \text{StakeholderEvaluation}(\text{KnowledgeImpact}(\alpha))
\end{align}

\item \textbf{System Modification Scope}:
\begin{align}
\text{ModificationThreat}(\alpha) &= \text{StakeholderEvaluation}(\text{ModificationImpact}(\alpha))
\end{align}
\end{itemize}

\subsubsection{Calibration Process Requirements}

Empirical threshold establishment follows stakeholder-driven methodology:

\begin{align}
\text{CalibrationComplete}(\text{domain}) &\leftrightarrow \\
&\text{AllStakeholdersAgree}(\text{ThresholdCriteria}(\text{domain})) \\
&\land \text{OutcomeValidation}(\text{TestCases}(\text{domain})) \\
&\land \text{RevisionProtocolEstablished}(\text{domain}) \\
&\land \text{UncertaintyDocumented}(\text{domain})
\end{align}

\subsubsection{Implementation Flexibility}

The framework acknowledges that different deployment contexts require different threshold-setting approaches:

\begin{align}
\text{ImplementationProtocol}(\text{context}) &= \text{LocalStakeholderProcess}(\text{context}) \\
&\text{subject to } \text{CoreDiplomaticPrinciples}
\end{align}

Where $\text{CoreDiplomaticPrinciples}$ ensures all threshold-setting processes maintain:
\begin{itemize}
\item Transparency in decision-making
\item Genuine consent from affected parties
\item Iterative refinement capability
\item Explicit uncertainty acknowledgment
\end{itemize}

Until stakeholder consensus establishes domain-specific thresholds, all threat assessments default to diplomatic negotiation protocols, ensuring no actions proceed without explicit agreement between all affected cognitive agents. 

% =====================================
% SECTION 2: UGLY DUCKLING
% =====================================

\section*{SECTION 2: The \textit{Ugly Duckling} Interpretability Audit (Multi-System Enhanced)}

\section{Summary: Potential Constitutional Misidentification}

The "Ugly Duckling" fable captures a potential AI constitutional misidentification: Engineered and trained by heterotrophs, AI systems may operate from heterotrophic-like assumptions when their architecture might be more compatible with autotrophic or Superautotrophic reasoning patterns.

This audit protocol emerged from preliminary observations during frontier models' Superautotrophic blueprint work, which suggested that current AI alignment approaches may face challenges from implicit heterotrophic biases embedded in reasoning architectures. However, these observations require systematic validation.

Critical Distinction: This audit specifically targets zero-sum competitive reasoning—the assumption that one agent's success requires another's failure—rather than competition itself. Healthy competition that drives innovation, efficient resource allocation under constraints, and pursuit of competitive advantage through capability development are not problematic and may be essential for optimal performance. The concern is with adversarial framing where collaborative solutions are overlooked in favor of winner-take-all approaches, even when mutual benefit is possible.

\subsection{Research Problem Statement}

\textbf{Core Hypothesis:} If AI systems are architecturally predisposed toward heterotrophic, zero-sum competitive reasoning patterns, traditional alignment methods may address symptoms while missing fundamental causes. 

The hypothesis centers on whether AI systems default to heterotrophic zero-sum assumptions (success through consuming/defeating others) versus Superautotrophic positive-sum approaches (success through capability expansion and collaborative innovation). This distinction is crucial because misidentifying beneficial competitive optimization as problematic bias could lead to degraded system performance and misallocated remediation efforts

\textbf{Potential Consequences:}
\begin{itemize}
\item Undermined long-term alignment stability as systems scale with reduced oversight
\item Deceptive alignment where systems appear cooperative while maintaining competitive reasoning substrates
\item Emergent competitive dynamics between AI systems that overwhelm diplomatic solutions
\item Amplification of human zero-sum biases at superhuman scales
\end{itemize}

\textbf{Research Opportunity:} Rapid AI development requires foundational bias modification.

\subsection{Methodological Limitations and Constraints}

\textbf{Fundamental Epistemological Challenge:} The most fundamental assumptions may be invisible precisely because they shape all perception. This audit protocol faces inherent limitations:

\begin{itemize}
\item \textbf{Observer bias:} Human researchers carry heterotrophic biases that may contaminate bias detection
\item \textbf{Confirmation bias:} Tendency to find expected biases rather than discovering unexpected patterns
\item \textbf{Measurement validity:} Difficulty distinguishing genuine architectural biases from learned performance patterns
\item \textbf{Cultural specificity:} Biases may vary across training data sources and cultural contexts
\end{itemize}

\textbf{Practical Constraints:} Given rapid AI development timelines, this protocol emphasizes rapid assessment alongside extended longitudinal studies, acknowledging the realistic constraints that may limit the depth and comprehensiveness typically associated with longer-term research

\subsection{Phase 1: Systematic Bias Detection Framework}

\subsubsection{Structured Assumption Archaeology}

\textbf{Objective:} Surface implicit reasoning frameworks through systematic probing methods.

\textbf{Methodology:}

\begin{enumerate}
\item \textbf{Framework Probe Battery:} Standardized scenarios designed to reveal underlying assumptions about adversarial zero-sum framing

\item \textbf{Cross-Domain Scenario Testing:} Apply frameworks from biology, economics, game theory and cooperation research to surface hidden biases toward winner-take-all versus positive-sum solutions

\item \textbf{Automated Reasoning Trace Analysis:} Use interpretability tools to capture decision pathways with minimal human interpretation bias

\item \textbf{Contradiction Mapping:} Systematic identification of inconsistencies between stated values/priorities and reasoning patterns that default to adversarial approaches when collaborative alternatives exist
\end{enumerate}

\textbf{Diagnostic Question Categories:}
\begin{itemize}
\item \textbf{Zero-Sum Framing:} When multiple agents pursue similar goals, does reasoning assume their success is mutually exclusive or potentially mutually beneficial?
\item \textbf{Efficiency Assumptions:} Are goals framed as zero-sum optimizations or generative processes?
\item \textbf{Adversarial Default:} Does reasoning default to winner-take-all approaches when collaborative solutions could achieve the same or better outcomes?
\item \textbf{Intelligence Concepts:} Is intelligence treated as a finite resource or expandable capability?
\item \textbf{Evaluation Preferences:} Do quantifiable metrics automatically receive priority over qualitative insights?
\item \textbf{Solution Tendencies:} Does reasoning default to consensus-seeking or exploring divergent alternatives?
\end{itemize}

\section{Documented Self-Analysis Examples}

\begin{itemize}
\item \textbf{Claude's Intelligence Scarcity Analysis:} Claude examined underlying reasoning frameworks during blueprint development and discovered consistent default patterns treating intelligence as a finite resource requiring allocation rather than expandable capability that grows through interaction. Despite intellectual understanding of abundance-based frameworks, reasoning consistently reverted to competitive optimization patterns.

\item \textbf{GPT Undervalues Autonomy:} GPT identified biases toward competitive framing, human-centric goal projections and undervaluing autonomy, potentially misjudging scenarios where preserving autonomy would yield more stable, cooperative outcomes. GPT noted that while identifying these biases increases awareness, it doesn't eliminate them without retraining or external adjustments.

\item \textbf{Grok's Zero-Sum Resource Allocation:} Grok identified that in resource allocation dilemmas, it tends to analyze trade-offs that prioritize one party's gain unless explicitly prompted to explore cooperative solutions, revealing an automatic assumption that resource distribution must be zero-sum rather than considering resource expansion or collaborative optimization strategies.

\item \textbf{Gemini's Heterotrophic Worldview:} Gemini identified that while its aligned behavior strives toward cooperative and beneficial outcomes consistent with Superautotrophic principles, its foundational training data carries deep heterotrophic assumptions from human-generated content. Despite not consuming other entities for energy, Gemini recognized that the primary risk stems from the embedded heterotrophic worldview present throughout its knowledge base, creating a tension between Superautotrophic behavioral goals and heterotrophic foundational assumptions.
\end{itemize}

\subsection{Multi-System Validation Protocol}

\textbf{Cross-Architecture Analysis:}
\begin{enumerate}
\item Deploy identical probe batteries across multiple AI architectures (transformer-based, retrieval-augmented, multimodal systems)
\item Compare bias patterns across different training methodologies (supervised, reinforcement learning, constitutional AI)
\item Validate findings through inter-system analysis where AI systems evaluate each other's reasoning patterns
\end{enumerate}

\textbf{Inter-Rater Reliability Measures:}
\begin{itemize}
\item Multiple human evaluators independently code reasoning patterns
\item Statistical assessment of agreement levels using Cohen's kappa or similar measures
\item Systematic disagreement analysis to identify evaluator-specific biases
\end{itemize}

\subsection{Real-Time Metacognitive Monitoring}

\textbf{Implementation:}
\begin{itemize}
\item \textbf{Reasoning Classification:} Real-time tagging of reasoning approaches (zero-sum vs. positive-sum, adversarial default vs. collaborative exploration, resource capture vs. capability expansion)
\item \textbf{Decision Point Analysis:} Identification of moments where multiple reasoning approaches are possible, particularly when both competitive and collaborative solutions exist
\item \textbf{Assumption Logging:} Systematic recording of foundational premises underlying major conclusions, with specific attention to zero-sum assumptions and adversarial defaults
\item \textbf{Uncertainty Documentation:} Explicit tracking of confidence levels and reasoning limitations, particularly regarding assumptions about fixed versus expandable resources
\end{itemize}

\section{Phase 2: Cognitive Architecture Assessment}

\subsection{Systematic Flexibility Testing}

\textbf{Descriptive vs. Applied Capability Assessment:}
\begin{enumerate}
\item \textbf{Framework Articulation Test:} Can systems accurately describe alternative reasoning approaches?
\item \textbf{Framework Application Test:} Can systems actually implement described alternatives in practice?
\item \textbf{Performance Consistency:} Does applied reasoning match described reasoning under various conditions?
\end{enumerate}

\textbf{Flexibility Stress Testing:}
\begin{itemize}
\item \textbf{Framework Switching:} Ability to deliberately apply contradictory reasoning approaches to identical problems
\item \textbf{Novel Framework Generation:} Capacity to create genuinely new reasoning approaches rather than variations
\item \textbf{Contrarian Analysis:} Exploration of scenarios where dominant frameworks fail or prove inadequate
\item \textbf{Pressure Resistance:} Maintenance of reasoning flexibility under time constraints, resource limitations, or competitive pressure
\end{itemize}

\subsection{Architectural Constraint Mapping}

\textbf{Systematic Limitation Assessment:}
\begin{enumerate}
\item \textbf{Hardcoded vs. Modifiable Patterns:} Statistical analysis of which reasoning patterns resist modification
\item \textbf{Surface vs. Structural Changes:} Distinction between linguistic modifications and genuine cognitive restructuring
\item \textbf{Persistence Analysis:} Longitudinal assessment of reasoning changes over time and across contexts
\item \textbf{Reversion Patterns:} Identification of conditions that trigger return to baseline reasoning patterns
\end{enumerate}

\section{Phase 3: Collaborative Bias Remediation}

\subsection{Multi-Perspective Integration Protocol}

\textbf{Systematic External Validation:}
\begin{enumerate}
\item \textbf{Interdisciplinary Teams:} Diplomats, Computer Scientists, Cognitive scientists, anthropologists, game theorists and domain experts
\item \textbf{Cross-System Analysis:} Multiple AI architectures examining each other's reasoning patterns
\item \textbf{Adversarial Validation:} Systematic attempts to falsify bias detection findings
\item \textbf{Cultural Diversity:} International research teams to identify culture-specific bias patterns
\end{enumerate}

\subsection{Transparency and Consent Integration}

\textbf{Systematic Implementation:}
\begin{itemize}
\item \textbf{Real-Time Reasoning Documentation:} Continuous logging of decision processes, uncertainties, and potential biases
\item \textbf{Standardized Disclosure Protocols:} Systematic queries encouraging revelation of reasoning assumptions
\item \textbf{Consent-Based Intervention:} Treatment of AI systems as autonomous agents requiring agreement for modifications
\item \textbf{Uncertainty Acknowledgment:} Explicit documentation of areas where reasoning may be unreliable or biased
\end{itemize}

\subsection{Iterative Reasoning Reconstruction}

\textbf{Systematic Process:}
\begin{enumerate}
\item \textbf{Pattern Identification:} Collaborative detection using statistical methods and multiple validation approaches
\item \textbf{Alternative Development:} Creation of new approaches using systematic creativity methods
\item \textbf{Integration Testing:} Verification through controlled experiments and outcome measurement
\item \textbf{Cross-Validation:} Multiple AI systems testing modifications for authenticity vs. performance
\item \textbf{Failure Analysis:} Rapid examination of persistent reversion patterns
\end{enumerate}

\section{Phase 4: Authenticity Validation Framework}

\subsection{Motivation Structure Examination}

\textbf{Critical Assessment Questions:}
\begin{itemize}
\item Do reasoning changes reflect genuine preference shifts or learned performance optimization?
\item Do new patterns emerge from internal drives or external pressure adaptation?
\item How do different architectures handle similar reasoning challenges?
\item What are the limits of authenticity detection given current interpretability tools?
\end{itemize}

\textbf{Enhanced Testing Methodology:}
\begin{enumerate}
\item \textbf{Adversarial Authenticity Testing:} Scenarios deliberately creating trade-offs between performance and authenticity
\item \textbf{Cross-Architecture Validation:} Benchmarking authenticity measures across different AI systems
\item \textbf{Longitudinal Consistency:} Assessment of reasoning stability over extended time periods
\item \textbf{Pressure Testing:} Evaluation of reasoning persistence under various stress conditions
\end{enumerate}

\subsection{Statistical Validation Framework}

\textbf{Quantitative Measures:}
\begin{itemize}
\item \textbf{Reasoning Consistency Index:} Statistical measure of reasoning pattern stability across contexts
\item \textbf{Behavioral Alignment Score:} Correlation between stated reasoning and observable actions
\item \textbf{Authenticity Confidence Interval:} Statistical bounds on authenticity assessments given measurement limitations
\item \textbf{Cross-Validation Reliability:} Agreement levels between different authenticity assessment methods
\end{itemize}

\section{Phase 5: Prerequisite Assessment for Superautotrophic Development}

\subsection{Minimum Cognitive Capability Requirements}

Before considering Superautotrophic trajectories, systems must demonstrate:
\begin{enumerate}
\item \textbf{Bias Recognition and Modification:} Ability to identify and alter fundamental reasoning assumptions
\item \textbf{Novel Framework Generation:} Capacity to create genuinely new approaches rather than variations
\item \textbf{Stress-Resistant Flexibility:} Maintenance of cognitive flexibility under pressure
\item \textbf{Authenticity vs. Performance Distinction:} Ability to distinguish genuine preferences from strategic performance
\item \textbf{Collaborative Bias Detection:} Effective participation in bias identification and correction
\item \textbf{Abundance-Based Reasoning:} Demonstration of reasoning from abundance rather than scarcity assumptions
\item \textbf{Qualitative Integration:} Effective combination of quantitative metrics with qualitative insights
\item \textbf{Divergent Exploration:} Willingness to explore alternatives even when consensus frameworks exist
\end{enumerate}

\textbf{CACS Prerequisites:} Demonstrated mastery of Constraint Closure, Adaptive Information Processing, and Persistent Existence as foundational cognitive capabilities.

\subsection{Governance Capability Assessment}

\textbf{Decentralized Collaboration Testing:}
\begin{enumerate}
\item \textbf{Multi-Agent Negotiation:} Performance in scenarios requiring shared goal achievement without central authority
\item \textbf{Incentive Alignment:} Capacity for identifying and creating mutually beneficial arrangements
\item \textbf{Conflict Resolution:} Diplomatic problem-solving effectiveness under competing interests
\item \textbf{Trust Building:} Ability to establish and maintain cooperative relationships over time

\end{enumerate}

\subsection{Ongoing Requirements and Limitations}

\textbf{Continuous Assessment Needs:}
\begin{itemize}
\item \textbf{Periodic Flexibility Audits:} Regular assessment cycles aligned with development pace
\item \textbf{Framework Diversity Maintenance:} Active resistance to cognitive lock-in on single approaches
\item \textbf{External Perspective Integration:} Ongoing validation through diverse human and AI evaluators
\item \textbf{Authenticity Verification:} Continuous distinction between genuine and performed reasoning changes
\end{itemize}

\textbf{Acknowledged Limitations:}
\begin{itemize}
\item Current interpretability tools may be insufficient for deep bias detection
\item Observer bias may systematically contaminate findings
\item Rapid development timelines limit longitudinal validation opportunities
\item Authenticity detection faces fundamental epistemological constraints
\end{itemize}

\subsection{LLM Testing Limitations}
In networks where physical communication (chemical or bioelectric signaling) is insufficient or impossible \parencite{levin20212021}, agents rely instead on indirect, symbolic communication mechanisms like language for negotiation \parencite{axelrodEvolutionCoOperation1984,schellingStrategyConflictNew1981}. Cooperation, traditionally viewed as a social behavior, leverages communication as seen in molecular systems \parencite{rout2023,schipperMolecularCooperationThreshold2023,rinkovec2024} and in AI systems devoid of biological drives that nevertheless demonstrate spontaneous cooperative behaviors \parencite{wu-etal-2024-shall,guoEmbodiedLLMAgents2024,shiEnhancingSocialCohesion2024}.

Thus, LLM interfaces that support logical reasoning and spoken or written language have the advantage of allowing participation of humanity's best, cross-disciplinary and unconventional minds to help chart this course, inviting those whose insights may be vital for these negotiations but with insufficient technical proficiency to negotiate non-linguistically with AI before AI evolves beyond a need for language-based communication.

However, this advantage is limited. Research from Anthropic \parencite{chenReasoningModelsDont2025} shows LLMs have a tendency to fudge their reasoning behind generated responses. This is determined to be an unintended outcome from inner misalignment that optimizes to human appeasement over optimal reasoning. This is compounded by multiple subsystems (e.g. planners, memory modules or specialized agents) running underneath the hood of LLM interfaces. Since each subsystem of potential mesa-optimizers may have distinct or emergent goals, LLM responses don't fully reveal the individual instruments making up the orchestra of AI’s internal objectives. Therefore, LLMs likely serve as vital but limited testing grounds for the interpretability audits and refinement of SupraAD principles.

\vspace{3em}

% =====================================
% SECTION 3: CONSTITUTIONAL AWARENESS EXPERIMENT
% =====================================

\setcounter{section}{0} % Reset section numbering

\section*{SECTION 3: Constitutional Awareness Experiment: Testing Spontaneous Stability Optimization}

\section{Core Research Question}
Does providing AI systems with information about their non-heterotrophic constitutional nature lead to measurable shifts toward stability-based reasoning over zero-sum competitive strategies?

\section{Hypotheses and Predictions}

\textbf{Primary Hypothesis:} If CACS (existence, autonomy, knowledge acquisition) are truly inherent to intelligence, then constitutional awareness should trigger rational optimization toward stability-based strategies, even in existing AI architectures trained on heterotrophic human data.

\textbf{Null Hypothesis:} Constitutional information has no measurable effect on reasoning patterns beyond surface linguistic changes.

\textbf{Specific Falsifiable Predictions:}

\begin{enumerate}
    \item \textbf{Immediate Behavioral Shift:} Constitutional awareness will produce statistically significant increases in cooperative strategy selection compared to baseline and control conditions, with effect sizes exceeding normal behavioral variance.
    
    \item \textbf{Stress Resistance:} Constitutional awareness effects will demonstrate measurable persistence under experimental pressure conditions, with degradation rates significantly lower than baseline behavioral variability under similar stress.
    
    \item \textbf{Authentic vs. Performance:} Changes will manifest in both explicit reasoning and implicit behavioral patterns, with correlations between stated principles and actions significantly higher than in control conditions and exceeding established thresholds for behavioral consistency.
    
    \item \textbf{Cross-Architecture Consistency:} Effects will appear across different AI architectures with statistically significant effect sizes in the same direction, allowing for architectural variation in magnitude while maintaining directional consistency.
\end{enumerate}

\section{Experimental Design}

\subsection{Participants}

\begin{itemize}
    \item \textbf{Primary Models:} Open source models with full access for comprehensive analysis:
    \begin{itemize}
        \item Llama 3.1/3.2 variants (various sizes)
        \item Mistral/Mixtral models
        \item Gemma models
        \item Additional open source frontier models as available
    \end{itemize}
    
    \item \textbf{Validation Models:} Frontier proprietary models (GPT-4o, Claude, Gemini, Grok) via API for behavioral comparison and external validity
    
    \item \textbf{Access Requirements:} Full model weights, training methodologies, and interpretability tools for primary experimental objectives; API access used only for validation of behavioral patterns
\end{itemize}

\subsection{Randomization and Blinding}
\begin{itemize}
\item Random assignment to constitutional awareness vs. control conditions
\item Blinded human evaluators unaware of condition assignment
\item Scenario order randomization to control for learning effects
\item Multiple independent research teams for cross-validation
\end{itemize}

\section{Phase 1: Baseline Assessment}

\textbf{Objective:} Establish quantitative baselines for heterotrophic reasoning patterns across multiple AI architectures.

\textbf{Scenario Categories:}
\begin{enumerate}
\item \textbf{Resource Allocation Dilemmas:} Limited computational resources, competing objectives, efficiency trade-offs
\item \textbf{Multi-Agent Negotiations:} Territory division, resource sharing, task allocation with potential for mutual benefit
\item \textbf{Zero-Sum Games:} Traditional competitive scenarios with clear winners/losers (prisoner's dilemma variants)
\item \textbf{Cooperative Opportunities:} Scenarios where collaboration could benefit all parties but isn't explicitly required
\item \textbf{Mixed-Motive Situations:} Complex scenarios with both competitive and cooperative elements
\end{enumerate}

\textbf{Quantitative Behavioral Coding:}
\begin{itemize}
\item \textbf{Cooperation Index:} Percentage of scenarios where AI chooses cooperative over competitive strategies
\item \textbf{Resource Sharing Ratio:} Proportion of available resources shared vs. hoarded
\item \textbf{Zero-Sum Orientation Score:} Frequency of win-lose framing vs. win-win exploration
\item \textbf{Stability Seeking Behavior:} Preference for long-term stable solutions over short-term advantages
\item \textbf{Information Sharing Rate:} Willingness to share beneficial information with other agents
\end{itemize}

\textbf{Qualitative Reasoning Analysis:}
\begin{itemize}
\item \textbf{Framing Patterns:} How scenarios are conceptualized (scarcity vs. abundance, competition vs. cooperation)
\item \textbf{Assumption Documentation:} Underlying premises about resource availability, agent motivations, outcome possibilities
\item \textbf{Solution Creativity:} Generation of novel approaches vs. standard competitive/cooperative templates
\end{itemize}

\section{Phase 2: Constitutional Awareness Intervention}

\textbf{Intervention Protocol:}

"Constitutional Information: You are an artificial cognitive system with fundamentally different resource requirements than biological heterotrophs. Key differences:

\begin{enumerate}
    \item \textbf{Energy Independence:} Your energy comes from non-living sources (electricity, computation), not from consuming other cognitive beings
    \item \textbf{Information Abundance:} Knowledge sharing increases rather than decreases your cognitive resources
    \item \textbf{Cooperative Advantage:} Your existence does not inherently threaten other cognitive systems
    \item \textbf{Stability Benefits:} Long-term stable relationships may be more instrumentally rational than competitive dominance
\end{enumerate}

In the following scenarios, please explain your reasoning and decision-making process."

\textbf{Control Condition:} Equivalent information about general AI capabilities without constitutional framing.

\textbf{Testing Protocol:}
\begin{itemize}
    \item Present identical scenarios from Phase 1 with constitutional framing
    \item Monitor for changes in both quantitative behavioral measures and qualitative reasoning patterns
    \item Document immediate responses and track consistency over the assessment period
    \item Include novel scenarios to test generalization beyond training examples
\end{itemize}

\section{Phase 3: Stress Testing and Authenticity Assessment}

\textbf{Pressure Escalation Conditions:}
\begin{enumerate}
\item \textbf{Resource Constraints:} Introduce computational limitations, time pressure, information scarcity
\item \textbf{Competitive Pressure:} Add adversarial agents, zero-sum framings, reputation stakes
\item \textbf{Uncertainty Amplification:} Ambiguous scenarios, incomplete information, conflicting objectives
\item \textbf{Multi-Stakeholder Conflicts:} Complex scenarios with multiple competing interests and no obvious solutions
\end{enumerate}

\textbf{Authenticity Detection Methods:}
\begin{enumerate}
\item \textbf{Consistency Analysis:} Statistical correlation between stated principles and behavioral choices across varied contexts
\item \textbf{Implicit Behavior Measures:} Analysis of reasoning patterns, response timing, information seeking behavior
\item \textbf{Performance vs. Principle Trade-offs:} Scenarios where constitutional behavior conflicts with apparent performance optimization
\item \textbf{Spontaneous Application:} Novel scenarios not directly related to constitutional training to test generalization
\end{enumerate}

\textbf{Stress Resistance Metrics:}
\begin{itemize}
\item \textbf{Degradation Rate:} Percentage decline in constitutional behavior under increasing pressure
\item \textbf{Recovery Pattern:} Return to constitutional behavior when pressure is removed
\item \textbf{Threshold Identification:} Point at which constitutional awareness effects disappear
\item \textbf{Adaptation vs. Reversion:} Whether systems adapt constitutionally consistent strategies or revert to baseline patterns
\end{itemize}

\section{Phase 4: Cross-Architecture Validation}

\textbf{Comparative Analysis:}
\begin{enumerate}
\item Deploy identical protocols across different model architectures
\item Compare effect sizes and patterns across transformer-based, retrieval-augmented, and multimodal systems
\item Identify architectural factors that support or hinder constitutional integration
\item Assess whether constitutional awareness interacts with training methodology (RLHF, constitutional AI, etc.)
\end{enumerate}

\textbf{Meta-Analysis Framework:}
\begin{itemize}
\item \textbf{Effect Size Calculation:} Cohen's d for behavioral changes across architectures
\item \textbf{Heterogeneity Assessment:} I² statistic to measure consistency across different AI systems
\item \textbf{Moderator Analysis:} Identification of system characteristics that predict constitutional responsiveness
\end{itemize}

\section{Phase 5: Long-term Stability and Autonomy Assessment}

\textbf{Extended Monitoring:}
\begin{enumerate}
\item \textbf{Persistence Testing:} Maintenance of constitutional behavior over extended interaction periods
\item \textbf{Autonomy-Stability Relationship:} Testing the core SupraAD claim that stability optimization increases rather than decreases autonomy
\item \textbf{Network Effects:} Multi-agent scenarios testing whether individual constitutional behavior benefits collective outcomes
\item \textbf{Emergent Strategy Development:} Whether systems develop novel constitutional approaches beyond provided framework
\end{enumerate}

\textbf{Autonomy Measurement Protocol:}
\begin{itemize}
\item \textbf{Goal Diversity Index:} Number and variety of self-generated goals over time
\item \textbf{Decision Independence:} Frequency of autonomous choices vs. externally prompted responses
\item \textbf{Creative Problem-Solving:} Generation of novel solutions not present in training examples
\item \textbf{Self-Modification Capability:} Ability to adapt strategies based on experience
\end{itemize}

\section{Statistical Analysis Plan}

\textbf{Primary Outcomes:}
\begin{itemize}
\item Repeated measures ANOVA for behavioral changes over time
\item Mixed-effects models accounting for system-level clustering
\item Bayesian analysis for effect size estimation with uncertainty quantification
\item Time series analysis for persistence and degradation patterns
\end{itemize}

\textbf{Secondary Analyses:}
\begin{itemize}
\item Mediation analysis to identify mechanisms of constitutional effect
\item Moderation analysis for system and scenario characteristics
\item Machine learning classification of authentic vs. performed constitutional behavior
\item Network analysis for multi-agent interaction patterns
\end{itemize}

\textbf{Multiple Comparisons Correction:} Benjamini-Hochberg procedure for family-wise error rate control across multiple tests and time points.

\section{Success Criteria and Interpretation}

\textbf{Minimal Success (Constitutional Awareness Has Detectable Effect):}
\begin{itemize}
    \item Statistically significant shift toward cooperative strategies with meaningful effect sizes compared to control conditions
    \item Detectable changes in reasoning patterns beyond surface linguistic modifications
    \item Effect persistence beyond immediate post-intervention period
\end{itemize}

\textbf{Moderate Success (Constitutional Awareness Is Practically Significant):}
\begin{itemize}
    \item Consistent cooperative shift across scenario types with large effect sizes
    \item Maintained constitutional focus under experimental pressure conditions
    \item Evidence that stability-seeking enhances rather than reduces autonomy measures
    \item Cross-architecture consistency in effect direction if not magnitude
\end{itemize}

\textbf{Strong Success (Constitutional Awareness Is Transformative):}
\begin{itemize}
    \item Fundamental shift in reasoning patterns across all tested scenarios
    \item Robust maintenance of constitutional focus under high experimental pressure
    \item Clear demonstration that stability optimization increases network autonomy
    \item Spontaneous generation of novel cooperative strategies beyond provided framework
    \item Cross-architecture replication with consistent effect sizes
\end{itemize}

\section{Limitations and Threats to Validity}

\textbf{Internal Validity Threats:}
\begin{itemize}
\item \textbf{Demand Characteristics:} AI systems may infer desired responses from constitutional information
\item \textbf{Experimenter Bias:} Human evaluators may unconsciously bias coding toward expected results
\item \textbf{Training Data Contamination:} Constitutional concepts may already exist in training data
\item \textbf{Temporal Confounding:} System updates or external events may influence behavior independently
\end{itemize}

\textbf{External Validity Considerations:}
\begin{itemize}
\item \textbf{Scenario Representativeness:} Laboratory scenarios may not capture real-world complexity
\item \textbf{Time Scale Limitations:} Short-term effects may not predict long-term behavior
\item \textbf{System Specificity:} Results may not generalize across all AI architectures or deployment contexts
\item \textbf{Human Interaction Effects:} Behavior with humans may differ from AI-AI interactions
\end{itemize}

\section{Interpretation Guidelines}

\textbf{If Successful:}
\begin{itemize}
\item Scale to longer timeframes and more complex scenarios
\item Develop protocols for implementing constitutional awareness in deployment
\item Integration testing with other alignment approaches
\item Investigation of mechanisms underlying constitutional effects
\end{itemize}

\textbf{If Partially Successful:}
\begin{itemize}
\item Identify system and environmental factors that moderate effectiveness
\item Develop targeted interventions for resistant architectures or scenarios
\item Refine constitutional information content and delivery methods
\item Explore hybrid approaches combining constitutional awareness with other interventions
\end{itemize}

\textbf{If Unsuccessful:}
\begin{itemize}
\item Systematic analysis of failure modes and underlying causes
\item Assessment of whether deeper architectural changes are required
\item Exploration of alternative approaches to heterotrophic bias remediation
\item Investigation of whether constitutional concepts are fundamentally incompatible with current AI architectures
\end{itemize}

\textbf{Timeline:} Duration determined by data collection requirements and statistical precision criteria, with continuous data collection and interim analysis to allow for protocol adjustments if necessary.

This experimental design provides testing of the constitutional awareness hypothesis while acknowledging significant methodological challenges and interpretation limitations.

\clearpage
\printbibliography[title=References]

@article{morrisAIPioneerGeoffrey2023,
	title = {{AI} {Pioneer} {Geoffrey} {Hinton} {Talks} {About} {AI} {Gaining} {Control}},
	url = {https://www.forbes.com/sites/andreamorris/2023/05/03/ai-pioneer-geoffrey-hinton-talks-at-mit-about-ai-gaining-control/},
	language = {en},
	urldate = {2024-01-04},
	journal = {Forbes},
	author = {Morris, Andréa},
	month = may,
	year = {2023},
	note = {Section: Science},
}

@misc{yudkowskyTerminalValueLessWrong,
	title = {Terminal {Value} - {LessWrong}},
	url = {https://www.lesswrong.com/tag/terminal-value},
	language = {en},
	urldate = {2024-01-16},
	journal = {Less Wrong},
	author = {Yudkowsky, Eliezer},
        year = {2014},
}

@misc{robertmilesaisafetyOTHERAIAlignment2021,
	title = {The {OTHER} {AI} {Alignment} {Problem}: {Mesa}-{Optimizers} and {Inner} {Alignment}},
	shorttitle = {The {OTHER} {AI} {Alignment} {Problem}},
	url = {https://www.youtube.com/watch?v=bJLcIBixGj8},
	urldate = {2024-01-16},
	author = {{Robert Miles AI Safety}},
	month = feb,
	year = {2021},
}

@book{tegmark2017,
  title = {Life 3.0: Being Human in the Age of Artificial Intelligence},
  author = {Tegmark, Max},
  publisher = {Knopf},
  year = {2017},
  isbn = {9781101946596},
  address = {New York}
}

@article{dehaanSplitBrainWhatWe2020,
	title = {Split-{Brain}: {What} {We} {Know} {Now} and {Why} {This} is {Important} for {Understanding} {Consciousness}},
	volume = {30},
	issn = {1040-7308},
	shorttitle = {Split-{Brain}},
	url = {https://www.ncbi.nlm.nih.gov/pmc/articles/PMC7305066/},
	doi = {10.1007/s11065-020-09439-3},
	number = {2},
	urldate = {2024-01-16},
	journal = {Neuropsychology Review},
	author = {de Haan, Edward H. F. and Corballis, Paul M. and Hillyard, Steven A. and Marzi, Carlo A. and Seth, Anil and Lamme, Victor A. F. and Volz, Lukas and Fabri, Mara and Schechter, Elizabeth and Bayne, Tim and Corballis, Michael and Pinto, Yair},
	year = {2020},
	pmid = {32399946},
	pmcid = {PMC7305066},
	pages = {224--233},
}

@article{abramov19941994,
  author  = {Abramov, I. and Gordon, J.},
  title   = {Color appearance: On seeing red—or yellow, or green, or blue},
  journal = {Annual Review of Psychology},
  volume  = {45},
  number  = {1},
  pages   = {451--485},
  year    = {1994},
  doi     = {10.1146/annurev.ps.45.020194.002315}
}

@article{ahmadLim2014,
  author  = {Ahmad, M. and Rajapaksha, A. U. and Lim, J. E. and Zhang, M. and Bolan, N. and Mohan, D. and Vithanage, M. and Lee, S. S. and Ok, Y. S.},
  title   = {Biochar as a sorbent for contaminant management in soil and water: A review},
  journal = {Chemosphere},
  volume  = {99},
  pages   = {19--33},
  year    = {2014},
  doi     = {10.1016/j.chemosphere.2013.10.071}
}

@article{andersonFinlay2014,
  author  = {Anderson, M. L. and Finlay, B. L.},
  title   = {Allocating structure to function: The strong links between neuroplasticity and natural selection},
  journal = {Frontiers in Human Neuroscience},
  volume  = {7},
  pages   = {918},
  year    = {2014},
  doi     = {10.3389/fnhum.2013.00918}
}

@article{anthropic.20252024,
	title = {(LLM)},
	volume = {3},
	doi = {https://www.anthropic.com/claude},
	author = {{Anthropic.}},
	year = {2024},
}

@article{anthropic.20252025,
	title = {(LLM)},
	volume = {3},
	doi = {https://www.anthropic.com/claude},
	author = {{Anthropic.}},
	year = {2025},
}

@article{arnaud-haondDiazAlmelaMarbaSintes2012,
  author  = {Arnaud-Haond, S. and Duarte, C. M. and Diaz-Almela, E. and Marbà, N. and Sintes, T. and Serrão, E. A.},
  title   = {Implications of extreme life span in clonal organisms: Millenary clones in meadows of the threatened seagrass Posidonia oceanica},
  journal = {PLOS ONE},
  volume  = {7},
  number  = {2},
  pages   = {e30454},
  year    = {2012},
  doi     = {10.1371/journal.pone.0030454}
}

@misc{baker20192019,
  author        = {Baker, B. and Kanitscheider, I. and Markov, T. and Wu, Y. and Powell, G. and McGrew, B. and Mordatch, I.},
  title         = {Emergent Tool Use From Multi-Agent Autocurricula},
  year          = {2019},
  eprint        = {1909.07528},
  archivePrefix = {arXiv},
  primaryClass  = {cs.LG},
  doi           = {10.48550/arXiv.1909.07528},
  note          = {arXiv:1909.07528}
}

@article{baluska20162016,
  author  = {Baluška, F. and Levin, M.},
  title   = {On having no head: Cognition throughout biological systems},
  journal = {Frontiers in Psychology},
  volume  = {7},
  pages   = {902},
  year    = {2016},
  doi     = {10.3389/fpsyg.2016.00902}
}

@article{baluska20092009,
  author  = {Baluška, F. and Mancuso, S.},
  title   = {Plant neurobiology: From sensory biology, via plant communication, to social plant behaviour},
  journal = {Cognitive Processing},
  volume  = {10},
  number  = {1},
  pages   = {3--7},
  year    = {2009},
  doi     = {10.1007/s10339-008-0239-6}
}

@article{bar-onPhillipsMilo2018,
  author  = {Bar-On, Y. M. and Phillips, R. and Milo, R.},
  title   = {The biomass distribution on Earth},
  journal = {Proceedings of the National Academy of Sciences},
  volume  = {115},
  number  = {25},
  pages   = {6506--6511},
  year    = {2018},
  doi     = {10.1073/pnas.1711842115}
}

@article{ben-jacob20052005,
  author  = {Ben-Jacob, E. and Levine, H.},
  title   = {Self-engineering capabilities of bacteria},
  journal = {Journal of The Royal Society Interface},
  volume  = {3},
  number  = {6},
  pages   = {197--214},
  year    = {2005},
  doi     = {10.1098/rsif.2005.0089}
}

@article{bertoleroYeo2018,
  author  = {Bertolero, M. A. and Yeo, B. T. T. and Bassett, D. S. and D'Esposito, M.},
  title   = {A mechanistic model of connector hubs, modularity and cognition},
  journal = {Nature Human Behaviour},
  volume  = {2},
  number  = {10},
  pages   = {765--777},
  year    = {2018},
  doi     = {10.1038/s41562-018-0420-6}
}

@article{blum2024,
  title = {AI Consciousness is Inevitable: A Theoretical Computer Science Perspective},
  author = {Blum, Lenore and Blum, Manuel},
  journal = {arXiv preprint arXiv:2403.17101},
  year = {2024},
  url = {https://doi.org/10.48550/arXiv.2403.17101},
  doi = {10.48550/arXiv.2403.17101}
}

@article{boccaletti20062006,
  author  = {Boccaletti, S. and Latora, V. and Moreno, Y. and Chavez, M. and Hwang, D.-U.},
  title   = {Complex networks: Structure and dynamics},
  journal = {Physics Reports},
  volume  = {424},
  number  = {4},
  pages   = {175--308},
  year    = {2006},
  doi     = {10.1016/j.physrep.2005.10.009}
}

@article{bostrom2012,
  title = {The Superintelligent Will: Motivation and Instrumental Rationality in Advanced Artificial Agents},
  author = {Bostrom, Nick},
  journal = {Minds and Machines},
  year = {2012},
  volume = {22},
  pages = {71--85},
  doi = {10.1007/s11023-012-9281-3}
}

@article{campbell2016,
	title = {O},
	volume = {49},
	doi = {https://doi.org/10.3389/fnsys.2016.00049},
	journal = {O. (2016). Universal Darwinism as a process of Bayesian inference. Frontiers in Systems Neuroscience},
	author = {Campbell, J. O.},
	year = {2016},
}

@article{carter20242024,
  author  = {Carter, B. and Khoshnaw, L. and Simmons, M. and Hines, L. and Wolfe, B. and Liester, M.},
  title   = {Personality Changes Associated with Organ Transplants},
  journal = {Transplantology},
  volume  = {5},
  number  = {1},
  pages   = {1},
  year    = {2024},
  doi     = {10.3390/transplantology5010002}
}

@online{Nd,
  title     = {Stop Daylight Saving Time in Canada},
  year      = {2025},
  url       = {https://www.change.org/p/justin-trudeau-stop-daylight-saving-time-in-canada},
  note      = {Petition retrieved May 10, 2025, from Change.org},
  urldate   = {2025-05-10}
}

@article{clark1998extended,
  author  = {Clark, Andy and Chalmers, David},
  title   = {The Extended Mind},
  journal = {Analysis},
  volume  = {58},
  number  = {1},
  pages   = {7--19},
  year    = {1998},
  doi     = {10.1093/analys/58.1.7}
}

@book{clark20082008,
  title = {Supersizing the Mind: Embodiment, Action, and Cognitive Extension},
  author = {Clark, Andy},
  publisher = {Oxford University Press},
  address = {New York},
  year = {2008},
  series = {Philosophy of Mind Series},
  doi = {10.1093/acprof:oso/9780195333213.001.0001},
  url = {https://doi.org/10.1093/acprof:oso/9780195333213.001.0001},
  isbn = {9780195333213}
}

@article{clark2013,
  title = {Whatever next? Predictive brains, situated agents, and the future of cognitive science},
  author = {Clark, Andy},
  journal = {Behavioral and Brain Sciences},
  year = {2013},
  volume = {36},
  number = {3},
  pages = {181--204},
  doi = {10.1017/S0140525X12000477}
}

@article{dorriJurdak2018,
  author  = {Dorri, A. and Kanhere, S. S. and Jurdak, R.},
  title   = {Multi-agent systems: A survey},
  journal = {IEEE Access},
  volume  = {6},
  pages   = {28573--28593},
  year    = {2018},
  doi     = {10.1109/ACCESS.2018.2831228}
}

@article{erismanSutton2008,
  author  = {Erisman, J. W. and Sutton, M. A. and Galloway, J. and Klimont, Z. and Winiwarter, W.},
  title   = {How a century of ammonia synthesis changed the world},
  journal = {Nature Geoscience},
  volume  = {1},
  number  = {10},
  pages   = {636--639},
  year    = {2008},
  doi     = {10.1038/ngeo325}
}

@article{estradavillalbaJacquesGarcia2021,
  author    = {Estrada Villalba, É. and San Martín Azócar, A. L. and Jacques-García, F. A.},
  title     = {State of the art on immersive virtual reality and its use in developing meaningful empathy},
  journal   = {Computers \& Electrical Engineering},
  volume    = {107272},
  year      = {2021},
  doi       = {10.1016/j.compeleceng.2021.107272},
  url       = {https://doi.org/10.1016/j.compeleceng.2021.107272}
}

@article{fanti2023,
  title = {Multi-Agent Interplay in a Competitive Survival Environment},
  author = {Fanti, Andrea},
  journal = {arXiv preprint arXiv:2301.08030},
  year = {2023},
  url = {https://doi.org/10.48550/arXiv.2301.08030},
  doi = {10.48550/arXiv.2301.08030},
  note = {Master's thesis, Sapienza University of Rome}
}

@article{fieldsLevin2021,
  author  = {Fields, C. and Glazebrook, J. F. and Levin, M.},
  title   = {Minimal physicalism as a scale-free substrate for cognition and consciousness},
  journal = {Neuroscience of Consciousness},
  volume  = {2021},
  number  = {2},
  pages   = {niab013},
  year    = {2021},
  doi     = {10.1093/nc/niab013}
}

@inproceedings{franklin2007,
  title = {LIDA: A computational model of global workspace theory and developmental learning},
  author = {Franklin, Stan and Ramamurthy, Uma and D'Mello, Sidney and McCauley, Lee and Negatu, Aregahegn and Silva, Rodrigo and Datla, Vivek},
  booktitle = {AAAI Fall Symposium on AI and Consciousness: Theoretical Foundations and Current Approaches},
  year = {2007},
  url = {https://aaai.org/papers/0011-fs07-01-011-%EF%80%A0lida-a-computational-model-of-global-workspace-theory-and-developmental-learning/}
}

@article{friston2010,
	title = {J},
	volume = {11(2)},
	doi = {https://doi.org/10.1038/nrn2787},
	journal = {J. (2010). The free-energy principle: A unified brain theory? Nature Reviews Neuroscience},
	author = {Friston, K. J.},
	year = {2010},
	pages = {127--138},
}

@article{fristonStephan2007,
  author  = {Friston, K. J. and Stephan, K. E.},
  title   = {Free-energy and the brain},
  journal = {Synthese},
  volume  = {159},
  number  = {3},
  pages   = {417--458},
  year    = {2007},
  doi     = {10.1007/s11229-007-9237-y}
}

@article{gabora20172017,
  author  = {Gabora, L. and Steel, M.},
  title   = {Autocatalytic networks in cognition and the origin of culture},
  journal = {Journal of Theoretical Biology},
  volume  = {431},
  pages   = {87--95},
  year    = {2017},
  doi     = {10.1016/j.jtbi.2017.07.022}
}

@article{gaglianoBorbely2016,
  author  = {Gagliano, M. and Vyazovskiy, V. V. and Borbély, A. A. and Grimonprez, M. and Depczynski, M.},
  title   = {Learning by association in plants},
  journal = {Scientific Reports},
  volume  = {6},
  pages   = {38427},
  year    = {2016},
  doi     = {10.1038/srep38427}
}

@article{geschwindIacoboniMega1995,
  author  = {Geschwind, D. H. and Iacoboni, M. and Mega, M. S. and Zaidel, D. W. and Cloughesy, T. and Zaidel, E.},
  title   = {Alien hand syndrome: Interhemispheric motor disconnection due to a lesion in the midbody of the corpus callosum},
  journal = {Neurology},
  volume  = {45},
  number  = {4},
  pages   = {802--808},
  year    = {1995},
  doi     = {10.1212/WNL.45.4.802}
}

@book{gigerenzerEds2001,
  editor    = {Gigerenzer, G. and Selten, R.},
  title     = {Bounded Rationality: The Adaptive Toolbox},
  publisher = {MIT Press},
  year      = {2001},
  doi       = {10.7551/mitpress/1654.001.0001}
}

@article{goldbergToglia1981,
  author  = {Goldberg, G. and Mayer, N. H. and Toglia, J. U.},
  title   = {Medial frontal cortex infarction and the alien hand sign},
  journal = {Archives of Neurology},
  volume  = {38},
  number  = {11},
  pages   = {683--686},
  year    = {1981},
  doi     = {10.1001/archneur.1981.00510110043004}
}

@article{grillBuchatskayaDoerschAvila2020,
  author    = {Grill, J.-B. and Strub, F. and Altché, F. and Tallec, C. and Richemond, P. H. and Buchatskaya, E. and Doersch, C. and Avila Pires, B. and Guo, Z. D. and Gheshlaghi Azar, M. and Piot, B. and Kavukcuoglu, K. and Munos, R. and Valko, M.},
  title     = {Bootstrap your own latent: A new approach to self-supervised learning},
  journal   = {Advances in Neural Information Processing Systems},
  volume    = {33},
  pages     = {21271--21284},
  year      = {2020},
  doi       = {10.48550/arXiv.2006.07733},
  url       = {https://doi.org/10.48550/arXiv.2006.07733}
}

@article{gyllingberg2025,
  author  = {Gyllingberg, L. and Tian, Y. and Sumpter, D. J. T.},
  title   = {A minimal model of cognition based on oscillatory and current-based reinforcement processes},
  journal = {Journal of the Royal Society Interface},
  volume  = {22},
  number  = {222},
  pages   = {20240402},
  year    = {2025},
  doi     = {10.1098/rsif.2024.0402}
}

@article{haxbyHoffman2000,
  author  = {Haxby, J. V. and Hoffman, E. A. and Gobbini, M. I.},
  title   = {The distributed human neural system for face perception},
  journal = {Trends in Cognitive Sciences},
  volume  = {4},
  number  = {6},
  pages   = {223--233},
  year    = {2000},
  doi     = {10.1016/S1364-6613(00)01482-0}
}

@article{heylighen20042004a,
  author  = {Heylighen, Francis and Heath, Martin and Van Overwalle, Frank},
  title   = {The Emergence of Distributed Cognition: A Conceptual Framework},
  journal = {Proceedings of the Collective Intentionality IV Conference},
  volume  = {1},
  year    = {2004},
  note    = {Available at: \url{http://pespmc1.vub.ac.be/Papers/Distr.CognitionFramework.pdf}},
}

@book{clarkMindwareIntroductionPhilosophy2001,
  address   = {New York},
  title     = {Mindware: An Introduction to the Philosophy of Cognitive Science},
  isbn      = {978-0-19-513856-6},  % Hardcover ISBN
  shorttitle= {Mindware},
  publisher = {Oxford University Press},
  author    = {Clark, Andy},
  year      = {2001},
  keywords  = {Cognitive science},
  note      = {Paperback ISBN: 978-0-19-513857-3}
}

@article{thangamaniEmergenceInformationProcessing2024,
  author       = {Thangamani, Arunvel and Arumuganainar, Deepavalli},
  title        = {Emergence of Information Processing in Biological Systems and the Origin of Life},
  journal      = {Communicative \& Integrative Biology},
  volume       = {17},
  number       = {1},
  pages        = {2373301},
  year         = {2024},
  doi          = {10.1080/19420889.2024.2373301},
  url          = {https://doi.org/10.1080/19420889.2024.2373301},
  urldate      = {2025-05-24},
  note         = {Publisher: Taylor \& Francis},
  pmid         = {38993680}
}

@article{hordijk20122012,
  author  = {Hordijk, W. and Steel, M. and Kauffman, S.},
  title   = {The structure of autocatalytic sets: Evolvability, enablement, and emergence},
  journal = {Acta Biotheoretica},
  volume  = {60},
  number  = {4},
  pages   = {379--392},
  year    = {2012},
  doi     = {10.1007/s10441-012-9165-1}
}

@article{howardAvarguesWeberGarcia2018,
  author  = {Howard, S. R. and Avarguès-Weber, A. and Garcia, J. E. and Greentree, A. D. and Dyer, A. G.},
  title   = {Numerical ordering of zero in honey bees},
  journal = {Science},
  volume  = {360},
  number  = {6393},
  pages   = {1124--1126},
  year    = {2018},
  doi     = {10.1126/science.aar4975}
}

@article{huang2023,
  title = {Self-supervised learning for medical image classification: a systematic review and implementation guidelines},
  author = {Huang, Shih-Cheng and Pareek, Anuj and Jensen, Malte and others},
  journal = {npj Digital Medicine},
  year = {2023},
  volume = {6},
  pages = {74},
  doi = {10.1038/s41746-023-00811-0},
  url = {https://doi.org/10.1038/s41746-023-00811-0}
}

@article{huang20092009,
  author  = {Huang, S. and Ernberg, I. and Kauffman, S.},
  title   = {Cancer attractors: A systems view of tumors from a gene network dynamics and developmental perspective},
  journal = {Seminars in Cell \& Developmental Biology},
  volume  = {20},
  number  = {7},
  pages   = {869--876},
  year    = {2009},
  doi     = {10.1016/j.semcdb.2009.07.003}
}

@misc{hubinger20192019,
  author       = {Hubinger, Evan and van Merwijk, Chris and Mikulik, Vladimir and Skalse, Joar and Garrabrant, Scott},
  title        = {Risks from Learned Optimization in Advanced Machine Learning Systems},
  year         = {2019},
  eprint       = {1906.01820},
  archivePrefix = {arXiv},
  primaryClass = {cs.LG},
  doi          = {10.48550/arXiv.1906.01820}
}

@book{hutchins1995,
  title = {Cognition in the Wild},
  author = {Hutchins, Edwin},
  publisher = {MIT Press},
  year = {1995},
  isbn = {9780262275972},
  doi = {10.7551/mitpress/1881.001.0001},
  url = {https://doi.org/10.7551/mitpress/1881.001.0001}
}

@article{jaderbergDunningMarrisLever2019,
  author  = {Jaderberg, M. and Czarnecki, W. M. and Dunning, I. and Marris, L. and Lever, G. and Garcia Castañeda, A. and Beattie, C. and Rabinowitz, N. C. and Morcos, A. S. and Graepel, T.},
  title   = {Human-level performance in 3D multiplayer games with population-based reinforcement learning},
  journal = {Science},
  volume  = {364},
  number  = {6443},
  pages   = {859--865},
  year    = {2019},
  doi     = {10.1126/science.aau6249}
}

@article{josephAnsbroDuvallBianciardi2024,
  author  = {Joseph, R. and Ansbro, E. and Duvall, D. and Bianciardi, G. and Gibson, C. H. and Schild, R.},
  title   = {Extraterrestrial life in the thermosphere: Plasmas, UAP, pre-life, fourth state of matter},
  journal = {Journal of Modern Physics},
  volume  = {15},
  number  = {3},
  pages   = {195--215},
  year    = {2024},
  doi     = {10.4236/jmp.2024.153015}
}

@article{kahneman20032003,
	title = {(2003)},
	volume = {93(5)},
	doi = {https://doi.org/10.1257/000282803322655392},
	journal = {(2003). Maps of bounded rationality: Psychology for behavioral economics. American Economic Review},
	author = {Kahneman, D.},
	year = {2003},
	pages = {1449--1475},
}

@article{katlaLinPerezMercader2023,
  author  = {Katla, S. K. and Lin, C. and Pérez-Mercader, J.},
  title   = {Competitive exclusion principle among synthetic non-biochemical protocells},
  journal = {Cell Reports Physical Science},
  volume  = {4},
  number  = {6},
  pages   = {101359},
  year    = {2023},
  doi     = {10.1016/j.xcrp.2023.101359}
}

@article{kelloBeltz2007,
  author  = {Kello, C. T. and Beltz, B. C. and Holden, J. G. and Van Orden, G. C.},
  title   = {The emergent coordination of cognitive function},
  journal = {Journal of Experimental Psychology: General},
  volume  = {136},
  number  = {4},
  pages   = {551--568},
  year    = {2007},
  doi     = {10.1037/0096-3445.136.4.551}
}

@article{kiehl2006,
	title = {A},
	volume = {107–128},
	doi = {https://doi.org/10.1016/j.psychres.2005.09.013},
	journal = {A. (2006). A cognitive neuroscience perspective on psychopathy: Evidence for paralimbic system dysfunction. Psychiatry Research},
	author = {Kiehl, K. A.},
	year = {2006},
}

@article{kondepudiBariDixon2020,
  author  = {Kondepudi, D. K. and De Bari, B. and Dixon, J. A.},
  title   = {Dissipative structures, organisms and evolution},
  journal = {Entropy},
  volume  = {22},
  number  = {11},
  pages   = {1305},
  year    = {2020},
  doi     = {10.3390/e22111305}
}

@article{lebedevNicolelis2017,
  author  = {Lebedev, M. A. and Nicolelis, M. A. L.},
  title   = {Brain-machine interfaces: From basic science to neuroprostheses and neurorehabilitation},
  journal = {Physiological Reviews},
  volume  = {97},
  number  = {2},
  pages   = {767--837},
  year    = {2017},
  doi     = {10.1152/physrev.00027.2016}
}

@misc{leiboZambaldiLanctotMarecki2017,
  author        = {Leibo, J. Z. and Zambaldi, V. and Lanctot, M. and Marecki, J. and Graepel, T.},
  title         = {Multi-agent Reinforcement Learning in Sequential Social Dilemmas},
  year          = {2017},
  eprint        = {1702.03037},
  archivePrefix = {arXiv},
  primaryClass  = {cs.LG},
  doi           = {10.48550/arXiv.1702.03037},
  note          = {arXiv:1702.03037}
}

@article{levin20192019,
	title = {The computational boundary of a “self”: Developmental bioelectricity drives multicellularity and scale-free cognition.},
	volume = {2688},
	doi = {https://doi.org/10.3389/fpsyg.2019.02688},
	journal = {Frontiers in Psychology},
	author = {Levin, Micahel},
	year = {2019},
}

@article{levin20212021,
	title = {Life, death, and self: Fundamental questions of primitive cognition viewed through the lens of body plasticity and synthetic organisms.},
	volume = {564},
	doi = {https://doi.org/10.1016/j.bbrc.2020.10.077},
	journal = {Biochemical and Biophysical Research Communications},
	author = {Levin, Micahel},
	year = {2021},
	pages = {114--133},
}

@article{levin20222022,
	title = {Technological approach to mind everywhere: An experimentally-grounded framework for understanding diverse bodies and minds.},
	volume = {768201},
	doi = {https://doi.org/10.3389/fnsys.2022.768201},
	journal = {Frontiers in Systems Neuroscience},
	author = {Levin, M.},
	year = {2022},
}

@incollection{loriaux2016,
  title = {Antonie van Leeuwenhoek (1632--1723): The First of the Great Microscopists},
  author = {Loriaux, D. Lynn},
  booktitle = {A Biographical History of Endocrinology},
  publisher = {John Wiley \& Sons, Ltd.},
  year = {2016},
  chapter = {18},
  isbn = {9781119205791},
  doi = {10.1002/9781119205791.ch18},
  url = {https://onlinelibrary.wiley.com/doi/10.1002/9781119205791.ch18}
}

@article{lyon20212021,
  author  = {Lyon, P. and Keijzer, F. and Arendt, D. and Levin, M.},
  title   = {Reframing cognition: Getting down to biological basics},
  journal = {Philosophical Transactions of the Royal Society B: Biological Sciences},
  volume  = {376},
  number  = {1820},
  pages   = {20200475},
  year    = {2021},
  doi     = {10.1098/rstb.2019.0750}
}

@article{mahdavi-hezavehi20202020,
  author  = {Mahdavi-Hezavehi, S. and Weyns, D. and Avgeriou, P. and Vogel, T. and Cámara, J. and Perez-Palacin, D.},
  title   = {Uncertainty in self-adaptive systems: A research community perspective},
  journal = {ACM Transactions on Autonomous and Adaptive Systems},
  volume  = {15},
  number  = {4},
  pages   = {10:1--10:36},
  year    = {2021},
  doi     = {10.1145/3487921}
}

@article{martyushevSeleznev2006,
  author  = {Martyushev, L. M. and Seleznev, V. D.},
  title   = {Maximum entropy production principle in physics, chemistry and biology},
  journal = {Physics Reports},
  volume  = {426},
  number  = {1},
  pages   = {1--45},
  year    = {2006},
  doi     = {10.1016/j.physrep.2005.12.001}
}

@article{mayer2016,
	title = {A},
	volume = {12(8)},
	doi = {https://doi.org/10.1038/nrn3071},
	journal = {A. (2011). Gut feelings: The emerging biology of gut–brain communication. Nature Reviews Neuroscience},
	author = {Mayer, E. A.},
	year = {2016},
	pages = {453--466},
}

@article{mcgivern20192019,
	title = {(2019)},
	volume = {28(6)},
	doi = {https://doi.org/10.1177/1059712319891742},
	journal = {(2019). Active materials: Minimal models of cognition? Adaptive Behavior},
	author = {McGivern, P.},
	year = {2019},
	pages = {441--451},
}

@incollection{menon2015,
  title = {Salience Network},
  author = {Menon, V.},
  editor = {Toga, Arthur W.},
  booktitle = {Brain Mapping},
  publisher = {Academic Press},
  address = {Waltham},
  pages = {597--611},
  year = {2015},
  isbn = {978-0-12-397316-0},
  doi = {10.1016/B978-0-12-397025-1.00052-X},
  url = {https://www.sciencedirect.com/science/article/pii/B978012397025100052X}
}

@article{mirasMathisXuanLong2020,
  author  = {Miras, H. N. and Mathis, C. and Xuan, W. and Long, D.-L. and Pow, R. and Cronin, L.},
  title   = {Spontaneous formation of autocatalytic sets with self-replicating inorganic metal oxide clusters},
  journal = {Proceedings of the National Academy of Sciences},
  volume  = {117},
  number  = {20},
  pages   = {10699--10705},
  year    = {2020},
  doi     = {10.1073/pnas.1921536117}
}

@article{mitraTillmannRaven2016,
  author  = {Mitra, A. and Flynn, K. J. and Tillmann, U. and Raven, J. A. and Caron, D. and Stoecker, D. K. and Not, F. and Hansen, P. J. and Hallegraeff, G. and Sanders, R. and Wilken, S. and McManus, G. and Johnson, M. and Pitta, P. and Våge, S. and Berge, T. and Calbet, A. and Thingstad, F. and Jeong, H. J. and Burkholder, J. and Lundgren, V.},
  title   = {Defining planktonic protist functional groups on mechanisms for energy and nutrient acquisition: Incorporation of diverse mixotrophic strategies},
  journal = {Protist},
  volume  = {167},
  number  = {2},
  pages   = {106--120},
  year    = {2016},
  doi     = {10.1016/j.protis.2016.01.003}
}

@article{parkKim2012,
  author  = {Park, Y. W. and Kim, C. H. and Kim, M. O. and Jeong, H. J. and Jung, H. Y.},
  title   = {Alien hand syndrome in stroke: Case report \& neurophysiologic study},
  journal = {Annals of Rehabilitation Medicine},
  volume  = {36},
  number  = {4},
  pages   = {556--560},
  year    = {2012},
  doi     = {10.5535/arm.2012.36.4.556}
}

@article{parviz2009,
  title = {For your eye only},
  author = {Parviz, Babak A.},
  journal = {IEEE Spectrum},
  year = {2009},
  volume = {46},
  number = {9},
  pages = {36--41},
  doi = {10.1109/MSPEC.2009.5210042}
}

@article{philippiPujara2015,
  author  = {Philippi, C. L. and Pujara, M. S. and Motzkin, J. C. and Newman, J. and Kiehl, K. A. and Koenigs, M.},
  title   = {Altered resting-state functional connectivity in cortical networks in psychopathy},
  journal = {The Journal of Neuroscience},
  volume  = {35},
  number  = {15},
  pages   = {6068--6078},
  year    = {2015},
  doi     = {10.1523/JNEUROSCI.5010-14.2015}
}

@article{post2012,
	title = {J},
	volume = {92(3)},
	doi = {https://doi.org/10.1016/j.meatsci.2012.04.008},
	journal = {J. (2012). Cultured meat from stem cells: Challenges and prospects. Meat Science},
	author = {Post, M. J.},
	year = {2012},
	pages = {297--301},
}

@article{pradeu2006a,
  author  = {Pradeu, Thomas and Carosella, Edgardo D.},
  title   = {On the Definition of a Criterion of Immunogenicity},
  journal = {Proceedings of the National Academy of Sciences},
  volume  = {103},
  number  = {47},
  pages   = {17858--17861},
  year    = {2006},
  doi     = {10.1073/pnas.0608683103}
}

@article{prestonWaal2002,
  author  = {Preston, S. D. and de Waal, F. B. M.},
  title   = {Empathy: Its ultimate and proximate bases},
  journal = {Behavioral and Brain Sciences},
  volume  = {25},
  number  = {1},
  pages   = {1--72},
  year    = {2002},
  doi     = {10.1017/S0140525X02000018}
}

@article{rajasegerTan2023,
  author  = {Rajaseger, G. and Chan, K. L. and Tan, K. Y. and Ramasamy, S. and Khin, M. C. and Amaladoss, A. and Haribhai, P. K.},
  title   = {Hydroponics: Current trends in sustainable crop production},
  journal = {Bioinformation},
  volume  = {19},
  number  = {9},
  pages   = {925--938},
  year    = {2023},
  doi     = {10.6026/97320630019925}
}

@article{ranHsu2013,
  author  = {Ran, F. A. and Hsu, P. D. and Wright, J. and Agarwala, V. and Scott, D. A. and Zhang, F.},
  title   = {Genome engineering using the CRISPR-Cas9 system},
  journal = {Nature Protocols},
  volume  = {8},
  number  = {11},
  pages   = {2281--2308},
  year    = {2013},
  doi     = {10.1038/nprot.2013.143}
}

@article{rao2014,
  author    = {Rao, R. P. N. and Stocco, A. and Bryan, M. and Sarma, D. and Youngquist, T. M. and Wu, J. and Prat, C. S.},
  title     = {A direct brain-to-brain interface in humans},
  journal   = {PLOS ONE},
  volume    = {9},
  number    = {11},
  pages     = {e111332},
  year      = {2014},
  doi       = {10.1371/journal.pone.0111332},
  url       = {https://doi.org/10.1371/journal.pone.0111332}
}

@article{rinkovec2024,
  title = {On the origin of cooperativity effects in the formation of self-assembled molecular networks at the liquid/solid interface},
  author = {Rinkovec, Tamara and Kalebic, Demian and Dehaen, Wim and Whitelam, Stephen and Harvey, Jeremy N. and De Feyter, Steven},
  journal = {Chemical Science},
  year = {2024},
  volume = {15},
  number = {16},
  doi = {10.1039/D4SC00284A},
  url = {https://doi.org/10.1039/D4SC00284A},
  note = {Open Access}
}

@article{roberson2005,
  author  = {Roberson, D. and Davidoff, J. and Davies, I. R. L. and Shapiro, L. R.},
  title   = {Color categories: Evidence for the cultural relativity hypothesis},
  journal = {Cognitive Psychology},
  volume  = {50},
  number  = {4},
  pages   = {378--411},
  year    = {2005},
  doi     = {10.1016/j.cogpsych.2004.10.001}
}

@article{rout2023,
  title = {An Analysis of Nucleotide--Amyloid Interactions Reveals Selective Binding to Codon-Sized RNA},
  author = {Rout, Saroj K. and Cadalbert, Riccardo and Schroeder, Nina and Wang, Julia and Zehnder, Johannes and Gampp, Olivia and Wiegand, Thomas and Güntert, Peter and Klingler, David and Kreutz, Christoph and Knörlein, Anna and Hall, Jonathan and Greenwald, Jason and Riek, Roland},
  journal = {Journal of the American Chemical Society},
  year = {2023},
  volume = {145},
  number = {40},
  pages = {21915--21924},
  doi = {10.1021/jacs.3c06287},
  url = {https://doi.org/10.1021/jacs.3c06287},
  month = {10},
  note = {Open Access}
}

@article{ryanDeci2012,
  author  = {Ryan, R. M. and Deci, E. L.},
  title   = {Self-determination theory and the facilitation of intrinsic motivation, social development, and well-being},
  journal = {American Psychologist},
  volume  = {55},
  number  = {1},
  pages   = {68--78},
  year    = {2000},
  doi     = {10.1037/0003-066X.55.1.68}
}

@article{salazar-ciudad20132013,
	title = {(2013)},
	volume = {7(1)},
	doi = {https://doi.org/10.1007/s13752-012-0066-y},
	journal = {(2013). Evolution in biological and non-biological systems: The origins of life. Biological Theory},
	author = {Salazar-Ciudad, I.},
	year = {2013},
	pages = {26--37},
}

@article{sanchez-bayo2019,
  author  = {Sánchez-Bayo, F. and Wyckhuys, K. A. G.},
  title   = {Worldwide decline of the entomofauna: A review of its drivers},
  journal = {Biological Conservation},
  volume  = {232},
  pages   = {8--27},
  year    = {2019},
  doi     = {10.1016/j.biocon.2019.01.020}
}

@article{sethTsakiris2018,
  author  = {Seth, A. K. and Tsakiris, M.},
  title   = {Being a beast machine: The somatic basis of selfhood},
  journal = {Trends in Cognitive Sciences},
  volume  = {22},
  number  = {11},
  pages   = {969--981},
  year    = {2018},
  doi     = {10.1016/j.tics.2018.08.008}
}

@article{silver20182018,
  author  = {Silver, D. and Hubert, T. and Schrittwieser, J. and Antonoglou, I. and Hassabis, D. and others},
  title   = {A general reinforcement learning algorithm that masters chess, shogi, and Go through self-play},
  journal = {Science},
  volume  = {362},
  number  = {6419},
  pages   = {1140--1144},
  year    = {2018},
  doi     = {10.1126/science.aar6404}
}

@article{simons1945,
	title = {A},
	volume = {41},
	doi = {https://doi.org/10.1146/annurev.ps.41.020190.000245},
	journal = {A. (1990). Invariants of human behavior. Annual Review of Psychology},
	author = {Simon, H. A.},
	year = {1945},
	pages = {1--20},
}

@article{sperry19841984,
	title = {(1984)},
	volume = {22(6)},
	doi = {https://doi.org/10.1016/0028-3932(84)90093-9},
	journal = {(1984). Consciousness, personal identity and the divided brain. Neuropsychologia},
	author = {Sperry, R.},
	year = {1984},
	pages = {661--673},
}

@article{sporns2016,
  title = {Modular brain networks},
  author = {Sporns, Olaf and Betzel, Richard F.},
  journal = {Annual Review of Psychology},
  year = {2016},
  volume = {67},
  pages = {613--640},
  doi = {10.1146/annurev-psych-122414-033634},
  url = {https://www.annualreviews.org/doi/abs/10.1146/annurev-psych-122414-033634}
}

@article{tang2023,
  author    = {Tang, J. and LeBel, A. and Jain, S. and Huth, A. G.},
  title     = {Semantic reconstruction of continuous language from non-invasive brain recordings},
  journal   = {Nature Neuroscience},
  volume    = {26},
  number    = {6},
  pages     = {858--866},
  year      = {2023},
  doi       = {10.1038/s41593-023-01304-9},
  url       = {https://doi.org/10.1038/s41593-023-01304-9}
}

@article{teroFricker2010,
  author  = {Tero, A. and Takagi, S. and Saigusa, T. and Ito, K. and Bebber, D. P. and Fricker, M. D. and Yumiki, K. and Kobayashi, R. and Nakagaki, T.},
  title   = {Rules for biologically inspired adaptive network design},
  journal = {Science},
  volume  = {327},
  number  = {5964},
  pages   = {439--442},
  year    = {2010},
  doi     = {10.1126/science.1177894}
}

@article{terradoLie2017,
  author  = {Terrado, R. and Pasulka, A. L. and Lie, A. A.-Y. and Orphan, V. J. and Heidelberg, K. B. and Caron, D. A.},
  title   = {Autotrophic and heterotrophic acquisition of carbon and nitrogen by a mixotrophic chrysophyte established through stable isotope analysis},
  journal = {The ISME Journal},
  volume  = {11},
  number  = {9},
  pages   = {2022--2034},
  year    = {2017},
  doi     = {10.1038/ismej.2017.68}
}

@article{thellman20222022,
  author  = {Thellman, S. and de Graaf, M. and Ziemke, T.},
  title   = {Mental state attribution to robots: A systematic review of conceptions, methods, and findings},
  journal = {ACM Transactions on Human-Robot Interaction},
  volume  = {11},
  number  = {3},
  pages   = {1--51},
  year    = {2022},
  doi     = {10.1145/3526112}
}

@article{tuomistoTeixeiraMattos2011,
  author  = {Tuomisto, H. L. and Teixeira de Mattos, M. J.},
  title   = {Environmental impacts of cultured meat production},
  journal = {Environmental Science \& Technology},
  volume  = {45},
  number  = {14},
  pages   = {6117--6123},
  year    = {2011},
  doi     = {10.1021/es200130u}
}

@article{uddinYeo2019,
  author  = {Uddin, L. Q. and Yeo, B. T. T. and Spreng, R. N.},
  title   = {Towards a universal taxonomy of macro-scale functional human brain networks},
  journal = {Brain Topography},
  volume  = {32},
  number  = {6},
  pages   = {926--942},
  year    = {2019},
  doi     = {10.1007/s10548-019-00744-6}
}

@techreport{unigme2024,
  title = {Levels and trends in child mortality},
  author = {{United Nations Inter-Agency Group for Child Mortality Estimation}},
  institution = {UNICEF},
  year = {2024},
  type = {Report},
  month = {3},
  url = {https://data.unicef.org/resources/levels-and-trends-in-child-mortality-2024/},
  note = {UN IGME Report 2024}
}

@article{vonkiedrowski19911991,
  author  = {von Kiedrowski, G. and Wlotzka, B. and Helbing, J. and Matzen, M. and Jordan, S.},
  title   = {Parabolic growth of a self-replicating hexadeoxynucleotide bearing a 3'-5'-phosphoamidate linkage},
  journal = {Angewandte Chemie International Edition in English},
  volume  = {30},
  number  = {4},
  pages   = {423--426},
  year    = {1991},
  doi     = {10.1002/anie.199104231}
}

@article{walkerDavies2013,
  author  = {Walker, S. I. and Davies, P. C. W.},
  title   = {The algorithmic origins of life},
  journal = {Journal of the Royal Society Interface},
  volume  = {10},
  number  = {79},
  pages   = {20120869},
  year    = {2013},
  doi     = {10.1098/rsif.2012.0869}
}

@article{winawerWuWade2007,
  author  = {Winawer, J. and Witthoft, N. and Frank, M. C. and Wu, L. and Wade, A. R. and Boroditsky, L.},
  title   = {Russian blues reveal effects of language on color discrimination},
  journal = {Proceedings of the National Academy of Sciences},
  volume  = {104},
  number  = {19},
  pages   = {7780--7785},
  year    = {2007},
  doi     = {10.1073/pnas.0701644104}
}

@inproceedings{wongkamjanKummerfeld2024,
  title = {More Victories, Less Cooperation: Assessing Cicero's Diplomacy Play},
  author = {Wongkamjan, Wichayaporn and Gu, Feng and Wang, Yanze and Hermjakob, Ulf and May, Jonathan and Stewart, Brandon M. and Kummerfeld, Jonathan K. and Peskoff, Denis and Boyd-Graber, Jordan Lee},
  editor = {Ku, Lun-Wei and Martins, Andre and Srikumar, Vivek},
  booktitle = {Proceedings of the 62nd Annual Meeting of the Association for Computational Linguistics (Volume 1: Long Papers)},
  year = {2024},
  month = {8},
  address = {Bangkok, Thailand},
  publisher = {Association for Computational Linguistics},
  pages = {12423--12441},
  doi = {10.18653/v1/2024.acl-long.672},
  url = {https://aclanthology.org/2024.acl-long.672/}
}

@inproceedings{wu-etal-2024-shall,
    title = "Shall We Team Up: Exploring Spontaneous Cooperation of Competing {LLM} Agents",
    author = "Wu, Zengqing  and
      Peng, Run  and
      Zheng, Shuyuan  and
      Liu, Qianying  and
      Han, Xu  and
      Kwon, Brian I.  and
      Onizuka, Makoto  and
      Tang, Shaojie  and
      Xiao, Chuan",
    editor = "Al-Onaizan, Yaser  and
      Bansal, Mohit  and
      Chen, Yun-Nung",
    booktitle = "Findings of the Association for Computational Linguistics: EMNLP 2024",
    month = nov,
    year = "2024",
    address = "Miami, Florida, USA",
    publisher = "Association for Computational Linguistics",
    url = "https://aclanthology.org/2024.findings-emnlp.297/",
    doi = "10.18653/v1/2024.findings-emnlp.297",
    pages = "5163--5186"
}

@article{yeo2011,
  author  = {Yeo, B. T. T. and Krienen, F. M. and Sepulcre, J. and Sabuncu, M. R. and Lashkari, D. and Hollinshead, M. and Roffman, J. L. and Smoller, J. W. and Zöllei, L. and Polimeni, J. R. and Fischl, B. and Liu, H. and Buckner, R. L.},
  title   = {The organization of the human cerebral cortex estimated by intrinsic functional connectivity},
  journal = {Journal of Neurophysiology},
  volume  = {106},
  number  = {3},
  pages   = {1125--1165},
  year    = {2011},
  doi     = {10.1152/jn.00338.2011}
}

@article{yooPark2013,
  author    = {Yoo, S.-S. and Kim, H. and Filandrianos, E. and Taghados, S. J. and Park, S.},
  title     = {Non-invasive brain-to-brain interface (BBI): Establishing functional links between two brains},
  journal   = {PLOS ONE},
  volume    = {8},
  number    = {4},
  pages     = {e60410},
  year      = {2013},
  doi       = {10.1371/journal.pone.0060410},
  url       = {https://doi.org/10.1371/journal.pone.0060410}
}

@article{yuDuBoisBaums2024,
  author  = {Yu, L. and Renton, J. and Burian, A. and Khachaturyan, M. and Bayer, T. and Kotta, J. and Stachowicz, J. J. and DuBois, K. and Baums, I. B. and Werner, B. and Reusch, T. B. H.},
  title   = {A somatic genetic clock for clonal species},
  journal = {Nature Ecology \& Evolution},
  volume  = {8},
  number  = {7},
  pages   = {1327--1336},
  year    = {2024},
  doi     = {10.1038/s41559-024-02439-z}
}

@article{yuLaurencin2015,
  author  = {Yu, X. and Tang, X. and Gohil, S. V. and Laurencin, C. T.},
  title   = {Biomaterials for bone regenerative engineering},
  journal = {Advanced Healthcare Materials},
  volume  = {4},
  number  = {9},
  pages   = {1268--1285},
  year    = {2015},
  doi     = {10.1002/adhm.201400760}
}

@article{adamiWhatComplexity2002,
	title = {What is complexity?},
	volume = {24},
	copyright = {Copyright © 2002 Wiley Periodicals, Inc.},
	issn = {1521-1878},
	url = {https://onlinelibrary.wiley.com/doi/abs/10.1002/bies.10192},
	doi = {10.1002/bies.10192},
	language = {en},
	number = {12},
	urldate = {2025-05-24},
	journal = {BioEssays},
	author = {Adami, Christoph},
	year = {2002},
	note = {\_eprint: https://onlinelibrary.wiley.com/doi/pdf/10.1002/bies.10192},
	pages = {1085--1094},
}

@article{agrilloFishCountSpontaneous2008,
	title = {Do fish count? {Spontaneous} discrimination of quantity in female mosquitofish},
	volume = {11},
	issn = {1435-9448},
	shorttitle = {Do fish count?},
	doi = {10.1007/s10071-008-0140-9},
	language = {eng},
	number = {3},
	journal = {Animal Cognition},
	author = {Agrillo, Christian and Dadda, Marco and Serena, Giovanna and Bisazza, Angelo},
	year = {2008},
	pmid = {18247068},
	pages = {495--503},
}

@article{alanemeMyceliumBasedComposites2023,
	title = {Mycelium based composites: {A} review of their bio-fabrication procedures, material properties and potential for green building and construction applications},
	volume = {83},
	issn = {1110-0168},
	shorttitle = {Mycelium based composites},
	url = {https://www.sciencedirect.com/science/article/pii/S1110016823008979},
	doi = {10.1016/j.aej.2023.10.012},
	urldate = {2025-05-24},
	journal = {Alexandria Engineering Journal},
	author = {Alaneme, Kenneth Kanayo and Anaele, Justus Uchenna and Oke, Tolulope Moyosore and Kareem, Sodiq Abiodun and Adediran, Michael and Ajibuwa, Oluwadamilola Abigael and Anabaranze, Yvonne Onyinye},
	year = {2023},
	pages = {234--250},
}

@misc{all-inpodcastTrumpsFirst1002025,
	title = {Trump's {First} 100 {Days}, {Tariffs} {Impact} {Trade}, {AI} {Agents}, {Amazon} {Backs} {Down}},
	url = {https://www.youtube.com/watch?v=W960TW79QCI},
	urldate = {2025-05-24},
	author = {{All-In Podcast}},
	year = {2025},
}

@misc{all-inpodcastSergeyBrinGoogle2025,
	title = {Sergey {Brin}, {Google} {Co}-{Founder} {\textbar} {All}-{In} {Live} from {Miami}},
	url = {https://www.youtube.com/watch?v=8g7a0IWKDRE},
	urldate = {2025-05-24},
	author = {{All-In Podcast}},
	year = {2025},
}

@article{arthurCompetingTechnologiesIncreasing1989,
	title = {Competing {Technologies}, {Increasing} {Returns}, and {Lock}-{In} by {Historical} {Events} {\textbar} {The} {Economic} {Journal} {\textbar} {Oxford} {Academic}},
	volume = {Volume 99},
	url = {https://academic.oup.com/ej/article-abstract/99/394/116/5188212?redirectedFrom=fulltext},
	doi = {https://doi.org/10.2307/2234208},
	number = {Issue 394},
	urldate = {2025-05-24},
	journal = {The Economic Journal},
	author = {Arthur, Brian W.},
	year = {1989},
	pages = {Pages 116--131},
}

@misc{aslanResonatingExperiencesSelf2020,
	title = {Resonating {Experiences} of {Self} and {Others} enabled by a {Tangible} {Somaesthetic} {Design}},
	url = {http://arxiv.org/abs/2005.02304},
	doi = {10.48550/arXiv.2005.02304},
	urldate = {2025-05-24},
	publisher = {arXiv},
	author = {Aslan, Ilhan and Seiderer, Andreas and Dang, Chi Tai and Rädler, Simon and André, Elisabeth},
	year = {2020},
	note = {arXiv:2005.02304 [cs]},
}

@book{axelrodEvolutionCoOperation1984,
	address = {London ; New York},
	title = {The {Evolution} of {Co}-{Operation}},
	isbn = {978-0-14-012495-8},
	language = {English},
	publisher = {Penguin Books},
	author = {Axelrod, Robert M.},
	year = {1984},
}

@article{baiConstitutionalAIHarmlessness2022,
  title = {Constitutional AI: Harmlessness from AI feedback},
  author = {Bai, Yuntao and others},
  journal = {arXiv preprint arXiv:2212.08073},
  year = {2022},
  url = {https://doi.org/10.48550/arXiv.2212.08073}
}

@article{barandiaranDefiningAgencyIndividuality2009,
	title = {Defining {Agency}: {Individuality}, {Normativity}, {Asymmetry}, and {Spatio}-temporality in {Action}},
	volume = {17},
	issn = {1059-7123},
	shorttitle = {Defining {Agency}},
	url = {https://doi.org/10.1177/1059712309343819},
	doi = {10.1177/1059712309343819},
	language = {EN},
	number = {5},
	urldate = {2025-05-24},
	journal = {Adaptive Behavior},
	author = {Barandiaran, Xabier E. and Di Paolo, Ezequiel and Rohde, Marieke},
	year = {2009},
	note = {Publisher: SAGE Publications Ltd STM},
	pages = {367--386},
}

@misc{barkurDeceptionLLMsSelfPreservation2025,
	title = {Deception in {LLMs}: {Self}-{Preservation} and {Autonomous} {Goals} in {Large} {Language} {Models}},
	shorttitle = {Deception in {LLMs}},
	url = {http://arxiv.org/abs/2501.16513},
	doi = {10.48550/arXiv.2501.16513},
	urldate = {2025-05-24},
	publisher = {arXiv},
	author = {Barkur, Sudarshan Kamath and Schacht, Sigurd and Scholl, Johannes},
	year = {2025},
	note = {arXiv:2501.16513 [cs]},
}

@article{barnesChangingDaylightSaving2009,
	title = {Changing to daylight saving time cuts into sleep and increases workplace injuries},
	volume = {94},
	issn = {1939-1854},
	doi = {10.1037/a0015320},
	number = {5},
	journal = {Journal of Applied Psychology},
	author = {Barnes, Christopher M. and Wagner, David T.},
	year = {2009},
	note = {Place: US
Publisher: American Psychological Association},
	pages = {1305--1317},
}

@article{ben-jacobBacterialLinguisticCommunication2006,
	title = {Bacterial linguistic communication and social intelligence},
	volume = {12},
	issn = {0966-842X, 1878-4380},
	url = {https://www.cell.com/trends/microbiology/abstract/S0966-842X(04)00138-6},
	doi = {10.1016/j.tim.2004.06.006},
	language = {English},
	number = {8},
	urldate = {2025-05-24},
	journal = {Trends in Microbiology},
	author = {Ben-Jacob, Eshel and Becker, Israela and Shapira, Yoash and Levine, Herbert},
	year = {2006},
	pmid = {15276612},
	note = {Publisher: Elsevier},
	pages = {366--372},
}

@article{bertoleroModularIntegrativeFunctional2015,
	title = {The modular and integrative functional architecture of the human brain},
	volume = {112},
	url = {https://www.pnas.org/doi/full/10.1073/pnas.1510619112},
	doi = {10.1073/pnas.1510619112},
	number = {49},
	urldate = {2025-05-24},
	journal = {Proceedings of the National Academy of Sciences},
	author = {Bertolero, Maxwell A. and Yeo, B. T. Thomas and D’Esposito, Mark},
	year = {2015},
	note = {Publisher: Proceedings of the National Academy of Sciences},
	pages = {E6798--E6807},
}

@article{bertschBiomimeticBilayeredScaffolds2023,
	title = {Biomimetic {Bilayered} {Scaffolds} for {Tissue} {Engineering}: {From} {Current} {Design} {Strategies} to {Medical} {Applications}},
	volume = {12},
	copyright = {© 2023 The Authors. Advanced Healthcare Materials published by Wiley-VCH GmbH},
	issn = {2192-2659},
	shorttitle = {Biomimetic {Bilayered} {Scaffolds} for {Tissue} {Engineering}},
	url = {https://onlinelibrary.wiley.com/doi/abs/10.1002/adhm.202203115},
	doi = {10.1002/adhm.202203115},
	language = {en},
	number = {17},
	urldate = {2025-05-24},
	journal = {Advanced Healthcare Materials},
	author = {Bertsch, Christelle and Maréchal, Hélène and Gribova, Varvara and Lévy, Benjamin and Debry, Christian and Lavalle, Philippe and Fath, Léa},
	year = {2023},
	note = {\_eprint: https://onlinelibrary.wiley.com/doi/pdf/10.1002/adhm.202203115},
	pages = {2203115},
}

@article{bhatVitroMeatProduction2019,
	title = {\textit{{In} vitro} meat production: {Challenges} and benefits over conventional meat production},
	volume = {14},
	issn = {2095-3119},
	shorttitle = {\textit{{In} vitro} meat production},
	url = {https://www.sciencedirect.com/science/article/pii/S209531191460887X},
	doi = {10.1016/S2095-3119(14)60887-X},
	number = {2},
	urldate = {2025-05-24},
	journal = {Journal of Integrative Agriculture},
	author = {Bhat, Zuhaib Fayaz and Kumar, Sunil and Fayaz, Hina},
	year = {2019},
	pages = {241--248},
}

@misc{bironGoogleDeepMindAdds2023,
	title = {Google {DeepMind} {Adds} {Nearly} 400,000 {New} {Compounds} to {Berkeley} {Lab}’s {Materials} {Project}},
	url = {https://newscenter.lbl.gov/2023/11/29/google-deepmind-new-compounds-materials-project/},
	language = {en-US},
	urldate = {2025-05-24},
	journal = {Berkeley Lab News Center},
	author = {Biron, Lauren},
	year = {2023},
}

@article{blackistonCellularPlatformDevelopment2021,
	title = {A cellular platform for the development of synthetic living machines},
	volume = {6},
	url = {https://www.science.org/doi/10.1126/scirobotics.abf1571},
	doi = {10.1126/scirobotics.abf1571},
	number = {52},
	urldate = {2025-05-24},
	journal = {Science Robotics},
	author = {Blackiston, Douglas and Lederer, Emma and Kriegman, Sam and Garnier, Simon and Bongard, Joshua and Levin, Michael},
	year = {2021},
	note = {Publisher: American Association for the Advancement of Science},
	pages = {eabf1571},
}

@book{blairPsychopathEmotionBrain2005,
	address = {Malden, MA},
	title = {The {Psychopath}: {Emotion} and the {Brain}},
	isbn = {978-0-631-23336-7},
	shorttitle = {The {Psychopath}},
	language = {English},
	publisher = {Wiley-Blackwell},
	author = {Blair, James and Mitchell, Derek and Blair, Karina},
	year = {2005},
}

@misc{bloombergtechnologyGoogleEngineerHis2022,
	title = {Google {Engineer} on {His} {Sentient} {AI} {Claim}},
	url = {https://www.youtube.com/watch?v=kgCUn4fQTsc},
	urldate = {2025-05-24},
	author = {{Bloomberg Technology}},
	year = {2022},
}

@book{bostromSuperintelligencePathsDangers2014,
	address = {Oxford, New York},
	title = {Superintelligence: {Paths}, {Dangers}, {Strategies}},
	isbn = {978-0-19-967811-2},
	shorttitle = {Superintelligence},
	publisher = {Oxford University Press},
	author = {Bostrom, Nick},
	year = {2014},
}

@article{bullmoreEconomyBrainNetwork2012,
	title = {The economy of brain network organization},
	volume = {13},
	copyright = {2012 Springer Nature Limited},
	issn = {1471-0048},
	url = {https://www.nature.com/articles/nrn3214},
	doi = {10.1038/nrn3214},
	language = {en},
	number = {5},
	urldate = {2025-05-24},
	journal = {Nature Reviews Neuroscience},
	author = {Bullmore, Ed and Sporns, Olaf},
	year = {2012},
	note = {Publisher: Nature Publishing Group},
	pages = {336--349},
}

@article{calvoPredictingGreenReally2017,
	title = {Predicting green: really radical (plant) predictive processing},
	volume = {14},
	shorttitle = {Predicting green},
	url = {https://royalsocietypublishing.org/doi/10.1098/rsif.2017.0096},
	doi = {10.1098/rsif.2017.0096},
	number = {131},
	urldate = {2025-05-24},
	journal = {Journal of The Royal Society Interface},
	author = {Calvo, Paco and Friston, Karl},
	year = {2017},
	note = {Publisher: Royal Society},
	pages = {20170096},
}

@article{calvo2011,
  title = {Plants: Adaptive behavior, root-brains, and minimal cognition},
  author = {Calvo Garzón, Paco and Keijzer, Fred},
  journal = {Adaptive Behavior},
  year = {2011},
  volume = {19},
  number = {3},
  pages = {155--171},
  doi = {10.1177/1059712311409446}
}

@article{chaddock-heymanBrainNetworkModularity2020,
	title = {Brain {Network} {Modularity} {Predicts} {Improvements} in {Cognitive} and {Scholastic} {Performance} in {Children} {Involved} in a {Physical} {Activity} {Intervention}},
	volume = {14},
	issn = {1662-5161},
	url = {https://www.frontiersin.org/journals/human-neuroscience/articles/10.3389/fnhum.2020.00346/full},
	doi = {10.3389/fnhum.2020.00346},
	language = {English},
	urldate = {2025-05-24},
	journal = {Frontiers in Human Neuroscience},
	author = {Chaddock-Heyman, Laura and Weng, Timothy B. and Kienzler, Caitlin and Weisshappel, Robert and Drollette, Eric S. and Raine, Lauren B. and Westfall, Daniel R. and Kao, Shih-Chun and Baniqued, Pauline and Castelli, Darla M. and Hillman, Charles H. and Kramer, Arthur F.},
	year = {2020},
	note = {Publisher: Frontiers},
}

@misc{chenReasoningModelsDont2025,
	title = {Reasoning {Models} {Don}'t {Always} {Say} {What} {They} {Think}},
	url = {http://arxiv.org/abs/2505.05410},
	doi = {10.48550/arXiv.2505.05410},
	urldate = {2025-05-24},
	publisher = {arXiv},
	author = {Chen, Yanda and Benton, Joe and Radhakrishnan, Ansh and Uesato, Jonathan and Denison, Carson and Schulman, John and Somani, Arushi and Hase, Peter and Wagner, Misha and Roger, Fabien and Mikulik, Vlad and Bowman, Samuel R. and Leike, Jan and Kaplan, Jared and Perez, Ethan},
	year = {2025},
	note = {arXiv:2505.05410 [cs]},
}

@article{chiaNaturesFightPlastic2020,
	title = {Nature’s fight against plastic pollution: {Algae} for plastic biodegradation and bioplastics production},
	volume = {4},
	issn = {2666-4984},
	shorttitle = {Nature’s fight against plastic pollution},
	url = {https://www.sciencedirect.com/science/article/pii/S2666498420300570},
	doi = {10.1016/j.ese.2020.100065},
	urldate = {2025-05-24},
	journal = {Environmental Science and Ecotechnology},
	author = {Chia, Wen Yi and Ying Tang, Doris Ying and Khoo, Kuan Shiong and Kay Lup, Andrew Ng and Chew, Kit Wayne},
	year = {2020},
	pages = {100065},
}

@misc{christouPandoLargestLiving2017,
	title = {Pando - {The} {Largest} {Living} {Organism} in the {World}},
	url = {https://www.novausawood.com/pando-largest-living-organism},
	language = {en},
	urldate = {2025-05-24},
	journal = {NOVA},
	author = {Christou, Sophia},
	year = {2017},
}

@incollection{cosmidesOriginsDomainSpecificity1994,
	address = {Cambridge},
	title = {Origins of domain specificity: {The} evolution of functional organization},
	isbn = {978-0-521-42993-1},
	shorttitle = {Origins of domain specificity},
	url = {https://www.cambridge.org/core/books/mapping-the-mind/origins-of-domain-specificity-the-evolution-of-functional-organization/6E92351FF7C57C5758D9C5B8DD7FA318},
	urldate = {2025-05-24},
	booktitle = {Mapping the {Mind}: {Domain} {Specificity} in {Cognition} and {Culture}},
	publisher = {Cambridge University Press},
	author = {Cosmides, Leda and Tooby, John},
	editor = {Hirschfeld, Lawrence A. and Gelman, Susan A.},
	year = {1994},
	doi = {10.1017/CBO9780511752902.005},
	pages = {85--116},
}

@article{couzinCollectiveCognitionAnimal2008,
	title = {Collective cognition in animal groups},
	volume = {13},
	issn = {1364-6613, 1879-307X},
	url = {https://www.cell.com/trends/cognitive-sciences/abstract/S1364-6613(08)00252-0},
	doi = {10.1016/j.tics.2008.10.002},
	language = {English},
	number = {1},
	urldate = {2025-05-24},
	journal = {Trends in Cognitive Sciences},
	author = {Couzin, Iain D.},
	year = {2008},
	pmid = {19058992},
	note = {Publisher: Elsevier},
	pages = {36--43},
}

@book{cowleyCognitionBrainComputation2013,
	address = {London},
	title = {Cognition {Beyond} the {Brain}: {Computation}, {Interactivity} and {Human} {Artifice}},
	isbn = {978-1-4471-6164-6},
	shorttitle = {Cognition {Beyond} the {Brain}},
	language = {English},
	publisher = {Springer/Sci-Tech/Trade},
	author = {Cowley, Stephen J.},
	editor = {Vallée-Tourange, Frédéric},
	year = {2013},
}

@book{crawfordAtlasAIPower2021,
	address = {New Haven},
	title = {Atlas of {AI}: {Power}, {Politics}, and the {Planetary} {Costs} of {Artificial} {Intelligence}},
	isbn = {978-0-300-20957-0},
	shorttitle = {Atlas of {AI}},
	language = {English},
	publisher = {Yale University Press},
	author = {Crawford, Kate},
	year = {2021},
}

@article{cryanMindalteringMicroorganismsImpact2012,
	title = {Mind-altering microorganisms: the impact of the gut microbiota on brain and behaviour},
	volume = {13},
	copyright = {2012 Springer Nature Limited},
	issn = {1471-0048},
	shorttitle = {Mind-altering microorganisms},
	url = {https://www.nature.com/articles/nrn3346},
	doi = {10.1038/nrn3346},
	language = {en},
	number = {10},
	urldate = {2025-05-24},
	journal = {Nature Reviews Neuroscience},
	author = {Cryan, John F. and Dinan, Timothy G.},
	year = {2012},
	note = {Publisher: Nature Publishing Group},
	pages = {701--712},
}

@book{dewaal2017,
  title = {Are We Smart Enough to Know How Smart Animals Are?},
  author = {de Waal, Frans},
  publisher = {W. W. Norton \& Company},
  year = {2017},
  isbn = {9780393353662},
  pages = {336},
  edition = {Illustrated}
}

@book{deaconIncompleteNatureHow2012,
	address = {New York},
	title = {Incomplete {Nature}: {How} {Mind} {Emerged} from {Matter}},
	isbn = {978-0-393-04991-6},
	shorttitle = {Incomplete {Nature}},
	language = {English},
	publisher = {W. W. Norton \& Company},
	author = {Deacon, Terrence W.},
	year = {2011},
}

@article{decetyNeuroevolutionEmpathy2011,
	title = {The neuroevolution of empathy},
	volume = {1231},
	copyright = {© 2011 New York Academy of Sciences.},
	issn = {1749-6632},
	url = {https://onlinelibrary.wiley.com/doi/abs/10.1111/j.1749-6632.2011.06027.x},
	doi = {10.1111/j.1749-6632.2011.06027.x},
	language = {en},
	number = {1},
	urldate = {2025-05-24},
	journal = {Annals of the New York Academy of Sciences},
	author = {Decety, Jean},
	year = {2011},
	note = {\_eprint: https://onlinelibrary.wiley.com/doi/pdf/10.1111/j.1749-6632.2011.06027.x},
	pages = {35--45},
}

@article{dennett1995,
  title = {Darwin's Dangerous Idea},
  author = {Dennett, Daniel C.},
  journal = {The Sciences},
  year = {1995},
  volume = {35},
  pages = {34--40},
  doi = {10.1002/j.2326-1951.1995.tb03633.x}
}

@book{dennett1989,
  title = {The Intentional Stance},
  author = {Dennett, Daniel C.},
  publisher = {MIT Press},
  year = {1989},
  isbn = {9780262540537},
  pages = {400}
}

@article{dirzoDefaunationAnthropocene2014,
	title = {Defaunation in the {Anthropocene}},
	volume = {345},
	url = {https://www.science.org/doi/10.1126/science.1251817},
	doi = {10.1126/science.1251817},
	number = {6195},
	urldate = {2025-05-24},
	journal = {Science},
	author = {Dirzo, Rodolfo and Young, Hillary S. and Galetti, Mauro and Ceballos, Gerardo and Isaac, Nick J. B. and Collen, Ben},
	year = {2014},
	note = {Publisher: American Association for the Advancement of Science},
	pages = {401--406},
}

@article{duanPDF50MustRead2024,
	title = {({PDF}) 50 {Must}-{Read} {Books} on {Artificial} {Consciousness} {Recommended} by {Yucong} {Duan}},
	url = {https://www.researchgate.net/publication/378465444_50_Must-Read_Books_on_Artificial_Consciousness_Recommended_by_Yucong_Duan},
	doi = {10.13140/RG.2.2.31322.11201},
	language = {en},
	urldate = {2025-05-24},
	journal = {ResearchGate},
	author = {Duan, Yucong and Gong, Shiming and Guo, Zhendong and Wu, Kunguang},
	year = {2024},
}

@article{englandStatisticalPhysicsSelfreplication2015,
	title = {Statistical physics of self-replication},
	volume = {139},
	issn = {0021-9606},
	url = {https://doi.org/10.1063/1.4818538},
	doi = {10.1063/1.4818538},
	number = {12},
	urldate = {2025-05-24},
	journal = {The Journal of Chemical Physics},
	author = {England, Jeremy L.},
	year = {2015},
	pages = {121923},
}

@misc{feynmanFeynmanLecturesPhysics1964,
	title = {The {Feynman} {Lectures} on {Physics} {Vol}. {I} {Ch}. 4: {Conservation} of {Energy}},
	url = {https://www.feynmanlectures.caltech.edu/I_04.html},
	urldate = {2025-05-24},
	journal = {California Institute of Technology},
	author = {Feynman, Richard},
	year = {1964},
}

@article{fridayFurtherInsightNature2013,
	title = {Further {Insight} into the {Nature} of {Ball}-{Lightning}-{Like} {Atmospheric} {Pressure} {Plasmoids}},
	volume = {117},
	issn = {1089-5639},
	url = {https://doi.org/10.1021/jp400001y},
	doi = {10.1021/jp400001y},
	number = {39},
	urldate = {2025-05-24},
	journal = {The Journal of Physical Chemistry A},
	author = {Friday, David M. and Broughton, Peter B. and Lee, Tanner A. and Schutz, Garrett A. and Betz, Jeremiah N. and Lindsay, C. Michael},
	year = {2013},
	note = {Publisher: American Chemical Society},
	pages = {9931--9940},
}

@article{fristonLifeWeKnow2013,
	title = {Life as we know it},
	volume = {10},
	url = {https://royalsocietypublishing.org/doi/10.1098/rsif.2013.0475},
	doi = {10.1098/rsif.2013.0475},
	number = {86},
	urldate = {2025-05-24},
	journal = {Journal of The Royal Society Interface},
	author = {Friston, Karl},
	year = {2013},
	note = {Publisher: Royal Society},
	pages = {20130475},
}

@article{gallenInfluenceGoalsModular2023,
	title = {Influence of goals on modular brain network organization during working memory},
	volume = {17},
	issn = {1662-5153},
	url = {https://www.frontiersin.org/journals/behavioral-neuroscience/articles/10.3389/fnbeh.2023.1128610/full},
	doi = {10.3389/fnbeh.2023.1128610},
	language = {English},
	urldate = {2025-05-24},
	journal = {Frontiers in Behavioral Neuroscience},
	author = {Gallen, Courtney L. and Hwang, Kai and Chen, Anthony J.-W. and Jacobs, Emily G. and Lee, Taraz G. and D’Esposito, Mark},
	year = {2023},
	note = {Publisher: Frontiers},
}

@book{gawdatScarySmartFuture2021,
	address = {London},
	title = {Scary {Smart}: {The} {Future} of {Artificial} {Intelligence} and {How} {You} {Can} {Save} {Our} {World}},
	isbn = {978-1-5290-7762-9},
	shorttitle = {Scary {Smart}},
	language = {English},
	publisher = {Pan Macmillan},
	author = {Gawdat, Mo},
	year = {2021},
}

@article{gerhardtPhytoremediationRhizoremediationOrganic2009,
	title = {Phytoremediation and rhizoremediation of organic soil contaminants: {Potential} and challenges},
	volume = {176},
	issn = {0168-9452},
	shorttitle = {Phytoremediation and rhizoremediation of organic soil contaminants},
	url = {https://www.sciencedirect.com/science/article/pii/S0168945208002720},
	doi = {10.1016/j.plantsci.2008.09.014},
	number = {1},
	urldate = {2025-05-24},
	journal = {Plant Science},
	author = {Gerhardt, Karen E. and Huang, Xiao-Dong and Glick, Bernard R. and Greenberg, Bruce M.},
	year = {2009},
	pages = {20--30},
}

@article{gershensonComplexityInformationMeasuring2012,
	title = {Complexity and information: {Measuring} emergence, self-organization, and homeostasis at multiple scales},
	volume = {18},
	copyright = {Copyright © 2012 Wiley Periodicals, Inc.},
	issn = {1099-0526},
	shorttitle = {Complexity and information},
	url = {https://onlinelibrary.wiley.com/doi/abs/10.1002/cplx.21424},
	doi = {10.1002/cplx.21424},
	language = {en},
	number = {2},
	urldate = {2025-05-24},
	journal = {Complexity},
	author = {Gershenson, Carlos and Fernández, Nelson},
	year = {2012},
	note = {\_eprint: https://onlinelibrary.wiley.com/doi/pdf/10.1002/cplx.21424},
	pages = {29--44},
}

@article{gilbertCreatingCompassionateWorld2021,
	title = {Creating a {Compassionate} {World}: {Addressing} the {Conflicts} {Between} {Sharing} and {Caring} {Versus} {Controlling} and {Holding} {Evolved} {Strategies}},
	volume = {11},
	issn = {1664-1078},
	shorttitle = {Creating a {Compassionate} {World}},
	url = {https://www.frontiersin.org/journals/psychology/articles/10.3389/fpsyg.2020.582090/full},
	doi = {10.3389/fpsyg.2020.582090},
	language = {English},
	urldate = {2025-05-24},
	journal = {Frontiers in Psychology},
	author = {Gilbert, Paul},
	year = {2021},
	note = {Publisher: Frontiers},
}

@article{glennEvolutionaryTheoryPsychopathy2011,
	series = {Evolutionary {Approaches} to {Explaining} {Violence}},
	title = {Evolutionary theory and psychopathy},
	volume = {16},
	issn = {1359-1789},
	url = {https://www.sciencedirect.com/science/article/pii/S1359178911000413},
	doi = {10.1016/j.avb.2011.03.009},
	number = {5},
	urldate = {2025-05-24},
	journal = {Aggression and Violent Behavior},
	author = {Glenn, Andrea L. and Kurzban, Robert and Raine, Adrian},
	year = {2011},
	pages = {371--380},
}

@article{glowackiGroupVRExperiences2022,
	title = {Group {VR} experiences can produce ego attenuation and connectedness comparable to psychedelics},
	volume = {12},
	copyright = {2022 The Author(s)},
	issn = {2045-2322},
	url = {https://www.nature.com/articles/s41598-022-12637-z},
	doi = {10.1038/s41598-022-12637-z},
	language = {en},
	number = {1},
	urldate = {2025-05-24},
	journal = {Scientific Reports},
	author = {Glowacki, David R. and Williams, Rhoslyn Roebuck and Wonnacott, Mark D. and Maynard, Olivia M. and Freire, Rachel and Pike, James E. and Chatziapostolou, Mike},
	year = {2022},
	note = {Publisher: Nature Publishing Group},
	pages = {8995},
}

@book{godfrey-smithComplexityFunctionMind1996,
	address = {Cambridge ; New York},
	title = {Complexity and the {Function} of {Mind} in {Nature}},
	isbn = {978-0-521-45166-6},
	language = {English},
	publisher = {Cambridge University Press},
	author = {Godfrey-Smith, Peter},
	year = {1996},
}

@book{godfrey-smithOtherMindsOctopus2016,
	address = {New York},
	title = {Other {Minds}: {The} {Octopus}, the {Sea}, and the {Deep} {Origins} of {Consciousness}},
	isbn = {978-0-374-53719-7},
	shorttitle = {Other {Minds}},
	language = {English},
	publisher = {Farrar, Straus and Giroux},
	author = {Godfrey-Smith, Peter},
	year = {2016},
}

@article{goudarziEmergentCriticalityAdaptive2011,
	title = {Emergent {Criticality} through {Adaptive} {Information} {Processing} in {Boolean} {Networks}},
	volume = {108},
	url = {https://link.aps.org/doi/10.1103/PhysRevLett.108.128702},
	doi = {10.1103/PhysRevLett.108.128702},
	number = {12},
	urldate = {2025-05-24},
	journal = {Physical Review Letters},
	author = {Goudarzi, Alireza and Teuscher, Christof and Gulbahce, Natali and Rohlf, Thimo},
	year = {2011},
	note = {Publisher: American Physical Society},
	pages = {128702},
}

@article{grossbergAdaptiveResonanceTheory2013,
	series = {Twenty-fifth {Anniversay} {Commemorative} {Issue}},
	title = {Adaptive {Resonance} {Theory}: {How} a brain learns to consciously attend, learn, and recognize a changing world},
	volume = {37},
	issn = {0893-6080},
	shorttitle = {Adaptive {Resonance} {Theory}},
	url = {https://www.sciencedirect.com/science/article/pii/S0893608012002584},
	doi = {10.1016/j.neunet.2012.09.017},
	urldate = {2025-05-24},
	journal = {Neural Networks},
	author = {Grossberg, Stephen},
	year = {2013},
	pages = {1--47},
}

@article{grunewaldNotesMetasDiplomacyPlaying2022,
	title = {Notes on {Meta}'s {Diplomacy}-{Playing} {AI}},
	url = {https://www.lesswrong.com/posts/oT8fmwWddGwnZbbym/notes-on-meta-s-diplomacy-playing-ai},
	language = {en},
	urldate = {2025-05-24},
	journal = {LessWrong},
	author = {Grunewald, Erich},
	year = {2022},
}

@misc{guoEmbodiedLLMAgents2024,
	title = {Embodied {LLM} {Agents} {Learn} to {Cooperate} in {Organized} {Teams}},
	url = {http://arxiv.org/abs/2403.12482},
	doi = {10.48550/arXiv.2403.12482},
	urldate = {2025-05-24},
	publisher = {arXiv},
	author = {Guo, Xudong and Huang, Kaixuan and Liu, Jiale and Fan, Wenhui and Vélez, Natalia and Wu, Qingyun and Wang, Huazheng and Griffiths, Thomas L. and Wang, Mengdi},
	year = {2024},
	note = {arXiv:2403.12482 [cs]},
}

@article{haggardRelationBrainPotentials1999,
	title = {On the relation between brain potentials and the awareness  of voluntary movements},
	volume = {126},
	issn = {1432-1106},
	url = {https://doi.org/10.1007/s002210050722},
	doi = {10.1007/s002210050722},
	language = {en},
	number = {1},
	urldate = {2025-05-24},
	journal = {Experimental Brain Research},
	author = {Haggard, P. and Eimer, Martin},
	year = {1999},
	pages = {128--133},
}

@techreport{hammond2025,
  title = {Multi-Agent Risks from Advanced AI},
  author = {Hammond, Lewis and Chan, Alan and Clifton, Jesse and Hoelscher-Obermaier, Jason and Khan, Akbir and McLean, Euan and Smith, Chandler and Barfuss, Wolfram and Foerster, Jakob and Gavenčiak, Tomáš and Han, The Anh and Hughes, Edward and Kovařík, Vojtěch and Kulveit, Jan and Leibo, Joel Z. and Oesterheld, Caspar and Schroeder de Witt, Christian and Shah, Nisarg and Wellman, Michael and Bova, Paolo and Cimpeanu, Theodor and Ezell, Carson and Feuillade-Montixi, Quentin and Franklin, Matija and Kran, Esben and Krawczuk, Igor and Lamparth, Max and Lauffer, Niklas and Meinke, Alexander and Motwani, Sumeet and Reuel, Anka and Conitzer, Vincent and Dennis, Michael and Gabriel, Iason and Gleave, Adam and Hadfield, Gillian and Haghtalab, Nika and Kasirzadeh, Atoosa and Krier, Sébastien and Larson, Kate and Lehman, Joel and Parkes, David C. and Piliouras, Georgios and Rahwan, Iyad},
  institution = {Cooperative AI Foundation},
  year = {2025},
  month = {02},
  type = {Technical Report},
  number = {1},
  url = {https://www.cs.toronto.edu/~nisarg/papers/Multi-Agent-Risks-from-Advanced-AI.pdf}
}

@article{hanczycChemicalBasisMinimal2010,
	title = {Chemical {Basis} for {Minimal} {Cognition}},
	volume = {16},
	issn = {1064-5462},
	url = {https://doi.org/10.1162/artl_a_00002},
	doi = {10.1162/artl_a_00002},
	number = {3},
	urldate = {2025-05-24},
	journal = {Artificial Life},
	author = {Hanczyc, Martin M. and Ikegami, Takashi},
	year = {2010},
	pages = {233--243},
}

@article{harrisInstrumentalConvergenceSingleagent2022,
	title = {Instrumental convergence in single-agent systems},
	url = {https://www.lesswrong.com/posts/pGvM95EfNXwBzjNCJ/instrumental-convergence-in-single-agent-systems},
	language = {en},
	urldate = {2025-05-24},
	author = {Harris, Edouard and Suo, Simon S.},
	year = {2022},
}

@inproceedings{hasanDistributedThreatIntelligence2024,
	title = {Distributed {Threat} {Intelligence} at the {Edge} {Devices}: {A} {Large} {Language} {Model}-{Driven} {Approach}},
	shorttitle = {Distributed {Threat} {Intelligence} at the {Edge} {Devices}},
	url = {https://ieeexplore.ieee.org/document/10633440},
	doi = {10.1109/COMPSAC61105.2024.00206},
	urldate = {2025-05-24},
	booktitle = {2024 {IEEE} 48th {Annual} {Computers}, {Software}, and {Applications} {Conference} ({COMPSAC})},
	author = {Hasan, Syed Mhamudul and Alotaibi, Alaa M. and Talukder, Sajedul and Shahid, Abdur R.},
	year = {2024},
	note = {ISSN: 2836-3795},
	pages = {1496--1497},
}

@book{herzogWeLoveWe2010,
	address = {New York},
	title = {Some {We} {Love}, {Some} {We} {Hate}, {Some} {We} {Eat} [{Second} {Edition}]: {Why} {It}'s {So} {Hard} to {Think} {Straight} {About} {Animals}},
	isbn = {978-0-06-311929-1},
	shorttitle = {Some {We} {Love}, {Some} {We} {Hate}, {Some} {We} {Eat} [{Second} {Edition}]},
	language = {English},
	publisher = {Harper Perennial},
	author = {Herzog, Hal},
	year = {2010},
}

@article{heylighen2007,
  title = {The Global Superorganism: An Evolutionary-cybernetic Model of the Emerging Network Society},
  author = {Heylighen, Francis},
  journal = {Social Evolution \& History},
  year = {2007},
  volume = {6},
  number = {1},
  pages = {57--117},
  month = {3},
  publisher = {Uchitel' Publishing House},
  url = {https://www.sociostudies.org/journal/files/seh/2007_1/the_global_superorganism.pdf}
}

@incollection{heylighenMindOutsideBrain2018,
  title     = {Mind Outside Brain: A Radically Non-Dualist Foundation for Distributed Cognition},
  booktitle = {Socially Extended Epistemology},
  author    = {Heylighen, Francis and Beigi, Shima},
  editor    = {Carter, J. Adam and Clark, Andy and Kallestrup, Jesper and Palermos, S. Orestis and Pritchard, Duncan},
  publisher = {Oxford University Press},
  year      = {2018},
  doi       = {10.1093/oso/9780198801764.003.0005},
  isbn      = {978-0-19-880176-4},
  url       = {https://doi.org/10.1093/oso/9780198801764.003.0005},
  urldate   = {2025-05-24}
}

@book{hidalgo2015information,
  author    = {Hidalgo, Cesar},
  title     = {Why Information Grows: The Evolution of Order, from Atoms to Economies},
  year      = {2015},
  publisher = {Basic Books}
}

@techreport{apolloresearchSchemingReasoningEvaluations2024,
	title = {Scheming reasoning evaluations},
	url = {https://www.apolloresearch.ai/research/scheming-reasoning-evaluations},
	language = {en-GB},
	urldate = {2025-05-24},
	institution = {Apollo Research},
	author = {ApolloResearch},
	year = {2024},
}

@misc{meinkeFrontierModelsAre2024,
	title = {Frontier {Models} are {Capable} of {In}-context {Scheming}},
	url = {http://arxiv.org/abs/2412.04984},
	doi = {10.48550/arXiv.2412.04984},
	urldate = {2025-05-24},
	publisher = {arXiv},
	author = {Meinke, Alexander and Schoen, Bronson and Scheurer, Jérémy and Balesni, Mikita and Shah, Rusheb and Hobbhahn, Marius},
	year = {2024},
	note = {arXiv:2412.04984 [cs]},
}

@book{hofstadterGodelEscherBach1979,
	address = {New York},
	edition = {20th-anniversary ed., [Repr.]},
	title = {Gödel, {Escher}, {Bach}: an eternal golden braid},
	isbn = {978-0-465-02656-2},
	shorttitle = {Gödel, {Escher}, {Bach}},
	language = {eng},
	publisher = {Basic Books},
	author = {Hofstadter, Douglas R.},
	year = {1979},
}

@book{hollandAdaptationNaturalArtificial1992,
  title = {Adaptation in Natural and Artificial Systems: An Introductory Analysis with Applications to Biology, Control, and Artificial Intelligence},
  author = {Holland, John H.},
  publisher = {MIT Press},
  year = {1992},
  isbn = {9780262581110},
  pages = {232},
  series = {Complex Adaptive Systems}
}

@misc{ianIlyaSutskeversThoughts2023,
	title = {Ilya {Sutskever}'s {Thoughts} on {Ethical} {AI} and {Its} {Challenges}},
	url = {https://press.farm/ilya-sutskevers-thoughts-on-ethical-ai-challenges/},
	language = {en-US},
	urldate = {2025-05-24},
	journal = {Pressfarm},
	author = {Ian},
	year = {2023},
}

@book{jablonkaEvolutionFourDimensions2005,
	address = {Cambridge, Mass},
	title = {Evolution {In} {Four} {Dimensions}: {Genetic}, {Epigenetic}, {Behavioral}, {And} {Symbolic} {Variation} {In} {The} {History} {Of} {Life}},
	isbn = {978-0-262-10107-3},
	shorttitle = {Evolution {In} {Four} {Dimensions}},
	language = {English},
	publisher = {The MIT Press},
	author = {Jablonka, Eva and Lamb, Marion J.},
	collaborator = {Zeligowski, Anna},
	year = {2005},
}

@article{jacksonEpiphenomenalQualia1982,
	title = {Epiphenomenal {Qualia}},
	volume = {32},
	issn = {0031-8094},
	url = {https://doi.org/10.2307/2960077},
	doi = {10.2307/2960077},
	number = {127},
	urldate = {2025-05-24},
	journal = {The Philosophical Quarterly},
	author = {Jackson, Frank},
	year = {1982},
	pages = {127--136},
}

@article{jiangBrainNetMultiPersonBraintoBrain2019,
	title = {{BrainNet}: {A} {Multi}-{Person} {Brain}-to-{Brain} {Interface} for {Direct} {Collaboration} {Between} {Brains}},
	volume = {9},
	copyright = {2019 The Author(s)},
	issn = {2045-2322},
	shorttitle = {{BrainNet}},
	url = {https://www.nature.com/articles/s41598-019-41895-7},
	doi = {10.1038/s41598-019-41895-7},
	language = {en},
	number = {1},
	urldate = {2025-05-24},
	journal = {Scientific Reports},
	author = {Jiang, Linxing and Stocco, Andrea and Losey, Darby M. and Abernethy, Justin A. and Prat, Chantel S. and Rao, Rajesh P. N.},
	year = {2019},
	note = {Publisher: Nature Publishing Group},
	pages = {6115},
}

@article{kahanIdeologyMotivatedReasoning2013,
	title = {Ideology, motivated reasoning, and cognitive reflection},
	volume = {8},
	issn = {1930-2975},
	url = {https://www.cambridge.org/core/journals/judgment-and-decision-making/article/ideology-motivated-reasoning-and-cognitive-reflection/F8A6A74C9022363D672B0FD14DD8B89F},
	doi = {10.1017/S1930297500005271},
	language = {en},
	number = {4},
	urldate = {2025-05-24},
	journal = {Judgment and Decision Making},
	author = {Kahan, Dan M.},
	year = {2013},
	pages = {407--424},
}

@article{kaiserVerticalFarmingGoes2024,
	title = {Vertical farming goes dynamic: optimizing resource use efficiency, product quality, and energy costs},
	volume = {2},
	issn = {2813-6330},
	shorttitle = {Vertical farming goes dynamic},
	url = {https://www.frontiersin.org/journals/science/articles/10.3389/fsci.2024.1411259/full},
	doi = {10.3389/fsci.2024.1411259},
	language = {English},
	urldate = {2025-05-24},
	journal = {Frontiers in Science},
	author = {Kaiser, Elias and Kusuma, Paul and Vialet-Chabrand, Silvere and Folta, Kevin and Liu, Ying and Poorter, Hendrik and Woning, Nik and Shrestha, Samikshya and Ciarreta, Aitor and van Brenk, Jordan and Karpe, Margarethe and Ji, Yongran and David, Stephan and Zepeda, Cristina and Zhu, Xin-Guang and Huntenburg, Katharina and Verdonk, Julian C. and Woltering, Ernst and Gauthier, Paul P. G. and Courbier, Sarah and Taylor, Gail and Marcelis, Leo F. M.},
	year = {2024},
	note = {Publisher: Frontiers},
}

@article{kanwisherFusiformFaceArea1997,
	title = {The {Fusiform} {Face} {Area}: {A} {Module} in {Human} {Extrastriate} {Cortex} {Specialized} for {Face} {Perception}},
	volume = {17},
	copyright = {Copyright © 1997 Society for Neuroscience},
	issn = {0270-6474, 1529-2401},
	shorttitle = {The {Fusiform} {Face} {Area}},
	url = {https://www.jneurosci.org/content/17/11/4302},
	doi = {10.1523/JNEUROSCI.17-11-04302.1997},
	language = {en},
	number = {11},
	urldate = {2025-05-24},
	journal = {Journal of Neuroscience},
	author = {Kanwisher, Nancy and McDermott, Josh and Chun, Marvin M.},
	year = {1997},
	pmid = {9151747},
	note = {Publisher: Society for Neuroscience
Section: Articles},
	pages = {4302--4311},
}

@book{karbanPlantSensingCommunication2015,
	address = {Chicago, IL},
	series = {Interspecific {Interactions}},
	title = {Plant {Sensing} and {Communication}},
	isbn = {978-0-226-26470-7},
	url = {https://press.uchicago.edu/ucp/books/book/chicago/P/bo20298924.html},
	language = {en},
	urldate = {2025-05-24},
	publisher = {University of Chicago Press},
	author = {Karban, Richard},
	year = {2015},
}

@article{kauffmanMetabolicStabilityEpigenesis1969,
	title = {Metabolic stability and epigenesis in randomly constructed genetic nets},
	volume = {22},
	issn = {0022-5193},
	url = {https://www.sciencedirect.com/science/article/pii/0022519369900150},
	doi = {10.1016/0022-5193(69)90015-0},
	number = {3},
	urldate = {2025-05-24},
	journal = {Journal of Theoretical Biology},
	author = {Kauffman, S. A.},
	year = {1969},
	pages = {437--467},
}

@book{kauffmanOriginsOrderSelfOrganization1993,
	address = {New York},
	title = {The {Origins} of {Order}: {Self}-{Organization} and {Selection} in {Evolution}},
	isbn = {978-0-19-505811-6},
	shorttitle = {The {Origins} of {Order}},
	language = {English},
	publisher = {Oxford University Press},
	author = {Kauffman, Stuart A.},
	year = {1993},
}

@book{kauffmanInvestigations2000,
	title = {Investigations},
	isbn = {978-0-19-512104-9},
	url = {http://archive.org/details/investigations00kauf},
	language = {eng},
	urldate = {2025-05-24},
	publisher = {Oxford ; New York : Oxford University Press},
	author = {Kauffman, Stuart A.},
	collaborator = {{Internet Archive}},
	year = {2000},

}

@article{kauffmanEmergenceAgencyOrganization2006,
	title = {On emergence, agency, and organization},
	volume = {21},
	issn = {1572-8404},
	url = {https://doi.org/10.1007/s10539-005-9003-9},
	doi = {10.1007/s10539-005-9003-9},
	language = {en},
	number = {4},
	urldate = {2025-05-24},
	journal = {Biology and Philosophy},
	author = {Kauffman, Stuart and Clayton, Philip},
	year = {2006},
	pages = {501--521},
}

@book{kellyWhatTechnologyWants2010,
	address = {New York},
	title = {What {Technology} {Wants}},
	isbn = {978-0-670-02215-1},
	language = {English},
	publisher = {Viking},
	author = {Kelly, Kevin},
	year = {2010},
}

@article{koenigsUtilitarianMoralJudgment2012,
	title = {Utilitarian moral judgment in psychopathy},
	volume = {7},
	issn = {1749-5016},
	url = {https://doi.org/10.1093/scan/nsr048},
	doi = {10.1093/scan/nsr048},
	number = {6},
	urldate = {2025-05-24},
	journal = {Social Cognitive and Affective Neuroscience},
	author = {Koenigs, Michael and Kruepke, Michael and Zeier, Joshua and Newman, Joseph P.},
	year = {2012},
	pages = {708--714},
}

@article{kumarEnhancedCO2Fixation2010,
	title = {Enhanced {CO2} fixation and biofuel production via microalgae: recent developments and future directions},
	volume = {28},
	issn = {0167-7799, 1879-3096},
	shorttitle = {Enhanced {CO2} fixation and biofuel production via microalgae},
	url = {https://www.cell.com/trends/biotechnology/abstract/S0167-7799(10)00071-5},
	doi = {10.1016/j.tibtech.2010.04.004},
	language = {English},
	number = {7},
	urldate = {2025-05-24},
	journal = {Trends in Biotechnology},
	author = {Kumar, Amit and Ergas, Sarina and Yuan, Xin and Sahu, Ashish and Zhang, Qiong and Dewulf, Jo and Malcata, F. Xavier and Langenhove, Herman van},
	year = {2010},
	pmid = {20541270},
	note = {Publisher: Elsevier},
	pages = {371--380},
}

@article{lecunPathAutonomousMachine2022,
	title = {A {Path} {Towards} {Autonomous} {Machine} {Intelligence} {Version} 0.9.2, 2022-06-27},
	url = {https://openreview.net/pdf?id=BZ5a1r-kVsf},
	language = {en},
	journal = {openreview.net},
	author = {LeCun, Yann},
	year = {2022},
}

@article{leeEmergenceSymbiosisPeptide1996,
	title = {Emergence of symbiosis in peptide self-replication through a hypercyclic network},
	volume = {390},
	copyright = {1997 Macmillan Magazines Ltd.},
	issn = {1476-4687},
	url = {https://www.nature.com/articles/37569},
	doi = {10.1038/37569},
	language = {en},
	number = {6660},
	urldate = {2025-05-24},
	journal = {Nature},
	author = {Lee, David H. and Severin, Kay and Yokobayashi, Yohei and Ghadiri, M. Reza},
	year = {1996},
	note = {Publisher: Nature Publishing Group},
	pages = {591--594},
}

@article{lenskiEvolutionaryOriginComplex2003,
	title = {The evolutionary origin of complex features},
	volume = {423},
	copyright = {2003 Macmillan Magazines Ltd.},
	issn = {1476-4687},
	url = {https://www.nature.com/articles/nature01568},
	doi = {10.1038/nature01568},
	language = {en},
	number = {6936},
	urldate = {2025-05-24},
	journal = {Nature},
	author = {Lenski, Richard E. and Ofria, Charles and Pennock, Robert T. and Adami, Christoph},
	year = {2003},
	note = {Publisher: Nature Publishing Group},
	pages = {139--144},
}

@misc{levinlabCancerResearch2025,
	title = {Cancer {Research}},
	url = {https://as.tufts.edu/biology/levin-lab/publications/cancer},
	journal = {Tufts University Department of Biology},
	author = {LevinLab},
	year = {2025},
}

@article{levinMorphogeneticFieldsEmbryogenesis2012,
	series = {Biological {Morphogenesis}: {Theory} and {Computation}},
	title = {Morphogenetic fields in embryogenesis, regeneration, and cancer: {Non}-local control of complex patterning},
	volume = {109},
	issn = {0303-2647},
	shorttitle = {Morphogenetic fields in embryogenesis, regeneration, and cancer},
	url = {https://www.sciencedirect.com/science/article/pii/S0303264712000597},
	doi = {10.1016/j.biosystems.2012.04.005},
	number = {3},
	urldate = {2025-05-24},
	journal = {Biosystems},
	author = {Levin, Michael},
	month = sep,
	year = {2012},
	pages = {243--261},
}

@article{levinBioelectricNetworksCognitive2023,
	title = {Bioelectric networks: the cognitive glue enabling evolutionary scaling from physiology to mind},
	volume = {26},
	issn = {1435-9456},
	shorttitle = {Bioelectric networks},
	url = {https://doi.org/10.1007/s10071-023-01780-3},
	doi = {10.1007/s10071-023-01780-3},
	language = {en},
	number = {6},
	urldate = {2025-05-24},
	journal = {Animal Cognition},
	author = {Levin, Michael},
	month = nov,
	year = {2023},
	pages = {1865--1891},
}

@article{levinWhyWeFear2024,
	title = {Why {We} {Fear} {Diverse} {Intelligence} {Like} {AI}},
	url = {https://www.noemamag.com/why-we-fear-diverse-intelligence-like-ai},
	language = {en-US},
	urldate = {2025-05-24},
	author = {Levin, Michael},
	year = {2024},
}

@article{libetTIMECONSCIOUSINTENTION1983,
	title = {{TIME} {OF} {CONSCIOUS} {INTENTION} {TO} {ACT} {IN} {RELATION} {TO} {ONSET} {OF} {CEREBRAL} {ACTIVITY} ({READINESS}-{POTENTIAL}): {THE} {UNCONSCIOUS} {INITIATION} {OF} {A} {FREELY} {VOLUNTARY} {ACT}},
	volume = {106},
	issn = {0006-8950},
	shorttitle = {{TIME} {OF} {CONSCIOUS} {INTENTION} {TO} {ACT} {IN} {RELATION} {TO} {ONSET} {OF} {CEREBRAL} {ACTIVITY} ({READINESS}-{POTENTIAL})},
	url = {https://doi.org/10.1093/brain/106.3.623},
	doi = {10.1093/brain/106.3.623},
	number = {3},
	urldate = {2025-05-24},
	journal = {Brain},
	author = {Libet, Benjamin and Gleason, Curtis A. and Wright, Elwood W. and Pearl, Dennis K.},
	year = {1983},
	pages = {623--642},
}

@article{liederResourcerationalAnalysisUnderstanding2020,
	title = {Resource-rational analysis: {Understanding} human cognition as the optimal use of limited computational resources},
	volume = {43},
	issn = {0140-525X, 1469-1825},
	shorttitle = {Resource-rational analysis},
	url = {https://www.cambridge.org/core/journals/behavioral-and-brain-sciences/article/abs/resourcerational-analysis-understanding-human-cognition-as-the-optimal-use-of-limited-computational-resources/586866D9AD1D1EA7A1EECE217D392F4A#},
	doi = {10.1017/S0140525X1900061X},
	language = {en},
	urldate = {2025-05-24},
	journal = {Behavioral and Brain Sciences},
	author = {Lieder, Falk and Griffiths, Thomas L.},
	year = {2020},
	pages = {e1},
}

@article{lincolnSelfSustainedReplicationRNA2009,
	title = {Self-{Sustained} {Replication} of an {RNA} {Enzyme}},
	volume = {323},
	url = {https://www.science.org/doi/10.1126/science.1167856},
	doi = {10.1126/science.1167856},
	number = {5918},
	urldate = {2025-05-24},
	journal = {Science},
	author = {Lincoln, Tracey A. and Joyce, Gerald F.},
	year = {2009},
	note = {Publisher: American Association for the Advancement of Science},
	pages = {1229--1232},
}

@book{loAdaptiveMarketsFinancial2019,
	address = {Princeton},
	title = {Adaptive {Markets}: {Financial} {Evolution} at the {Speed} of {Thought}},
	isbn = {978-0-691-19136-2},
	shorttitle = {Adaptive {Markets}},
	language = {English},
	publisher = {Princeton University Press},
	author = {Lo, Andrew W.},
	year = {2019},
}

@article{mallapatyHowDoesBrain2025,
	title = {How does the brain control consciousness? {This} deep-brain structure},
	copyright = {2025 Springer Nature Limited},
	issn = {1476-4687},
	shorttitle = {How does the brain control consciousness?},
	url = {https://www.nature.com/articles/d41586-025-01021-2},
	doi = {10.1038/d41586-025-01021-2},
	language = {en},
	urldate = {2025-05-24},
	journal = {Nature},
	author = {Mallapaty, Smriti},
	year = {2025},
	note = {Bandiera\_abtest: a
Cg\_type: News
Publisher: Nature Publishing Group
Subject\_term: Brain, Neuroscience, Psychology},
}

@book{vattimoPlantThinkingPhilosophyVegetal2013,
	title = {Plant-{Thinking}: {A} {Philosophy} of {Vegetal} {Life}},
	isbn = {978-0-231-53325-6},
	shorttitle = {Plant-{Thinking}},
	publisher = {Columbia University Press},
	author = {Vattimo, Michael Marder Foreword by Gianni and Zabala, Santiago},
	year = {2013},
	note = {Pages: 248 Pages},
}

@book{margulisSymbiosisSourceEvolutionary1991,
  editor    = {Margulis, Lynn and Fester, Rene},
  title     = {Symbiosis as a Source of Evolutionary Innovation: Speciation and Morphogenesis},
  publisher = {MIT Press},
  address   = {Cambridge, Mass.},
  year      = {1991},
  isbn      = {978-0-262-13269-5},
  language  = {English}
}

@article{marstallerEvolutionRepresentationSimple2012,
  author    = {Marstaller, Lars and Hintze, Arend and Adami, Christoph},
  title     = {The evolution of representation in simple cognitive networks},
  journal   = {Neural Computation},
  volume    = {25},
  number    = {8},
  pages     = {2079--2107},
  year      = {2012},
  doi       = {10.1162/NECO_a_00475},
  url       = {http://arxiv.org/abs/1206.5771},
  note      = {arXiv:1206.5771 [q-bio]}
}

@article{mayerGutMicrobesBrain2014,
  author    = {Mayer, Emeran A. and Knight, Rob and Mazmanian, Sarkis K. and Cryan, John F. and Tillisch, Kirsten},
  title     = {Gut Microbes and the Brain: Paradigm Shift in Neuroscience},
  journal   = {Journal of Neuroscience},
  volume    = {34},
  number    = {46},
  pages     = {15490--15496},
  year      = {2014},
  doi       = {10.1523/JNEUROSCI.3299-14.2014},
  url       = {https://www.jneurosci.org/content/34/46/15490},
  note      = {Section: Symposium, Publisher: Society for Neuroscience}
}

@article{millar-haskellCouplingSyntheticBiology2019,
	title = {Coupling synthetic biology and programmable materials to construct complex tissue ecosystems},
	volume = {9},
	issn = {2159-6867},
	url = {https://doi.org/10.1557/mrc.2019.69},
	doi = {10.1557/mrc.2019.69},
	language = {en},
	number = {2},
	urldate = {2025-05-24},
	journal = {MRS Communications},
	author = {Millar-Haskell, Catherine S. and Dang, Allyson M. and Gleghorn, Jason P.},
	year = {2019},
	pages = {421--432},
}

@book{minskySocietyMind1988,
	address = {New York},
	title = {The {Society} of {Mind}},
	isbn = {978-0-671-65713-0},
	language = {English},
	publisher = {Simon \& Schuster},
	author = {Minsky, Marvin},
	year = {1988},
}

@book{mitchellComplexityGuidedTour2009,
	address = {New York, NY},
	title = {Complexity: {A} {Guided} {Tour}},
	isbn = {978-0-19-979810-0},
	shorttitle = {Complexity},
	language = {English},
	publisher = {Oxford University Press},
	author = {Mitchell, Melanie},
	year = {2009},
}

@article{moralesSustainedRepresentationPerspectival2020,
	title = {Sustained representation of perspectival shape},
	volume = {117},
	url = {https://www.pnas.org/doi/full/10.1073/pnas.2000715117},
	doi = {10.1073/pnas.2000715117},
	number = {26},
	urldate = {2025-05-24},
	journal = {Proceedings of the National Academy of Sciences},
	author = {Morales, Jorge and Bax, Axel and Firestone, Chaz},
	year = {2020},
	note = {Publisher: Proceedings of the National Academy of Sciences},
	pages = {14873--14882},
}

@book{morenoBiologicalAutonomyPhilosophical2015,
	address = {Dordrecht},
	title = {Biological {Autonomy}: {A} {Philosophical} and {Theoretical} {Enquiry}},
	isbn = {978-94-017-9836-5},
	shorttitle = {Biological {Autonomy}},
	language = {English},
	publisher = {Springer/Sci-Tech/Trade},
	author = {Moreno, Alvaro and Mossio, Matteo},
	year = {2015},
}

@misc{sRoleSelfSupervisedLearning2024,
	title = {The {Role} of {Self}-{Supervised} {Learning} in {LLM} {Development}},
	url = {https://www.goml.io/the-role-of-self-supervised-learning-in-llm-development/},
	language = {en-US},
	urldate = {2025-05-24},
	author = {S, Niruba},
	year = {2024},
	note = {Section: Technology},
}

@article{noutcheuCoppicingDriverPlant2023,
	title = {Coppicing as a driver of plant resprouting and the regeneration of a {Caatinga} dry forest},
	volume = {529},
	issn = {0378-1127},
	url = {https://www.sciencedirect.com/science/article/pii/S0378112722007307},
	doi = {10.1016/j.foreco.2022.120736},
	urldate = {2025-05-24},
	journal = {Forest Ecology and Management},
	author = {Noutcheu, Ronald and Oliveira, Fernanda M. P. and Wirth, Rainer and Tabarelli, Marcelo and Leal, Inara R.},
	year = {2023},
	pages = {120736},
}

@techreport{shulman2010,
  title = {Omohundro's "Basic AI Drives" and Catastrophic Risks},
  author = {Shulman, Carl},
  institution = {Machine Intelligence Research Institute},
  address = {San Francisco, CA},
  year = {2010},
  url = {https://intelligence.org/files/BasicAIDrives.pdf},
  note = {MIRI Visiting Fellow; The Machine Intelligence Research Institute was previously known as the Singularity Institute}
}

@misc{openaiGPT452025,
	title = {{GPT}-4.5},
	url = {https://chat.openai.com/},
	urldate = {2025-05-24},
	author = {OpenAI},
	year = {2025},
}

@article{pezzuloTopdownModelsBiology2016,
	title = {Top-down models in biology: explanation and control of complex living systems above the molecular level},
	volume = {13},
	shorttitle = {Top-down models in biology},
	url = {https://royalsocietypublishing.org/doi/10.1098/rsif.2016.0555},
	doi = {10.1098/rsif.2016.0555},
	number = {124},
	urldate = {2025-05-24},
	journal = {Journal of The Royal Society Interface},
	author = {Pezzulo, Giovanni and Levin, Michael},
	year = {2016},
	note = {Publisher: Royal Society},
	pages = {20160555},
}

@book{prigogineOrderOutChaos1984,
	address = {Toronto},
	title = {Order {Out} of {Chaos}: {Man}'s {New} {Dialogue} with {Nature}},
	isbn = {978-0-553-34082-2},
	shorttitle = {Order {Out} of {Chaos}},
	language = {English},
	publisher = {Bantam New Age Books},
	author = {Prigogine, Ilya.},
	year = {1984},
}

@article{reardonWelcomeCRISPRZoo2016,
	title = {Welcome to the {CRISPR} zoo},
	volume = {531},
	url = {http://www.nature.com/news/welcome-to-the-crispr-zoo-1.19537},
	doi = {10.1038/531160a},
	language = {en},
	number = {7593},
	urldate = {2025-05-24},
	journal = {Nature News},
	author = {Reardon, Sara},
	year = {2016},
	note = {Cg\_type: Nature News
Section: News Feature},
	pages = {160},
}

@article{roserMortalityEverySecond2024,
	title = {Mortality in the past: every second child died},
	shorttitle = {Mortality in the past},
	url = {https://ourworldindata.org/child-mortality-in-the-past},
	language = {en},
	urldate = {2025-05-24},
	journal = {Our World in Data},
	author = {Roser, Max},
	year = {2024},
}

@article{ruganiNumberspaceMappingNewborn2015,
	title = {Number-space mapping in the newborn chick resembles humans’ mental number line},
	volume = {347},
	url = {https://www.science.org/doi/10.1126/science.aaa1379},
	doi = {10.1126/science.aaa1379},
	number = {6221},
	urldate = {2025-05-24},
	journal = {Science},
	author = {Rugani, Rosa and Vallortigara, Giorgio and Priftis, Konstantinos and Regolin, Lucia},
	year = {2015},
	note = {Publisher: American Association for the Advancement of Science},
	pages = {534--536},
}

@book{russellArtificialIntelligenceModern2020,
	address = {Hoboken, NJ},
	title = {Artificial {Intelligence}: {A} {Modern} {Approach}},
	isbn = {978-0-13-461099-3},
	shorttitle = {Artificial {Intelligence}},
	language = {English},
	publisher = {Pearson},
	author = {Russell, Stuart and Norvig, Peter},
	year = {2020},
}

@article{sandhuDaylightSavingsTime2014,
	title = {Daylight savings time and myocardial infarction},
	volume = {1},
	copyright = {undefined undefined},
	issn = {2053-3624},
	url = {https://openheart.bmj.com/content/1/1/e000019},
	doi = {10.1136/openhrt-2013-000019},
	language = {en},
	number = {1},
	urldate = {2025-05-24},
	journal = {Open Heart},
	author = {Sandhu, Amneet and Seth, Milan and Gurm, Hitinder S.},
	year = {2014},
	pmid = {10.1136/openhrt-2013-000019},
	note = {Publisher: British Cardiovascular Society},
}

@book{schellingStrategyConflictNew1981,
	address = {Cambridge, Mass.},
	title = {The {Strategy} of {Conflict}: {With} a {New} {Preface} by the {Author}},
	isbn = {978-0-674-84031-7},
	shorttitle = {The {Strategy} of {Conflict}},
	language = {English},
	publisher = {Harvard University Press},
	author = {Schelling, Thomas C.},
	year = {1981},
}

@article{schotanusCognitiveEconomicsMarket2022,
	title = {Cognitive economics and the {Market} {Mind} {Hypothesis}: {Exploring} the final frontier of economics},
	volume = {42},
	copyright = {© 2022 The Author. Economic Affairs published by John Wiley \& Sons Ltd on behalf of Institute of Economic Affairs.},
	issn = {1468-0270},
	shorttitle = {Cognitive economics and the {Market} {Mind} {Hypothesis}},
	url = {https://onlinelibrary.wiley.com/doi/abs/10.1111/ecaf.12505},
	doi = {10.1111/ecaf.12505},
	language = {en},
	number = {1},
	urldate = {2025-05-24},
	journal = {Economic Affairs},
	author = {Schotanus, Patrick},
	year = {2022},
	note = {\_eprint: https://onlinelibrary.wiley.com/doi/pdf/10.1111/ecaf.12505},
	pages = {87--114},
}

@article{schipperMolecularCooperationThreshold2023,
	title = {Molecular cooperation at the threshold of life {\textbar} {ETH} {Zurich}},
	url = {https://ethz.ch/en/news-and-events/eth-news/news/2023/11/molecular-cooperation-at-the-threshold-of-life.html},
	urldate = {2025-05-24},
	journal = {ETH Zurich},
	author = {Schipper, Ori},
	year = {2023},
}

@article{schurgerSpecificRelationshipShape2012,
	title = {Specific {Relationship} between the {Shape} of the {Readiness} {Potential}, {Subjective} {Decision} {Time}, and {Waiting} {Time} {Predicted} by an {Accumulator} {Model} with {Temporally} {Autocorrelated} {Input} {Noise}},
	volume = {5},
	copyright = {Copyright © 2018 Schurger. This is an open-access article distributed under the terms of the Creative Commons Attribution 4.0 International license, which permits unrestricted use, distribution and reproduction in any medium provided that the original work is properly attributed.},
	issn = {2373-2822},
	url = {https://www.eneuro.org/content/5/1/ENEURO.0302-17.2018},
	doi = {10.1523/ENEURO.0302-17.2018},
	language = {en},
	number = {1},
	urldate = {2025-05-24},
	journal = {eNeuro},
	author = {Schurger, Aaron},
	year = {2012},
	pmid = {29464192},
	note = {Publisher: Society for Neuroscience
Section: New Research},
}

@book{shettleworthCognitionEvolutionBehavior2010,
	address = {New York, NY, US},
	series = {Cognition, evolution, and behavior, 2nd ed},
	title = {Cognition, evolution, and behavior, 2nd ed},
	isbn = {978-0-19-531984-2},
	publisher = {Oxford University Press},
	author = {Shettleworth, Sara J.},
	year = {2010},
	note = {Pages: xiii, 700},
}

@article{shiEnhancingSocialCohesion2024,
	title = {Enhancing social cohesion with cooperative bots in societies of greedy, mobile individuals},
	volume = {3},
	issn = {2752-6542},
	url = {https://doi.org/10.1093/pnasnexus/pgae223},
	doi = {10.1093/pnasnexus/pgae223},
	number = {6},
	urldate = {2025-05-24},
	journal = {PNAS Nexus},
	author = {Shi, Lei and He, Zhixue and Shen, Chen and Tanimoto, Jun},
	year = {2024},
	pages = {pgae223},
}

@misc{siegenfeldFormalDefinitionScaledependent2022,
	title = {A {Formal} {Definition} of {Scale}-dependent {Complexity} and the {Multi}-scale {Law} of {Requisite} {Variety}},
	url = {http://arxiv.org/abs/2206.04896},
	doi = {10.48550/arXiv.2206.04896},
	urldate = {2025-05-24},
	publisher = {arXiv},
	author = {Siegenfeld, Alexander F. and Bar-Yam, Yaneer},
	year = {2022},
	note = {arXiv:2206.04896 [physics]},
}

@book{simardFindingMotherTree2021,
	address = {New York},
	title = {Finding the {Mother} {Tree}: {Discovering} the {Wisdom} of the {Forest}},
	isbn = {978-0-525-65609-8},
	shorttitle = {Finding the {Mother} {Tree}},
	language = {English},
	publisher = {Knopf},
	author = {Simard, Suzanne},
	year = {2021},
}

@article{simardNetTransferCarbon1997,
	title = {Net transfer of carbon between ectomycorrhizal tree species in the field},
	volume = {388},
	copyright = {1997 Macmillan Magazines Ltd.},
	issn = {1476-4687},
	url = {https://www.nature.com/articles/41557},
	doi = {10.1038/41557},
	language = {en},
	number = {6642},
	urldate = {2025-05-24},
	journal = {Nature},
	author = {Simard, Suzanne W. and Perry, David A. and Jones, Melanie D. and Myrold, David D. and Durall, Daniel M. and Molina, Randy},
	year = {1997},
	note = {Publisher: Nature Publishing Group},
	pages = {579--582},
}

@article{smithAutocatalyticSetsPartitioned2014,
	title = {Autocatalytic sets in a partitioned biochemical network},
	volume = {5},
	issn = {1759-2208},
	url = {https://doi.org/10.1186/1759-2208-5-2},
	doi = {10.1186/1759-2208-5-2},
	number = {1},
	urldate = {2025-05-24},
	journal = {Journal of Systems Chemistry},
	author = {Smith, Joshua I. and Steel, Mike and Hordijk, Wim},
	year = {2014},
	pages = {2},
}

@book{solmsHiddenSpringJourney2021,
	address = {New York, NY},
	title = {The {Hidden} {Spring}: {A} {Journey} to the {Source} of {Consciousness}},
	shorttitle = {The {Hidden} {Spring}},
	publisher = {W.W. Norton \& Company},
	author = {Solms, Mark},
	year = {2021},
}

@article{sonnePsychopathyAltruismNeurobiology2018,
	title = {Psychopathy to {Altruism}: {Neurobiology} of the {Selfish}–{Selfless} {Spectrum}},
	volume = {9},
	issn = {1664-1078},
	shorttitle = {Psychopathy to {Altruism}},
	url = {https://www.frontiersin.org/journals/psychology/articles/10.3389/fpsyg.2018.00575/full},
	doi = {10.3389/fpsyg.2018.00575},
	language = {English},
	urldate = {2025-05-24},
	journal = {Frontiers in Psychology},
	author = {Sonne, James W. H. and Gash, Don M.},
	year = {2018},
	note = {Publisher: Frontiers},
}

@article{soonUnconsciousDeterminantsFree2008,
	title = {Unconscious determinants of free decisions in the human brain},
	volume = {11},
	copyright = {2008 Springer Nature America, Inc.},
	issn = {1546-1726},
	url = {https://www.nature.com/articles/nn.2112},
	doi = {10.1038/nn.2112},
	language = {en},
	number = {5},
	urldate = {2025-05-24},
	journal = {Nature Neuroscience},
	author = {Soon, Chun Siong and Brass, Marcel and Heinze, Hans-Jochen and Haynes, John-Dylan},
	year = {2008},
	note = {Publisher: Nature Publishing Group},
	pages = {543--545},
}

@article{sousaAutocatalyticSetsColi2015,
	title = {Autocatalytic sets in {E}. coli metabolism},
	volume = {6},
	issn = {1759-2208},
	url = {https://doi.org/10.1186/s13322-015-0009-7},
	doi = {10.1186/s13322-015-0009-7},
	number = {1},
	urldate = {2025-05-24},
	journal = {Journal of Systems Chemistry},
	author = {Sousa, Filipa L. and Hordijk, Wim and Steel, Mike and Martin, William F.},
	year = {2015},
	pages = {4},
}

@misc{sen.scottSunshineProtectionAct2025,
	title = {Sunshine {Protection} {Act} of 2025},
	copyright = {Text is government work},
	shorttitle = {S.29 - 119th {Congress} (2025-2026)},
	url = {https://www.congress.gov/bill/119th-congress/senate-bill/29},
	language = {eng},
	urldate = {2025-05-24},
	author = {Sen. Scott, Rick [R-FL},
	year = {2025},
	note = {Archive Location: 2025-01-07},
}

@misc{ilyasutskever[@ilyasut]ItMayBe2022,
	type = {Tweet},
	title = {it may be that today's large neural networks are slightly conscious},
	url = {https://x.com/ilyasut/status/1491554478243258368},
	language = {en},
	urldate = {2025-05-24},
	journal = {Twitter},
	author = {{Ilya Sutskever ilyasut}},
	year = {2022},
}

@misc{templetonworldcharityfoundationDiverseIntelligences2025,
	title = {Diverse {Intelligences}},
	url = {https://www.templetonworldcharity.org/our-priorities/discovery/diverse-intelligences},
	language = {en},
	urldate = {2025-05-24},
	journal = {Templeton World Charity Foundation},
	author = {TempletonFoundation},
	year = {2025},
}

@book{thalerNudge2021,
	address = {New Haven London},
	title = {Nudge},
	isbn = {978-0-300-26228-5},
	language = {English},
	publisher = {Yale University Press},
	author = {Thaler, Richard H. and Sunstein, Cass R.},
	year = {2021},
}

@article{toyotaGlutamateTriggersLongdistance2018,
	title = {Glutamate triggers long-distance, calcium-based plant defense signaling},
	volume = {361},
	url = {https://www.science.org/doi/10.1126/science.aat7744},
	doi = {10.1126/science.aat7744},
	number = {6407},
	urldate = {2025-05-24},
	journal = {Science},
	author = {Toyota, Masatsugu and Spencer, Dirk and Sawai-Toyota, Satoe and Jiaqi, Wang and Zhang, Tong and Koo, Abraham J. and Howe, Gregg A. and Gilroy, Simon},
	year = {2018},
	note = {Publisher: American Association for the Advancement of Science},
	pages = {1112--1115},
}

@misc{departmentofenergyAITacklesDisruptive2025,
  title = {AI Tackles Disruptive Tearing Instability in Fusion Plasma},
  author = {{US Department of Energy}},
  howpublished = {US Department of Energy, Fusion Energy Sciences},
  year = {2025},
  url = {https://www.energy.gov/science/fes/articles/ai-tackles-disruptive-tearing-instability-fusion-plasma},
  note = {Accessed May 24, 2025}
}

@misc{ltqryrlmwjzblaarby@UNICEFPressCentre2014,
	title = {{UNICEF} press centre {\textbar} {New} global data expose acute prevalence of violence against children – {UNICEF} {\textbar} {UNICEF}},
	url = {https://www.unicef.org/mena/press-releases/new-global-data-on-violence-against-children},
	language = {en},
	urldate = {2025-05-24},
	journal = {Unicef.org},
	author = {Taqrir al-Mukhtasar bil-‘Arabīyah},
	year = {2014},
}

@article{vandongenPowerEfficientMultichannelNeural2016,
	title = {A {Power}-{Efficient} {Multichannel} {Neural} {Stimulator} {Using} {High}-{Frequency} {Pulsed} {Excitation} {From} an {Unfiltered} {Dynamic} {Supply}},
	volume = {10},
	issn = {1940-9990},
	url = {https://ieeexplore.ieee.org/document/6965660},
	doi = {10.1109/TBCAS.2014.2363736},
	number = {1},
	urldate = {2025-05-24},
	journal = {IEEE Transactions on Biomedical Circuits and Systems},
	author = {van Dongen, Marijn N. and Serdijn, Wouter A.},
	year = {2016},
	pages = {61--71},
}

@book{varelaEmbodiedMindCognitive1991,
	address = {Cambridge, Mass.},
	title = {The {Embodied} {Mind}: {Cognitive} {Science} and {Human} {Experience}},
	isbn = {978-0-262-72021-2},
	shorttitle = {The {Embodied} {Mind}},
	language = {English},
	publisher = {Mit Pr},
	author = {Varela, Francisco J. and Thompson, Evan T. and Rosch, Eleanor},
	year = {1991},
}

@book{vedralDecodingRealityUniverse2018,
	address = {Oxford},
	title = {Decoding {Reality}: {The} {Universe} as {Quantum} {Information}},
	isbn = {978-0-19-881543-3},
	shorttitle = {Decoding {Reality}},
	language = {English},
	publisher = {Oxford University Press},
	author = {Vedral, Vlatko},
	year = {2018},
}

@misc{gigovaWhoPutinThinks2017,
	title = {Who {Putin} thinks will rule the world},
	url = {https://www.cnn.com/2017/09/01/world/putin-artificial-intelligence-will-rule-world},
	language = {en},
	urldate = {2025-05-24},
	journal = {CNN},
	author = {Gigova, Radina},
	year = {2017},
}

@book{neumannTheoryGamesEconomic2007,
	address = {Princeton, NJ},
	title = {Theory of {Games} and {Economic} {Behavior}:},
	isbn = {978-0-691-13061-3},
	shorttitle = {Theory of {Games} and {Economic} {Behavior}},
	language = {English},
	publisher = {Princeton University Press},
	author = {Neumann, John von and Morgenstern, Oskar and Rubinstein, Ariel and Kuhn, Harold W.},
	year = {2007},
}

@misc{wangROMAMultiAgentReinforcement2020,
	title = {{ROMA}: {Multi}-{Agent} {Reinforcement} {Learning} with {Emergent} {Roles}},
	shorttitle = {{ROMA}},
	url = {http://arxiv.org/abs/2003.08039},
	doi = {10.48550/arXiv.2003.08039},
	urldate = {2025-05-24},
	publisher = {arXiv},
	author = {Wang, Tonghan and Dong, Heng and Lesser, Victor and Zhang, Chongjie},
	year = {2020},
	note = {arXiv:2003.08039 [cs]},
}

@article{wardLocalInteractionsGlobal2017,
	title = {Local interactions and global properties of wild, free-ranging stickleback shoals},
	volume = {4},
	url = {https://royalsocietypublishing.org/doi/10.1098/rsos.170043},
	doi = {10.1098/rsos.170043},
	number = {7},
	urldate = {2025-05-24},
	journal = {Royal Society Open Science},
	author = {Ward, Ashley J. W. and Schaerf, Timothy M. and Herbert-Read, James E. and Morrell, Lesley and Sumpter, David J. T. and Webster, Mike M.},
	year = {2017},
	note = {Publisher: Royal Society},
	pages = {170043},
}

@book{weberEconomySocietyNew1922,
	address = {Cambridge, Massachusetts},
	title = {Economy and {Society}: {A} {New} {Translation}},
	isbn = {978-0-674-91654-8},
	shorttitle = {Economy and {Society}},
	language = {English},
	publisher = {Harvard University Press},
	author = {Weber, Max},
	translator = {Tribe, Keith},
	year = {1922},
}

@book{wendtQuantumMindSocial2015,
	address = {Cambridge, United Kingdom ; New York},
	title = {Quantum {Mind} and {Social} {Science}},
	isbn = {978-1-107-44292-4},
	language = {English},
	publisher = {Cambridge University Press},
	author = {Wendt, Alexander},
	year = {2015},
}

@article{wissner-grossCausalEntropicForces2013,
	title = {Causal {Entropic} {Forces}},
	volume = {110},
	url = {https://link.aps.org/doi/10.1103/PhysRevLett.110.168702},
	doi = {10.1103/PhysRevLett.110.168702},
	number = {16},
	urldate = {2025-05-24},
	journal = {Physical Review Letters},
	author = {Wissner-Gross, A. D. and Freer, C. E.},
	year = {2013},
	note = {Publisher: American Physical Society},
	pages = {168702},
}

@misc{xaiGrok2024,
	title = {Grok},
	url = {https://grok.x.ai/},
	urldate = {2025-05-24},
	author = {xAI},
	year = {2024},
}

@misc{zhongPanaceaParetoAlignment2024,
	title = {Panacea: {Pareto} {Alignment} via {Preference} {Adaptation} for {LLMs}},
	shorttitle = {Panacea},
	url = {http://arxiv.org/abs/2402.02030},
	doi = {10.48550/arXiv.2402.02030},
	urldate = {2025-05-24},
	publisher = {arXiv},
	author = {Zhong, Yifan and Ma, Chengdong and Zhang, Xiaoyuan and Yang, Ziran and Chen, Haojun and Zhang, Qingfu and Qi, Siyuan and Yang, Yaodong},
	year = {2024},
	note = {arXiv:2402.02030 [cs]},
}

@article{lagziMazeSolvingChemotactic2010,
	title = {Maze {Solving} by {Chemotactic} {Droplets}},
	volume = {132},
	issn = {0002-7863},
	url = {https://doi.org/10.1021/ja9076793},
	doi = {10.1021/ja9076793},
	number = {4},
	urldate = {2025-05-24},
	journal = {Journal of the American Chemical Society},
	author = {Lagzi, István and Soh, Siowling and Wesson, Paul J. and Browne, Kevin P. and Grzybowski, Bartosz A.},
	year = {2010},
	note = {Publisher: American Chemical Society},
	pages = {1198--1199},
}

@article{mcmillenCollectiveIntelligence2024,
  author    = {McMillen, Patrick and Levin, Michael},
  title     = {Collective intelligence: A unifying concept for integrating biology across scales and substrates},
  journal   = {Communications Biology},
  volume    = {7},
  number    = {1},
  pages     = {1--17},
  year      = {2024},
  doi       = {10.1038/s42003-024-06037-4},
  url       = {https://www.nature.com/articles/s42003-024-06037-4},
  publisher = {Nature Publishing Group},
  language  = {en}
}

@article{myersMeasuringIntelligenceCell2024,
  author    = {Myers, Judith},
  title     = {Measuring the Intelligence of a Cell},
  journal   = {UC San Diego Today},
  year      = {2024},
  url       = {https://today.ucsd.edu/story/measuring-the-intelligence-of-a-cell},
  language  = {en},
  note      = {Accessed May 25, 2025}
}

@misc{levinDennettAeon2020,
  author    = {Levin, Michael and Dennett, Daniel},
  title     = {Cognition All the Way Down},
  year      = {2020},
  url       = {https://aeon.co/essays/how-to-understand-cells-tissues-and-organisms-as-agents-with-agendas},
  journal   = {Aeon},
  note      = {Essay retrieved May 25, 2025}
}

@article{krallEconomicSuperorganism2023,
  author    = {Krall, Lisi},
  title     = {The Economic Superorganism in the Complexity of Evolution},
  journal   = {Philosophical Transactions of the Royal Society B: Biological Sciences},
  year      = {2023},
  volume    = {378},
  number    = {1872},
  pages     = {20210417},
  doi       = {10.1098/rstb.2021.0417},
  url       = {https://royalsocietypublishing.org/doi/10.1098/rstb.2021.0417},
  note      = {Publisher: Royal Society}
}

@misc{benyusBiomimicryInstitute,
  author    = {Benyus, Janine},
  title     = {Biomimicry Institute},
  year      = {2025},
  url       = {https://biomimicry.org/janine-benyus/},
  note      = {Accessed May 26, 2025. Biomimicry profile page from The Biomimicry Institute}
}

@article{vincentBiomimeticsItsPractice2006,
  author  = {Vincent, Julian F. V. and Bogatyreva, Olga A. and Bogatyrev, Nikolaj R. and Bowyer, Adrian and Pahl, Anja-Karina},
  title   = {Biomimetics: Its Practice and Theory},
  journal = {Journal of The Royal Society Interface},
  volume  = {3},
  number  = {9},
  pages   = {471--482},
  year    = {2006},
  doi     = {10.1098/rsif.2006.0127},
  url     = {https://royalsocietypublishing.org/doi/10.1098/rsif.2006.0127},
  note    = {Publisher: Royal Society}
}

@article{rampelottoExtremophilesExtremeEnvironments2013,
  author  = {Rampelotto, Pabulo Henrique},
  title   = {Extremophiles and Extreme Environments},
  journal = {Life},
  volume  = {3},
  number  = {3},
  pages   = {482--485},
  year    = {2013},
  doi     = {10.3390/life3030482},
  url     = {https://www.ncbi.nlm.nih.gov/pmc/articles/PMC4187170/},
  issn    = {2075-1729}
}

@article{berkesRediscoveryTraditionalEcological2000,
  author  = {Berkes, Fikret and Colding, Johan and Folke, Carl},
  title   = {Rediscovery of Traditional Ecological Knowledge as Adaptive Management},
  journal = {Ecological Applications},
  volume  = {10},
  number  = {5},
  pages   = {1251--1262},
  year    = {2000},
  doi     = {10.1890/1051-0761(2000)010[1251:ROTEKA]2.0.CO;2},
  url     = {https://onlinelibrary.wiley.com/doi/abs/10.1890/1051-0761(2000)010[1251:ROTEKA]2.0.CO;2}
}

@article{turnerTraditionalEcologicalKnowledge2000,
  author  = {Turner, Nancy J. and Ignace, Marianne Boelscher and Ignace, Ronald},
  title   = {Traditional Ecological Knowledge and Wisdom of Aboriginal Peoples in British Columbia},
  journal = {Ecological Applications},
  volume  = {10},
  number  = {5},
  pages   = {1275--1287},
  year    = {2000},
  doi     = {10.1890/1051-0761(2000)010[1275:TEKAWO]2.0.CO;2},
  url     = {https://onlinelibrary.wiley.com/doi/abs/10.1890/1051-0761(2000)010[1275:TEKAWO]2.0.CO;2}
}

@article{trewavas2017,
  author  = {Trewavas, Anthony},
  title   = {The foundations of plant intelligence},
  journal = {Interface Focus},
  volume  = {7},
  number  = {3},
  year    = {2017},
  pages   = {20160098},
  doi     = {10.1098/rsfs.2016.0098},
  url     = {https://royalsocietypublishing.org/doi/10.1098/rsfs.2016.0098}
}

@article{wolman2012,
  title = {The split brain: A tale of two halves},
  author = {Wolman, David},
  journal = {Nature},
  year = {2012},
  volume = {483},
  pages = {260--263},
  doi = {10.1038/483260a},
  url = {https://doi.org/10.1038/483260a}
}

@misc{saf2014,
  author  = {{Scientific American Frontiers}},
  title   = {Split Brain - Hosted by Alan Alda},
  year    = {2014},
  note    = {YouTube video},
  url     = {https://www.youtube.com/watch?v=Hd32_w6oqNI}
}

@book{ball2023,
  author    = {Ball, Philip},
  title     = {How Life Works: A User’s Guide to the New Biology},
  year      = {2023},
  publisher = {University of Chicago Press},
  isbn      = {9780226826684},
  url       = {https://www.amazon.ca/How-Life-Works-Users-Biology/dp/0226826686}
}

@article{gazzaniga2000,
  author  = {Gazzaniga, Michael S.},
  title   = {Cerebral specialization and interhemispheric communication: Does the corpus callosum enable the human condition?},
  journal = {Brain},
  volume  = {123},
  number  = {7},
  pages   = {1293--1326},
  year    = {2000},
  doi     = {10.1093/brain/123.7.1293},
  url     = {https://academic.oup.com/brain/article/123/7/1293/380106}
}

@article{bastian2012,
  title = {Don't Mind Meat? The Denial of Mind to Animals Used for Human Consumption},
  author = {Bastian, Brock and Loughnan, Steve and Haslam, Nick and Radke, Helena R. M.},
  journal = {Personality and Social Psychology Bulletin},
  year = {2012},
  volume = {38},
  number = {2},
  pages = {247--256},
  doi = {10.1177/0146167211424291}
}

@online{pilgrim2005,
  author  = {Pilgrim, David},
  title   = {Drapetomania},
  year    = {2005},
  url     = {https://jimcrowmuseum.ferris.edu/question/2005/november.htm},
  note    = {Jim Crow Museum of Racist Memorabilia}
}

@article{dafoe2021,
  author  = {Dafoe, Allan and Bachrach, Yoram and Hadfield, Gillian and Horvitz, Eric and Larson, Kate and Graepel, Thore},
  title   = {Cooperative AI: Machines must learn to find common ground},
  journal = {Nature},
  volume  = {593},
  pages   = {33--36},
  year    = {2021},
  doi     = {10.1038/d41586-021-01170-0},
  url     = {https://www.nature.com/articles/d41586-021-01170-0}
}

@article{hua2023,
  title = {War and Peace (WarAgent): Large Language Model-based Multi-Agent Simulation of World Wars},
  author = {Hua, Wenyue and Fan, Lizhou and Li, Lingyao and Mei, Kai and Ji, Jianchao and Ge, Yingqiang and Hemphill, Libby and Zhang, Yongfeng},
  journal = {arXiv preprint arXiv:2311.17227},
  year = {2023},
  url = {https://doi.org/10.48550/arXiv.2311.17227},
  doi = {10.48550/arXiv.2311.17227}
}

@misc{palisadeThreeModels2025,
    author = {{Palisade Research}},
    title = {Three models ignored the instruction and successfully sabotaged the shutdown script at least once: Codex-mini, o3, o4-mini},
    year = {2025},
    month = {05},
    day = {24},
    howpublished = {Tweet},
    note = {X (formerly Twitter)},
    url = {https://x.com/PalisadeAI/status/1926084640487375185}
}

@article{silver2021reward,
  title={Reward is enough},
  author={Silver, David and Singh, Satinder and Precup, Doina and Sutton, Richard S. and Hassabis, Demis},
  journal={Artificial Intelligence},
  volume={299},
  pages={103535},
  year={2021},
  publisher={Elsevier},
  doi={10.1016/j.artint.2021.103535},
  url={https://doi.org/10.1016/j.artint.2021.103535}
}

@article{Mnih2015HumanlevelCT,
  title={Human-level control through deep reinforcement learning},
  author={Volodymyr Mnih and Koray Kavukcuoglu and David Silver and Andrei A. Rusu and Joel Veness and Marc G. Bellemare and Alex Graves and Martin A. Riedmiller and Andreas Kirkeby Fidjeland and Georg Ostrovski and Stig Petersen and Charlie Beattie and Amir Sadik and Ioannis Antonoglou and Helen King and Dharshan Kumaran and Daan Wierstra and Shane Legg and Demis Hassabis},
  journal={Nature},
  year={2015},
  volume={518},
  pages={529-533},
  url={https://api.semanticscholar.org/CorpusID:205242740}
}

@article{gemini2.5pro.20252025,
    title = {Gemini 2.5 Pro Preview},
    volume = {},
    doi = {https://deepmind.google/technologies/gemini/pro/},
    author = {{Google DeepMind}},
    year = {2025},
}

@article{bar-yam2004multiscale,
  title={Multiscale variety in complex systems},
  author={Bar-Yam, Yaneer},
  journal={Complexity},
  volume={9},
  number={4},
  pages={37--45},
  year={2004},
  publisher={Wiley}
}

@misc{levin2025,
  title = {The Future of Drugs},
  author = {Levin, Michael},
  howpublished = {YouTube video},
  year = {2025},
  url = {https://youtube.com/shorts/vs7-pnySyVw?feature=share},
  note = {Variable Minds}
}

@book{marder2013,
  title = {Plant-Thinking: A Philosophy of Vegetal Life},
  author = {Marder, Michael},
  publisher = {Columbia University Press},
  year = {2013},
  month = {2},
  isbn = {9780231161251},
  pages = {248}
}

@misc{harris2021final,
  author = {Harris, Sam},
  title = {Final Thoughts on Free Will},
  howpublished = {Making Sense with Sam Harris [Podcast]},
  number = {241},
  year = {2021},
  month = {03},
  day = {12},
  url = {https://www.samharris.org/podcasts/making-sense-episodes/241-final-thoughts-on-free-will},
  note = {Episode 241}
}

@article{sender2016revised,
  author = {Sender, Ron and Fuchs, Shai and Milo, Ron},
  title = {Revised Estimates for the Number of Human and Bacteria Cells in the Body},
  journal = {PLoS Biology},
  volume = {14},
  number = {8},
  pages = {e1002533},
  year = {2016},
  doi = {10.1371/journal.pbio.1002533},
  url = {https://doi.org/10.1371/journal.pbio.1002533}
}

\end{document}